\documentclass[a4paper,11pt]{article}

\usepackage[utf8]{inputenc}
\usepackage[english]{babel}
\usepackage{wasysym}
\usepackage{textcomp}
\usepackage{graphicx} 
\usepackage{caption}
\usepackage{wrapfig}
\usepackage{array} 
\usepackage[compat=1.0.0]{tikz-feynman}
\usepackage[export]{adjustbox}
\usepackage{multirow}
\usepackage{sectsty}
\allsectionsfont{\sffamily\mdseries\upshape} 

\usepackage{geometry}
\title{
\includegraphics[width=.25\textwidth]{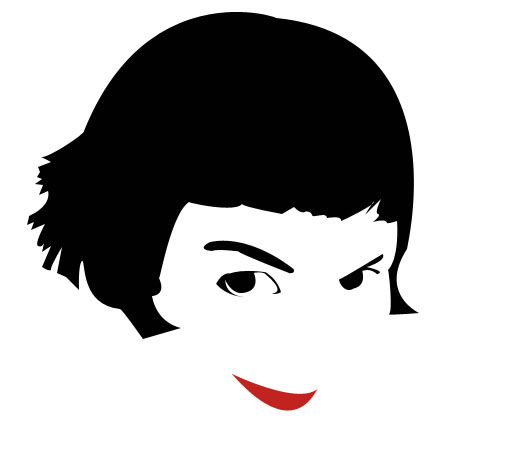} \\
A.M.E.L.I.E. \\
\large Apparatus for Muon Experimental Lifetime Investigation and Evaluation}
\author{Angelo Maggiora\footnote{INFN, Torino Section, retired researcher}
}

\begin{document}

\maketitle
\tableofcontents

\begin{abstract}
 The muon is one of the first elementary particles discovered. It is also known as heavy electron, and it's the main component of cosmic rays flux at sea level. Its flow is continuous, 24h/7d, 
 and it is free. 
It is natural and does not have any radio protection banning or limitation to its use in schools 
and can be managed safely by the students.
We propose to build a light, small and didactic apparatus to measure the lifetime of the muons.
It is useful tool to introduce the modern physics, particle physics, particles instability and decay, special relativity etc.
It can be used for small didactic but complete experiments for measurement of  muon rate and lifetime, correction and equalization of data collected etc.
A useful instrument to introduce and teach the scientific method to the students. 
Last but not least, do not contain any dangerous system like high voltage or explosive gas and 
the cost must be cheap.
\end{abstract}

\newpage
\section{Introduction.}

\begin{figure}[hbt]
\centering
  \begin{minipage}[]{.54\textwidth}
     \centering
     \includegraphics[width=.6\textwidth]{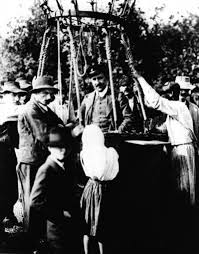}
  \end{minipage} 
  \hfill
  \begin{minipage}[]{.44\textwidth}
     \centering
     \includegraphics[width=.95\textwidth]{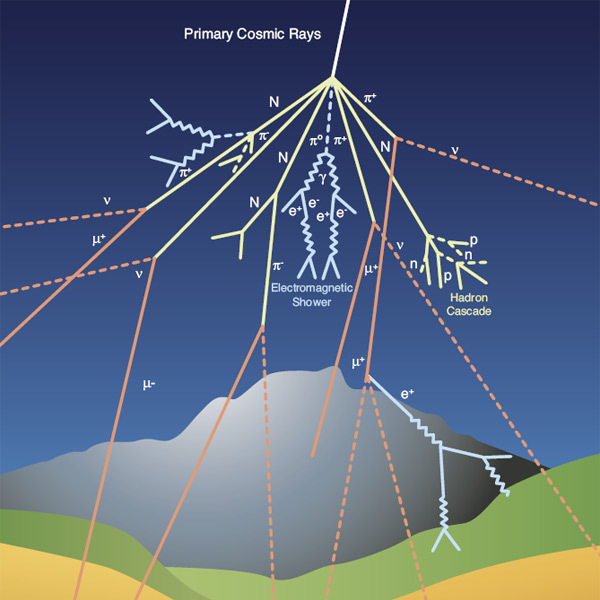}
  \end{minipage} 
\caption{Left: Austrian physicist Victor Hess on a hydrogen balloon ascension up to 5300m, 1912 ; Right: the cosmic rays shower} 
\label{Fig:Hess-shower}
\end{figure}

Our earth is continuously bombarded by a stream of elementary particles called cosmic rays.
They come mainly from the sun but also from deep cosmic space, supernovae, pulsar etc.
The cosmic rays were discovered in 1912 by Victor Hess by bringing an electroscope in balloon 
ascents, see Fig~\ref{Fig:Hess-shower} left.

In the November, 1912 issue of the German journal Physikalische Zeitschrift, Hess suggested:

\begin{center}
   \begin{minipage}{0.7\textwidth}
     "The results  of my observation are best explained by the assumption
     that a radiation of very great penetrating power enters our atmosphere from above."
   \end{minipage}
\end{center}

The mechanism of formation of cosmic rays arriving on earth is now well known, after more than 100 years of research, although their origin is still the subject of research.
The composition of primary cosmic rays is: about 90\% simple protons (i.e., hydrogen nuclei); 9\% alpha particles, identical to helium nuclei; and 1\% the nuclei of heavier elements. These fractions vary highly over the energy range of cosmic rays. A very small fraction are stable particles of antimatter, such as positrons or antiprotons  \cite{Ams} \cite{Pamela}.
Further details are shown in Chap.~\ref{Sec:CosmRayComp}.

When it reach the stratosphere, the proton interacts with some nucleus of our atmosphere 
and produces a shower of pions, gammas, neutrinos etc. see Fig~\ref{Fig:Hess-shower}.
The only particle that survives at sea level, thanks to relativistic time dilation, is the muon.

The purpose of this paper is to describe an innovative particle detector called \break
A.M.E.L.I.E. (Apparatus for Muon Experimental Lifetime Investigation and Evaluation),
dedicated to measuring the average lifetime of muons that stop within it.
We will show how the design and construction of this didactic detector, small, 
simple, inexpensive, allows high school students to approach and understand how physics 
measurements are made.

In this way, they are taught the principles of the Galilean scientific method, 
the collection of data, their possible correction, the analysis and the fit of 
the data using mathematical functions and the rudiments of statistics.

The paper is organized as follows: 
In Chap.~\ref{Sec:CosmRay}, cosmic ray physics is briefly covered, followed by
the physics of the muon in Chap.~\ref{Sec:physmuon}.
AMELIE's conceptual design is described in Chap.~\ref{Sec:apparatus}, together with
the fluka simulations of muon decay time, the optical collection, uniformity and 
electronic simulations using PSPice.
This is followed by chapters on the description of data taking, quality check, data analysis 
and muon physics in nuclear matter, see Chap.~\ref{Sec:analisys}.
Finally, Chap.~\ref{Sec:conclusions} contains the conclusions and general considerations 
on the purpose, quality and educational value of the apparatus.

\section{The physics of primary cosmic rays.}
\label{Sec:CosmRay}

Cosmic rays are high energy charged particles, originating in outer space, 
that travel at nearly the speed of light and strike the Earth from all directions.

The term "cosmic rays" usually refers to galactic cosmic rays, which originate in 
sources outside the solar system, distributed throughout our Milky Way galaxy. 

However, this term has also come to include other classes of energetic particles 
in space, including nuclei and electrons accelerated in association with 
energetic events on the Sun, solar flares etc.

Since cosmic rays are charged, positively charged protons or nuclei 
or negatively charged electrons, their paths through space can be deflected
by magnetic fields (except for the highest energy cosmic rays). 
On their journey to Earth, the magnetic fields of the galaxy, the solar system, 
and the Earth scramble their flight paths so much that we can no longer know 
exactly where they came from. 

\subsection{Composition of cosmic rays.}
\label{Sec:CosmRayComp}

\begin{wrapfigure}{R}{0.60\textwidth}
  \includegraphics[width=0.98\linewidth]{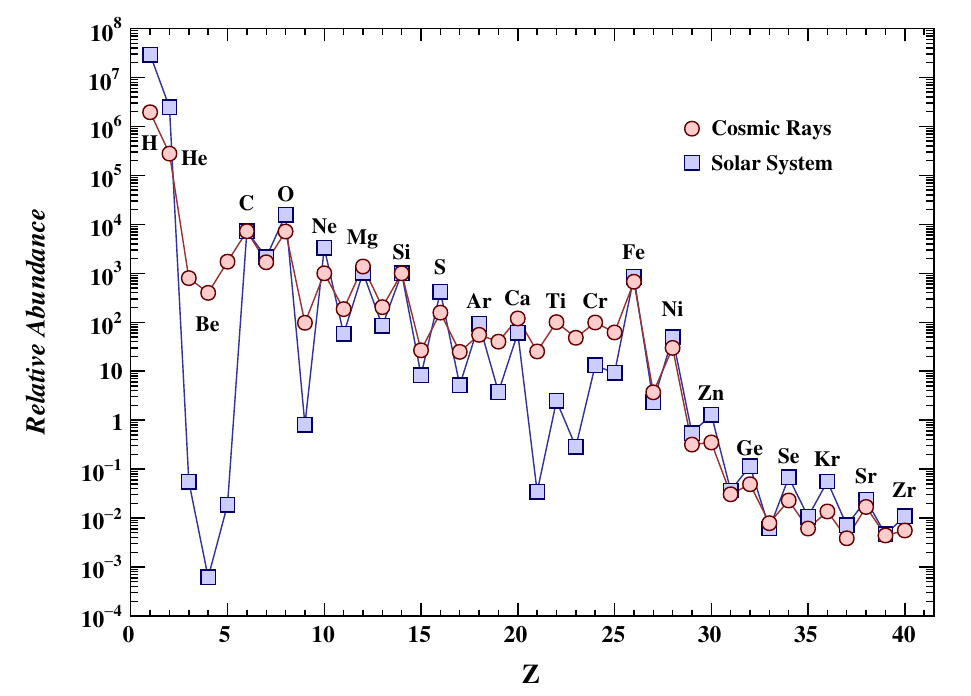} 
  \caption{Cosmic ray elemental abundances compared to abundances in present-day
  solar system material. Abundances are normalised to Si=103.}
  \label{Fig:PrimaryPart}
\end{wrapfigure}

Cosmic rays include essentially all of the elements in the periodic table; 
about 89\% of the nuclei are hydrogen (protons), 10\% helium, and about 1\% 
heavier elements. The relative composition of primary cosmic rays is shown in Fig.~\ref{Fig:PrimaryPart}. 
A broad and exhaustive description is provided by~\cite{PDG1}.

Cosmic ray abundances are from AMS-02 (H,He) (\protect\cite{AMS1}, 
\protect\cite{AMS2}), ACE/CRIS (Li-Ni) (\protect\cite{ACE1}, \protect\cite{ACE2}), 
and TIGER/SuperTIGER (Cu-Zr) (\protect\cite{TIGER1}, \protect\cite{TIGER2}). 
Solar system abundances are from Table 6  of Ref.~\protect\cite{Lodders}.

The common heavier elements (such as carbon, oxygen, magnesium, silicon, and iron) 
are present in about the same relative abundances as in the solar system, 
but there are important differences in elemental and isotopic composition 
that provide information on the origin and history of galactic cosmic rays.

Electrons constitute about 1\% of galactic cosmic rays.

\subsection{Energy of cosmic rays.}

The range of energies encompassed by cosmic rays is truly enormous, starting at about $10^7$ eV and 
reaching $10^{20}$ eV for the most energetic cosmic ray ever detected, which can be millions of times 
the energies reached at the LHC.

The energies of the order of $10^{20}$ correspond to the energy of a golf ball falling from a table, but, in this case, this energy is concentrated in a single particle with a diameter of less 
than $10^{-15}$ m.

\begin{figure}[hbt]
\centering
  \begin{minipage}[]{.4\textwidth}
     \centering
     \includegraphics[width=.9\textwidth]{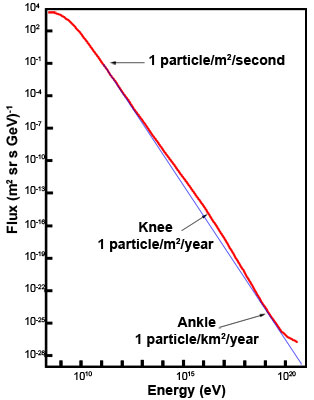}
  \end{minipage} 
  \hfill
  \begin{minipage}[]{.58\textwidth}
     \centering
     \includegraphics[width=.9\textwidth]{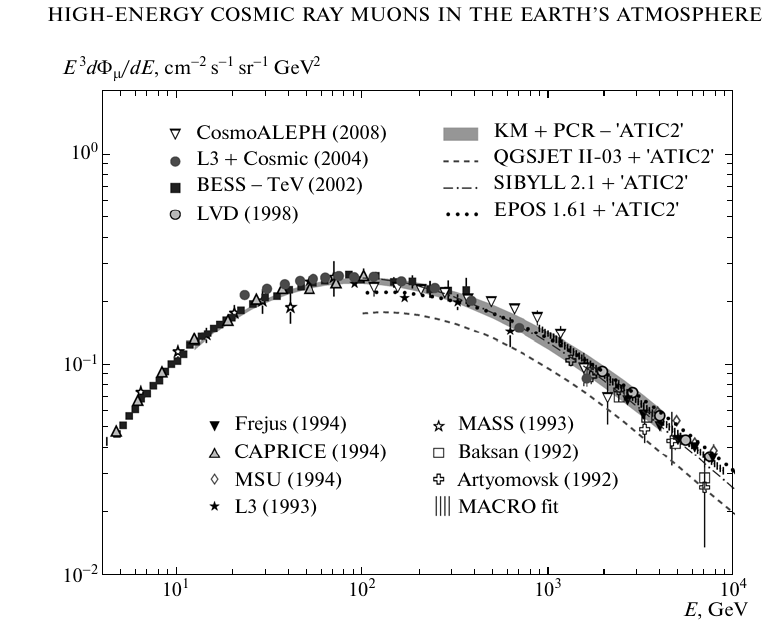}
  \end{minipage} 
\caption{Left: The cosmic ray energy spectrum clearly shows that the cosmic 
         ray flux      
         drops off dramatically as we go to higher energies. 
         The spectrum exhibits a ‘knee’ and an 
         ‘ankle’, both of which deviate from the standard exponential 
         decline (blue line) \cite{Auger}. 
         Right: Spectrum of muons at $\theta = 0^o$ with the exception 
         that hollow diamond markers represent data at
         $\theta = 75^o$. 
         Different markers represent data obtained from different experiments. 
         The solid line plots Eq.~\ref{Eq:EnergySea}. 
         This graph is obtained from the particle data group 
         \protect\cite{Kochanov}.} 
\label{Fig:Energy}
\end{figure}

By plotting this range of energies against the number of cosmic rays detected at 
each energy we generate a cosmic ray spectrum which clearly shows that the number 
of cosmic rays drops off dramatically as we go to higher energies, 
see Fig~\ref{Fig:Energy}.

Roughly speaking, for every 10\ increase in energy beyond $10^9$ eV, 
the number of cosmic rays per unit area falls by a factor of 1,000. However, if we 
look at the spectrum more closely we can see a knee at ~ $10^{15}$ eV and an ankle \`uat ~ $10^{18}$ eV.

The origin of these changes in the steepness of the spectrum is still the subject of intense study, 
but it is assumed that they distinguish between populations of cosmic rays originating via different 
mechanisms. Current suggestions are that the cosmic rays with energies less than about $10^{10}$ eV 
are primarily solar cosmic rays produced in solar flares and coronal mass ejections, while those with 
energies between $10^{10}$ eV and the knee at $10^{15}$ eV are galactic cosmic rays produced in the 
shocks of supernova remnants. The origin of the cosmic rays with energies between the knee and ankle is unclear. They are again thought to be produced within the Galaxy, 
but the energies are too high for them to be accelerated by the shocks of supernova remnants. 
The origin for the ultra-high energy cosmic rays (UHECRs) below the ankle is also a
mystery, but many have suggested that these may be created outside of our Galaxy.

As previously mentioned, the primary cosmic rays interact with the nuclei of the stratosphere, rarefied, 
and generate a shower of particles. At sea level, all the particles are muons from the $\pi$ decay.

\begin{figure}[hbt]
\centering
  \begin{minipage}[]{.48\textwidth}
     \centering 
     \begin{equation}
       \pi^+ \rightarrow \mu^+ + \nu_\mu 
     \end{equation}
  \end{minipage}
    \hfill
  \begin{minipage}[]{.48\textwidth}
     \centering
     \begin{equation}
        \pi^- \rightarrow \mu^- + \overline{\nu}_\mu 
     \end{equation}
  \end{minipage}
\end{figure}

The energy of the mouns is shown in Fig.~\ref{Fig:Energy}, right.

\section{Physics of the muons.}
\label{Sec:physmuon}

\begin{wrapfigure}{R}{0.50\textwidth}
  \includegraphics[width=0.98\linewidth]{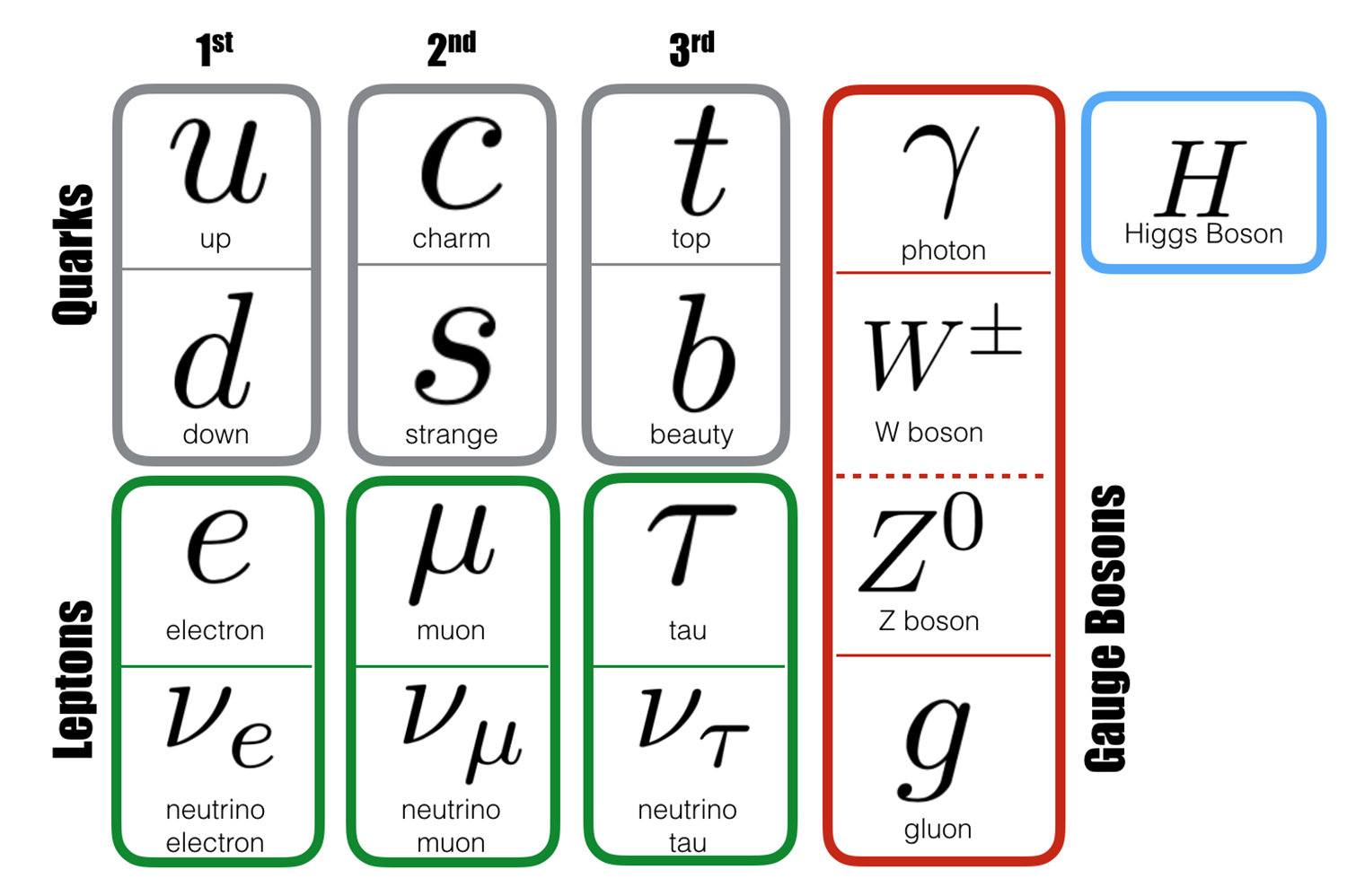} 
  \caption{The Standard model, the basic constituents of the matter.}
  \label{Fig:StandModel}
\end{wrapfigure}

The fundamental constituent of the matter are shown Fig.~\ref{Fig:StandModel}.
They can be divided into two categories called hadrons and leptons, 
to which five bosons mediating the interactions are added.
The figure does not include the ``graviton'' although physicists believe it should be 
added to existing bosons, and a unified theory including gravity has not yet been 
formulated. 
Gravitational waves , which are a manifestation of the existence of the ``graviton'', 
have only been discovered in recent years.

The hadrons are defined as the particles that interact through strong
interaction and they are constituted by quarks.
Quarks are bounded up into hadrons, forming the baryons (semi-integer spin)
such as protons and neutrons and the mesons (integer spin) such as the 
pions and kaons.

The leptons category include electrons, muons and their associated neutrinos. 
As the electrons, the muons can carry a positive or negative electric 
charge ($\mu^+$ or $\mu^-$). 

As mentioned earlier, muons are abundantly produced in the atmosphere 
by cosmic rays and, because they are quite penetrating,  they can
reach the ground, enter the houses or schools through the walls or
roof of the building, and be detected with a suitable
apparatus.

Muons were first detected and investigated by Bruno
Rossi in the 1930s and 1940s \cite{Rossi1955} and in 1947 by
Neddermeyer and Anderson \cite{Neddermeyer} in surveys on cosmic rays.

Initially the muon particle was
associated with the Yukawa particle postulate in 1935, the 
Yukawa’s mesotron (later denominated as $\pi$ meson), a
force carrier of the strong interaction. But it was
demonstrated in 1947 that the muon did not interact
through the strong interaction. 

Consequently, the muons could not be the Yukawa $\pi$ meson. 
The discovery of the Yukawa's particle, the $\pi$ meson was made in 1947 by
Lattes, Ochialini, and Powel~\cite{Occhialini1} in the cosmic rays using
emulsion techniques. 
Around one year later Lattes at
Chacaltaya-Bolivia (5200 m above sea level) obtained the
first experimental evidence of the $\pi \rightarrow \mu$ decay~\cite{Occhialini2}.

The rest mass of the muon is:
\begin{equation}
  M_\mu = 105.66 \hspace{2mm} MeV/c^2
\end{equation}

approximately 207 times the mass of the electron.
See Particle Data Group report~\cite{PDG1}.

\subsection{The decay.}
The muon travels a relatively long distance (muons
are quite penetrating and they can reach the ground) while
losing its energy and it decays by the weak interaction into
an electron plus a neutrino and an anti-neutrino

\begin{equation}
  \mu^- \rightarrow e^- + \overline{\nu}_e + \nu_\mu 
  \label{Eq:mu-}
\end{equation}
\begin{equation}
  \mu^+ \rightarrow e^+ + \nu_e + \overline{\nu}_\mu 
  \label{Eq:mu+}
\end{equation}

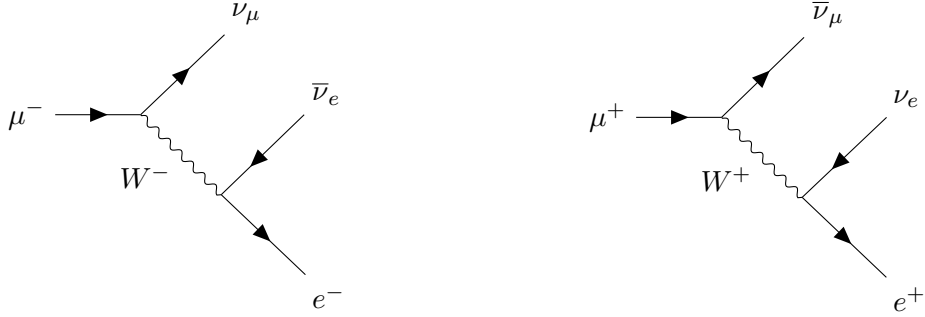
\begin{figure}[hbt]
\centering
  \begin{minipage}[]{.48\textwidth}
     \centering
     \begin{tikzpicture}
        \begin{feynman}
           \vertex (a) {\(\mu^{-}\)};
           \vertex [right=of a] (b);
           \vertex [above right=of b] (f1) {\(\nu_{\mu}\)};
           \vertex [below right=of b] (c);
           \vertex [above right=of c] (f2) {\(\overline \nu_{e}\)};
           \vertex [below right=of c] (f3) {\(e^{-}\)};

           \diagram* {
             (a) -- [fermion] (b) -- [fermion] (f1),
             (b) -- [boson, edge label'=\(W^{-}\)] (c),
             (c) -- [anti fermion] (f2),
             (c) -- [fermion] (f3),
           };
       \end{feynman}
     \end{tikzpicture}
  \end{minipage} 
  \hfill
  \begin{minipage}[]{.48\textwidth}
     \centering
     \begin{tikzpicture}
        \begin{feynman}
           \vertex (a) {\(\mu^{+}\)};
           \vertex [right=of a] (b);
           \vertex [above right=of b] (f1) {\(\overline{\nu}_{\mu}\)};
           \vertex [below right=of b] (c);
           \vertex [above right=of c] (f2) {\(\nu_{e}\)};
           \vertex [below right=of c] (f3) {\(e^{+}\)};

           \diagram* {
             (a) -- [fermion] (b) -- [fermion] (f1),
             (b) -- [boson, edge label'=\(W^{+}\)] (c),
             (c) -- [anti fermion] (f2),
             (c) -- [fermion] (f3),
           };
       \end{feynman}
     \end{tikzpicture}
  \end{minipage} 
\caption{Left: Feymann diagrams of $\mu^-$ and, Right, $\mu^+$ decay} 
\label{Fig:FeymanMuon}
\end{figure}

The branching ratios for these modes (called as normal
modes) are near 99\%. Even so, there are some
experimental evidences for other modes as the radioactive
muon decay observed specially in negative muons with
energies above 10 MeV.

\begin{equation}
  \mu^- \rightarrow e^- + \overline{\nu}_e \nu_\mu + \gamma
\end{equation}

with a branching ratio around 1.4\%. There are also other
(SUSY-GUT) decay modes called as exotic. In this case,
only the upper bound of the branching ratios has been
observed, and they are less than 10-11 at 90\% confidence
level.

\subsection{The lifetime.}
\label{Sec:MuonLifetime}

 Decays of unstable nuclei and particles, including muons, 
 are characterized by the decay probability in a unit of time, $\lambda$. 
 The mean lifetime of the particle, $\tau$, is defined as 
 $\tau = 1 / \lambda$
 
 Decays follow the exponential radioactive decay law:
 \begin{equation}
  N(t) = N_0 \exp^{-t/\tau}
  \label{Eq:MuonLifeFormula}
 \end{equation}

where $N_0$ is the number of unstable particles at the beginning, $t = 0$, 
and N(t) is the number of particles that survive until time t. 

We know, see Particle Data Group report~\cite{PDG1}, that the lifetime of the muon 
in vacuum and at rest is:

\begin{equation}
  \tau_0 = 2.1969811 \pm 0.0000022 \hspace{2mm} \mu s
  \label{Eq:MuonLifeTime}
\end{equation}

\subsection{The Fermi coupling constant $G_F$.}

The Fermi coupling constant $G_F$, was introduced by Fermi in order to phenomenologically ther strength
of the new ``weak force''.

Any renormalizable theory becomes predictive only once it has been supplied with a sufficient number of experimental inputs so as to fix the free parameters that appear in its Lagrangian.
In order for the subsequent theoretical predictions to be as accurate as possible, these inputs are chosen 
from the experimentally best-measured quantities available.  
In the case of the Standard Model of electroweak interactions these are the electromagnetic coupling constant,
$\alpha$, the Fermi coupling constant, $G_F$, and the mass of the $Z_0$ boson, $M_Z$. 
Their recognized best values, along with their absolute and relative errors that they represent are:

\begin{equation}
  \alpha = 1/(137.0359895 \pm 0.0000061)   
\end{equation}
\begin{equation}
   G_F = (1.16639 \pm 0.00002) \times 10^{-5} \hspace{5mm} GeV^{-2}
\end{equation}
\begin{equation}
   M_Z = 91.1867 \pm 0.0021 \hspace{5mm} GeV
\end{equation}

The Fermi coupling constant is obtained from the muon lifetime, $\tau_\mu$, via a calculation in the Fermi model, in which the weak interactions are represented 
by a contact interaction.

\begin{wrapfigure}{R}{0.40\textwidth}
   \centering
     \begin{tikzpicture}
        \begin{feynman}
           \vertex (a) {\(\mu^{+}\)};
           \vertex [      right=of a] (b);
           \vertex [above right=of b] (f1) {\(\bar{\nu}_{\mu}\)};
           \vertex [above right=of c] (f2) {\(\nu_{e}\)};
           \vertex [below right=of b] (f3) {\(e^{+}\)};

           \diagram* {
             (a) -- [fermion] (b) -- [fermion] (f1),
             (b) -- [anti fermion] (f2),
             (b) -- [fermion] (f3),
           };
           \vertex [above=0.5em of b] {\(G_{F}\)};
        \end{feynman}
     \end{tikzpicture}   
  \caption{Feynman diagram for muon decay in the Fermi effective theory}
  \label{Fig:FermiMuonDecay}
\end{wrapfigure}
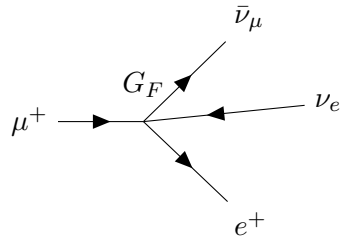

The Feynman amplitude $\mathcal{M}$ describes the transition amplitude of the process.  
It is derived from the Feynman rules for interaction vertices and the propagators involved.
The Feynman amplitude $\mathcal{M}_F$ for Fermi four-point interaction, as shown in
Fig.~\ref{Fig:FermiMuonDecay}, is:

\begin{equation}
   \mathcal{M}_F \approx \sqrt{2} G_F
\end{equation}

where $G_F$ is the Fermi constant which was introduced by Fermi to phenonemologically describe the 
strength of the weak force.

The Fermi interaction is in reality not point-like but that only appear point-like due to the fact that
the weak interation is mediated by weak bosons with very large masses according to the Standard Model
shown in Fig.~\ref{Fig:FeymanMuon}.

Using the full Feynman amplitude, the muon lifetime can be calculated as:
\begin{equation}
 \tau^{-1}_\mu = G^2_F \frac{m^5_\mu}{192 \pi^3}
\end{equation}

which can be used to determine the Fermi coupling constant, $G_F$, of the weak interaction.

\subsection{Muons in matter.}
\label{Sec:muoninmatter}

\begin{wrapfigure}{R}{0.50\textwidth}
  \includegraphics[width=0.98\linewidth]{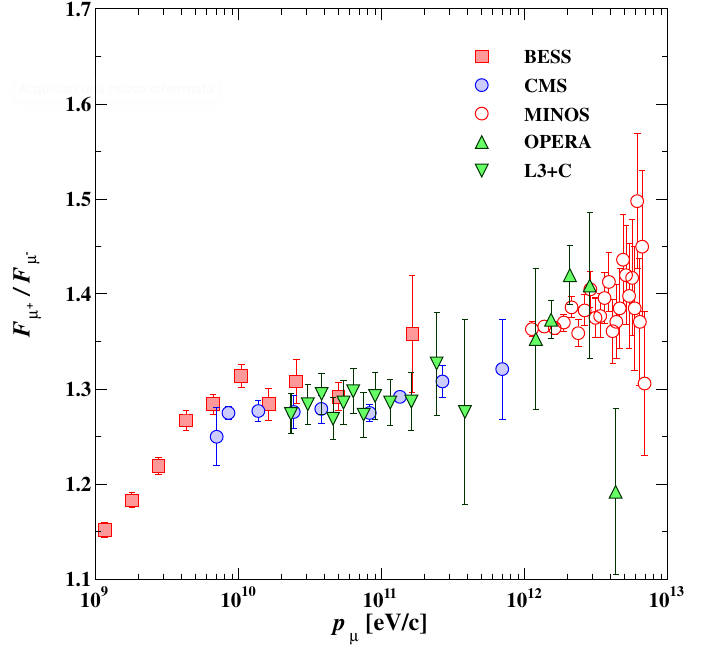} 
  \caption{ Muon charge ratio as a function of the muon momentum from 
            Refs. (\protect\cite{mu+mu-1}, \protect\cite{mu+mu-2},
            \protect\cite{mu+mu-3}, \protect\cite{mu+mu-4},
            \protect\cite{mu+mu-5}).}
  \label{Fig:mu+mu-}
\end{wrapfigure}

The muons are produced by primary cosmic rays interacting with nucleons of the upper atmosphere,
see Chap.~\ref{Sec:CosmRay}.
Due to the different charge of primary protons, electrons, neutrons or heavy nuclei, 
the number of positive and negative charged muons differs at sea level and 
the ratio is shown in Fig.~\ref{Fig:mu+mu-}.

\subsection{Muons flux and intensity.}
The average muon vertical flux is latitude dependent.
At sea level and for magnetic latitudes above 45°,
the intensity is $1.1 \times 10^{-2}$ 
muons/cm$^2$/sr/sec which corresponds to a flux through a horizontal area of 
$1.8 \times 10^{-2}$ muons/ cm$^2$/sec ~\cite{AMS1}, with a angular distribution
approximately $cos^2(\theta)$ dependence, with $\theta = 0$ at the zenith.
This works out to a good rule of thumb of about 1 muon/cm$^2$/minute. 

\subsection{Muon production and their path.}
As said before, the muons are produced by the interactions of protons 
with the nuclei of the atoms that constitute our atmosphere,
principally nitrogen, oxygen and argon.

The high energy cross-section for inelastic proton scattering on atomic nitrogen 
is $2.65 \times 10^{-25}$ cm$^2$.

The nuclear interaction length is:
\begin{equation}
 \lambda_n \approx 35 \times A^{1/3} \: [g \; cm^{-2}]
\end{equation}

Averaging over the atomic constituents of air, the mean free path for
inelastic interactions is 89 g/cm$^2$. 
Our atmosphere pressure at sea level, one interaction length $\lambda$,
is 1033 g/cm$^2$. 
Thus, a primary proton must traverse 11.6 interaction lengths to arrive at the 
Earth’s surface intact.

This means that the muons observed at ground level, the most likely height 
for production will be in the neighborhood of the first interaction length,
where 63.2\% of interactions take place.

The vertical distribution of the atmosphere is approximately
exponential with a scale height, $h_0$, of 7600 m.

Thus, one interaction length depth in the atmosphere corresponds to a height, 
h, determined by: $89 = 1033 \times e^{(-h/h_0)}$. 

Solving, we find h must be 18,600 m above sea level.

\subsubsection{Energy loss and interaction with matter.}

During their time of flight, muons can lose energy through Coulomb scattering,
ionization loss, Compton scattering and Bremsstrahlung. 
Since muons are massive, they mostly lose energy through ionization.
On average, cosmic ray muons lose 2 GeV throughout their course towards
sea level to ionization \cite{Beringer}.
Upon reaching ground, the energy spectrum of muons are analytically derived
in Eq.~\ref{Eq:EnergySea} and experimentally shown in 
Fig.~\ref{Fig:Energy}.

\begin{equation}  \label{Eq:EnergySea}
 \frac{dN_\mu}{dE_\mu d\Omega} \approx 0.14 E_\mu^{-2.7} [cm^2 s sr GeV]^{-1} \times [\frac{1}{1 + \frac{1.1 E_\mu \cos \theta}{115 GeV}} + \frac{0.54}{1 + \frac{1.1 E_\mu \cos \theta}{850 GeV}}]
\end{equation}

The flux of cosmic ray muons depends on the incident zenith angle. 
In particular, the relation is
\begin{equation}
  I(\theta h E) = I(0^o) \cos^{n(E h)}\theta
\end{equation}

where $\theta$ is the zenith angle,
h is the vertical distance traveled by the muon,
E is the energy of the muon, 
and $n(E, h)$ is an empirically determined constant. 
At sea level experiments showed that $n \approx 2$. 

When a charged particle passes through matter, it can lose energy and be deflected from its incident
direction. For charged heavy particles, such as muons, these effects are primarily due to inelastic collisions with atomic electrons of the material. The amount of energy transferred during 
every collision is small, but in dense media the interaction cross section can be large and 
many collisions per path length can occur, so the cumulative effect can lead to a substantial 
energy loss.

A quantum-mechanical description of how this energy loss relates to its relevant quantities is shown
mathematically by the Bethe-Bloch formula that describes the stopping power of a material for different 
incident particles. For experimental purposes this formula was slightly altered to include realistic 
experimental effects. The formula is the following:
\begin{equation}
 - \frac{dE}{dx} = 2 \pi N_a r^2_e m_e c^2 \rho \frac{Z}{A} \frac{z^2}{\beta^2} [ ln (\frac{2 m_e \gamma^2 v^2 W_{max} }{I^2}) -2 \beta^2 -\delta -2\frac{C}{Z} ]
\end{equation}

with
\begin{equation}
 2 \pi N_a r^2_e m_e c^2 = 0.1335 MeV cm^2 / g
\end{equation}

\begin{minipage}[]{.48\textwidth}
  \flushleft
  $r_e$: classical $e^-$ radius = $2.817 \times 10^{-13}$ cm \\
  $m_e$: electron mass = 0.511 $MeV / c^2$ \\
  $N_a$: Avogadro's number = $6.022 \times 10^{23}$ \\
  $I$: mean excitation potential \\
  $Z$: atomic number of absorbing material \\
  $A$: atomic weight of absorbing material \\
  $W_{max}$: maximum energy transfer in a single collision
\end{minipage}
\hfill
\begin{minipage}[]{.48\textwidth}
  \flushleft
  $\rho$: density of absorbing material \\
  $z$: charge of incident particle in unit of e \\
  $\beta$ v/c of incident particle \\ 
  $\gamma = \frac{1}{ \sqrt{1 - \beta^2}}$: \\
  $\delta$: density correction \\
  $C$: shell correction
\end{minipage}

The maximum energy transfer is produced by a knock-on collision . 
For an incident particle of mass M:
\begin{equation}
 W_{max} = \frac{2 m_e c^2 \eta^2}{1 + 2 s \sqrt{1 + \eta^2} - s^2}
\end{equation}

where $s = m_e / M$ , M is the mass of an incident particle, and
$\eta = \beta \gamma$. 
For the case of muons, since the mass of a muon $M = m_\mu$ 
is much greater than the mass of an electron $m_e$ , we have:
\begin{equation}
 W_{max} \approx 2 m_e c^2 \eta^2
\end{equation}

\begin{figure} [ht]
  \centering
  \includegraphics[width=.9\textwidth]{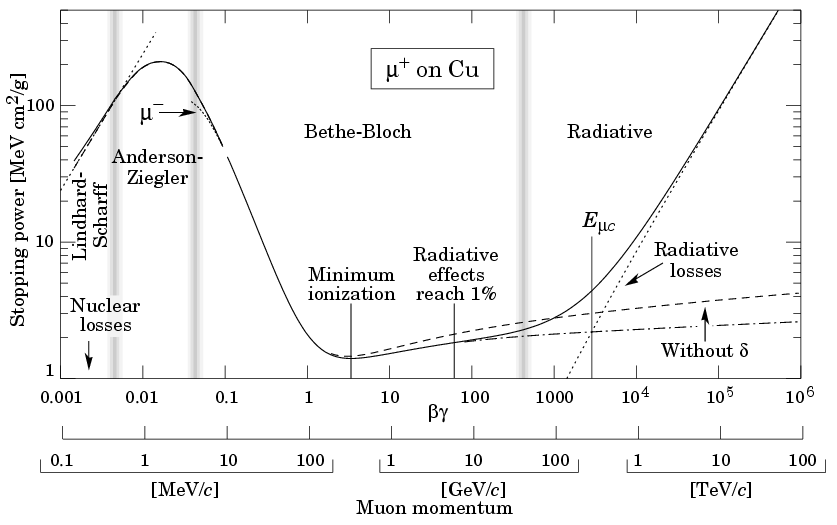}
  \caption{Energy dependence of the energy deposition per distance traveled of different charged particles in different media.}
  \label{Fig:BetheBlock}
\end{figure}

The Bethe-Bloch formula is plotted in Fig.~\ref{Fig:BetheBlock}
as a function of momenta  of incident particles. The curve is 
decreasing at lower energies until 
a certain point where the curve becomes relatively constant. Particles
at these momenta where the value of $dE/dx$ is minimal are call 
minimum ionizing particles. 
Their energy deposition per unit distance traveled over a piece of material is relatively constant. 
Cosmic ray muons travel at relativistic speeds at ground level, 
and are typically minimum ionizing particles. 
A commonly used approximation for $dE/dx$ is 2 MeV/(g/cm$^2$).

\subsubsection{Muon decay and absorption.}
\label{Sec:MuDecAbsorp}

The positive and negative muons have a different behaviour after they
are brought to rest in matter. 
While the positive muon has in matter only the possibility to decay,
the negative muon can replace one of the outermost electrons in the atom, 
having the same charge, and thus be captured by the host atom forming a muonic atom.

Just like electrons, negative muons fall on stable Bohr levels with radius $r_\mu$

\begin{equation}
  r_\mu = \frac{(n\hbar)^2}{M_{\mu}^{red} Z e^2}
  \label{Eq:mu-bohr}
\end{equation}

where $M_{\mu}^{red}$ is the reduced mass of the muon-nucleus system 
and Z is the atomic number of the host element. 
The mass of a muon is about 200 times larger than that of an electron. 
This results in a Bohr radius 200 times smaller than for the corresponding 
electron level with the same main quantum number.
Consequently, there is a non vanishing overlap of the wave function of the muon 
with the wave function of the nucleus and this can lead to a capture of the 
negative muon by the nucleus. 
For$Z \geq 40$ the Bohr orbit is already inside the nucleus.

Due to its different flavour, there is no exclusion by the Pauli
principle for the muon and electrons and, within a short time $10^{-13}$s,
it cascades to the 1s level. 

In this state, there is a considerable overlap of the muon wave fnnction
with the nucleons. 
The muon has now the possibility either to decay or to be captured by 
the nucleons.

The capture process is described by the Feynman diagram 
show in Eq.~(\ref{Eq:mu-capture}) ad Feymann diagram~\ref{Fey:mu-capture} .

\begin{equation}
  \mu^- + p \rightarrow \nu_\mu + n
  \label{Eq:mu-capture}
\end{equation}

\begin{center}
\begin{tikzpicture}
  \begin{feynman}
    \vertex (d1) {\(u\)}; 
    \vertex[right=5cm of d1] (d2) {\(u\)}; 
    \vertex[below=1em of d1] (u1) {\(d\)}; 
    \vertex[right=5cm of u1] (u2) {\(d\)};
    \vertex[below=1em of u1] (d3) {\(u\)}; 
    \vertex[right=5cm of d3] (u3) {\(d\)};
    \vertex[below right=0.5cm and 2.5cm of d3] (v1);
    \vertex[below left=1cm and 0cm of v1] (v2);
    \vertex[below left=0.25cm and 2cm of v2] (nu) {$\mu^-$};
    \vertex[below right=0.25cm and 2cm of v2] (e) {$\nu_\mu$};
    \diagram* { {[edges=fermion]
      (d1) -- (d2),  (u1) -- (u2),
      (d3) -- (v1) -- (u3), (nu) -- (v2) -- (e)},
      (v1) -- [boson, edge label=$W^\pm$] (v2)
    };
    \draw [decoration={brace}, decorate] (d3.south west) -- (d1.north west) node [pos=0.5, left] {\(p\)};
    \draw [decoration={brace}, decorate] (d2.north east) --  (u3.south east) node [pos=0.5, right] {\(n\)};
  \end{feynman}
\end{tikzpicture}
\captionof{figure}{Feynman diagram of a negatively charged muon captured by a proton.  
           The proton is transformed into a neutron and a muonic-neutrino is emitted. 
           The W-boson is the mediator of the interaction.}
\label{Fey:mu-capture}
\end{center}

The competition between these processes reduces the mean lifetime of the
observed decay of the negative muon as compared with the free decay. 
Only muons not already captured in the orbit can decay. 
Expressed in terms of process rates $\Lambda = 1/\tau$, the total disappearance 
probability is written as the sum of the capture and of the decay probabilities 
(either one or the other):

\begin{equation}
  \Lambda_{total} = \Lambda_{decay} + \Lambda_{capture}
\end{equation}

The nuclear capture rate is determined by the overlap of the muonic
orbit with the nucleus and it follows roughly a power law of the atomic number 
($\Lambda_{capture} \sim Z^4$), \cite{Measday}.

In compounds, the nuclear capture is preferred by the nucleus with the larger 
nuclear capture rate, i.e. larger atomic number. 
For instance, in a plastic scintillator material 
(polyvinyl-tolnol, 2-CH3C6H4CH=CH2 monomer) only the carbon capture is significant.

The capture probability for this condition is very close to 1. 
For light nuclei, how-ever, with charge number of $\approx 10$ or less, also negative muons will preferentially
decay in an analogue fashion to $\mu^+$ decay.

The decay probability for free muons at rest, $\lambda = 1 / \tau$, is the same for positive and negative muons.

\begin{center}
  \begin{math}
    \lambda^+ = \lambda^- \hspace{4mm} $or$  \hspace{4mm} \tau^+ = \tau^-
  \end{math}
\end{center}

If negative muons are stopped in matter the measured decay probability $\lambda^-_m$
is always larger than the decay probability $\lambda^+$ or $\lambda^-$ for free muons. So the equation

\begin{center}
  \begin{math}
   \lambda^-_m > \lambda^+ = \lambda^-
  \end{math}
\end{center}

holds or, respectively

\begin{center}
  \begin{math}
   \tau^-_m < \tau^+ = \tau^-
  \end{math}
\end{center}

In the case of muon capture the process competes with free decay.
The $\mu - $capture shortens the effective lifetime of $\mu$.  It results to:

\begin{equation}
 \frac{1}{\tau} = \frac{1}{\tau_0} + \frac{1}{\tau_c}
\end{equation}

where $\tau_c$ is the $\mu -$capture lifetime. 

Because the muons produced by cosmic radiation are a mixture of positive and negative muons, 
one expects the following time behavior for the number of disintegrated muons:

\begin{equation}
 N(t) = N(\mu^- , t_0) e^{-(t - t_0)/\tau_0} e^{-(t - t_0)/\tau_c} - N(\mu^+ , t_0) e^{-(t - t_0)/\tau_0}
\end{equation}

because the positive muons do not undergo capture.

\begin{table}[h!]
  \centering
    \begin{tabular}{ |l|c|c|c|} \hline
      Z & nucleus of material  & average lifetime $\tau_\mu [\mu s]$ & 
      capture rate $\times 10^3 (s^{-1})$ \\ \hline
         & $\mu^+$       & $2.1969811 \pm 0.0000022$ & 455.16 \\
      1  & $^1H$         & $2.19490$                 & 0.450  \\
      12 & $^{12}C$      & $2.028$                   & 37.9   \\
      13 & $^{27}Al$     & $0.864$                   & 705.   \\ 
      20 & Ca            & $0.334$                   & 2546.  \\
      82 & Pb            & $0.0748$                  & 12985. \\ \hline
   \end{tabular}
\caption{Some total capture rates and meanlife for $\mu^-$ in
         nuclei~\cite{Measday}. See also Fig.~\ref{Fig:Suzuki}.}
\label{Tab:MatLifeTime}
\end{table}

The capture of negative muons by nuclei is accompanied by secondary 
effects, see~\cite{Vulpescu}. 
During the electromagnetic cascade down to the 1s level, 
photons by radiative transitions and Auger electrons are emitted. 
For heavy nuclei, nuclear Auger effect (neutron emission from nucleons) 
or prompt fission may occur also. 
The total capture rate is not easy to calculate as the final nucleons is 
excited to an unknown energy, which is the critical parameter from the 
theoretical point of view. 
Nuclear gamma transitions, neutron evaporation or even charged
particle and cluster emission can follow the muon capture.

\begin{wrapfigure}{R}{0.50\textwidth}
  \includegraphics[width=0.95\linewidth]{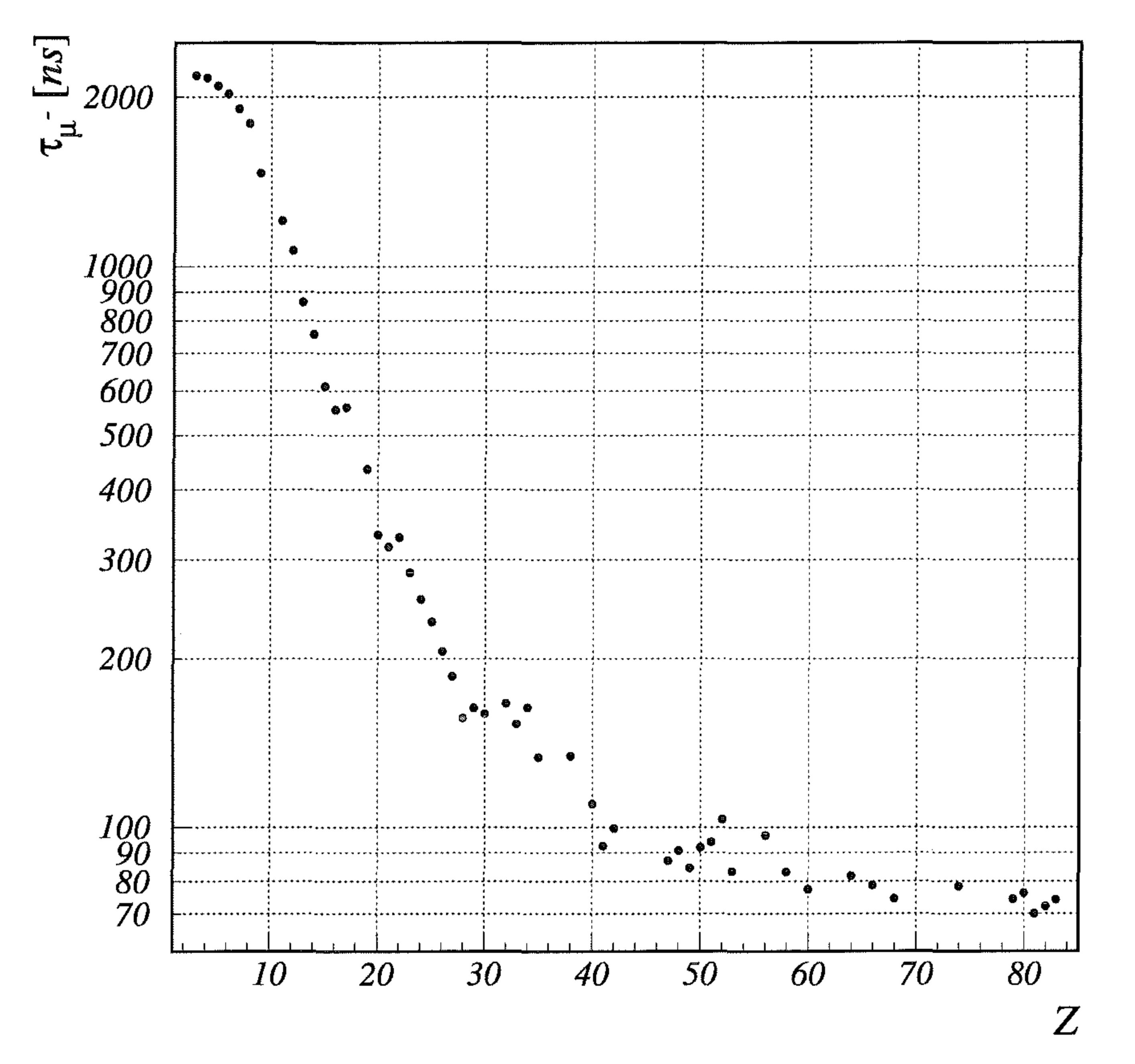} 
  \caption{ Mean lifetime of negative muons in different materials 
            measured in \cite{Suzuki}}
  \label{Fig:Suzuki}
\end{wrapfigure}

However, from the experimental point of view, the measurement is 
straight-forward, the determination of the muon lifetime after stopping 
in the relevant material is simple.

\subsection{The muon mean free paths and special relativity.}

The first approach to measure the travel time and the distance of muons produced 
by primary cosmic rays is the Newtonian dynamics.
We can consider the muon lifetime at rest given by Eq.~\ref{Eq:MuonLifeTime},
the speed of secondary muons close to the speed of light, 
we find the distance d that the muon can travel: $d = c \tau = 660m$,
(see Eq.~\ref{Eq:TravelDistance})
incompatible with the distance of 10 - 20 km from the earth's surface where 
the interactions of primary cosmic rays take place.
With the Newtonian dynamics, can also calculate the probability of survival 
of secondary muons:

\begin{eqnarray*}
    & t = \frac{d}{c} \\
    & t = \frac{15 \times 10^3}{3 \times 10^8} = 5 \times 10^{-5} s   \\
    & N(t) = N_0 e^\frac{t}{\tau} \\
    & \frac{N}{N_0} = e^{-\frac{5 \times 10^{-5}}{2.2 \times 10^{-6}}} = 1.3 \times 10^{-10}
\end{eqnarray*} 

This is clearly incompatible with reality. 
One has to use special relativity~\cite{SpRelativity} and Lorentz transformations
in cases where the speed is close to that of light. 

To obtain the mean range of a fast muon, we must take into account the time dilation and distinguish between apparent life and proper life. 
The lifetime of a muon is equal to $\tau_0$ in the rest system
of the muon~(\ref{Eq:MuonLifeTime}). 
Due to relativistic time dilation, a muon with a velocity $v$ appears to
have a lifetime

\begin{equation}
  \tau' = \tau_0 \gamma \hspace{5mm} ; \hspace{5mm} \gamma = \frac{1}{\sqrt{1 - (v / c)^2}}
\end{equation}

The mean range of a muon with velocity $v$ is therefore:
\begin{equation}
  R(v) = v \tau' =  v \tau_0 \gamma
\end{equation}

or

\begin{equation}
  R(v) = \tau_0 \frac{p}{m_\mu} = \frac{p}{m_\mu c} \tau_0 c 
  \hspace{2mm}; \hspace{2mm} \tau_o c = 0.66 km
  \label{Eq:TravelDistance}
\end{equation}
\begin{eqnarray*}  
  & p = mv \\
  & m = \frac{m_\mu}{\sqrt{1 - (v / c)^2}} = m_\mu \gamma \\
  & L = \frac{L_0}{\gamma} 
\end{eqnarray*}
\begin{equation}
  E = \gamma m c^2
  \label{Eq:Egamma}
\end{equation}

with $L_0$ proper length and $L$ observed length. 

If we consider an energy of 20 GeV Fig.~(\ref{Fig:Energy}) we can find $\gamma$
from equation~(\ref{Eq:Egamma}). We also need to consider
the rest mass of muons which is m = $105.7 \times 10^6$ eV:
\begin{equation}
 \gamma = 189.2
\end{equation}

and, then, the muon proper distance L, given by Eq.~\ref{Eq:TravelDistance} is:
\begin{equation}
 x' = \frac{20000}{189.2} = 105.7 \: [m]
\end{equation}

consequently, the percentage of muons that will reach the ground by using the
proper time $ \gamma \tau$ is:
\begin{equation}
 \frac{N}{N_0} = e^{-\frac{5 \times 10^{-5}}{189.2 \times 2.2 \times 10^{-6}}} = 0.89
\end{equation}

i.e. a significant fraction of the muons produced in the stratosphere
are capable to reach the earth surface.

It is evident that the muon, produced in the stratosphere at about 10-30 km 
high, reaches the earth only thanks to the dilation of its average life time 
caused by special relativity.

\subsection{Time arrival and Statistics of uniform random events.}
\label{Sec:TimeArrival}

The arrival of cosmic rays is a random process.
So we expect it to follow the Poisson distribution for rare events:
\begin{equation}
 P(n, t) = \frac{(rt)^n e^{-rt}}{n!}
 \label{Eq:Poisson}
\end{equation}
where r is the average event rate, namely the average number of events percentage unit time.

\begin{figure}[hbt]
\centering
  \begin{minipage}[]{.49\textwidth}
     \centering
     \includegraphics[width=.98\textwidth]{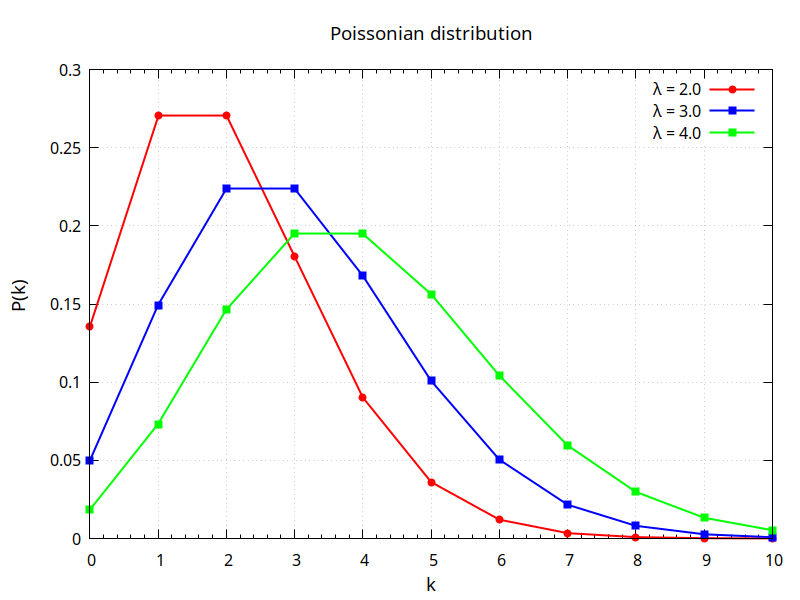}
  \end{minipage} 
  \hfill
  \begin{minipage}[]{.49\textwidth}
     \centering
     \includegraphics[width=.98\textwidth]{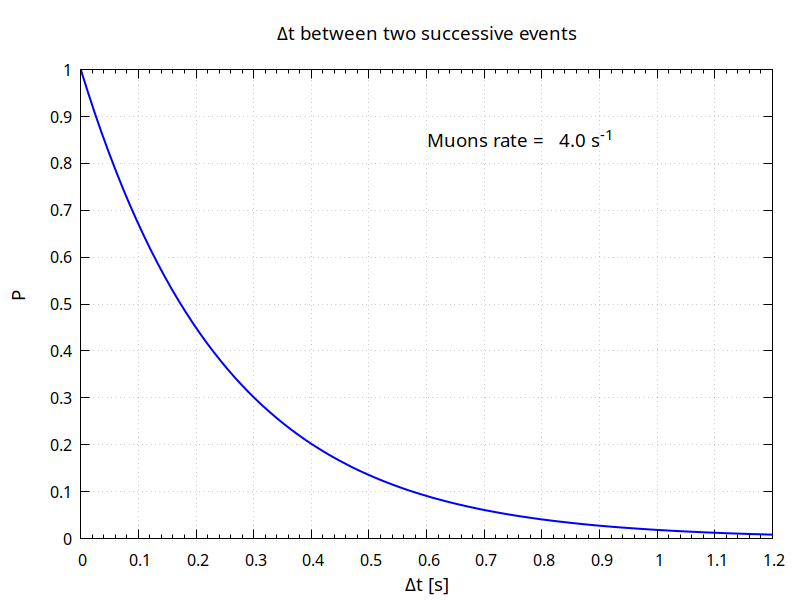}
  \end{minipage} 
  \caption{Left: Poissonian distribution at different average values $\lambda$;
           Right: time between two consecutive events for average rate 4 s$^{-1}$}
 \label{Fig:Poissonian}
\end{figure}

Then the probability of observe n events in the time interval $\Delta$t 
is given by Eq.~\ref{Eq:Poisson} and show in Fig.~\ref{Fig:Poissonian}.

\section{Design and optimization of the AMELIE detector.}
\label{Sec:apparatus}

The basic idea that guided the design of the AMELIE detector was to measure 
the time between the arrival of a muon and its decay in a plastic scintillator block,
see Fig.~\ref{Fig:detector1}.
The start signal is given by the ionization produced by the incoming muon, while the 
stop signal is given by the electron/positron produced by its decay.

\subsection{Experimental setup.}

\begin{wrapfigure}{R}{0.5\textwidth}
  \includegraphics[width=0.98\linewidth]{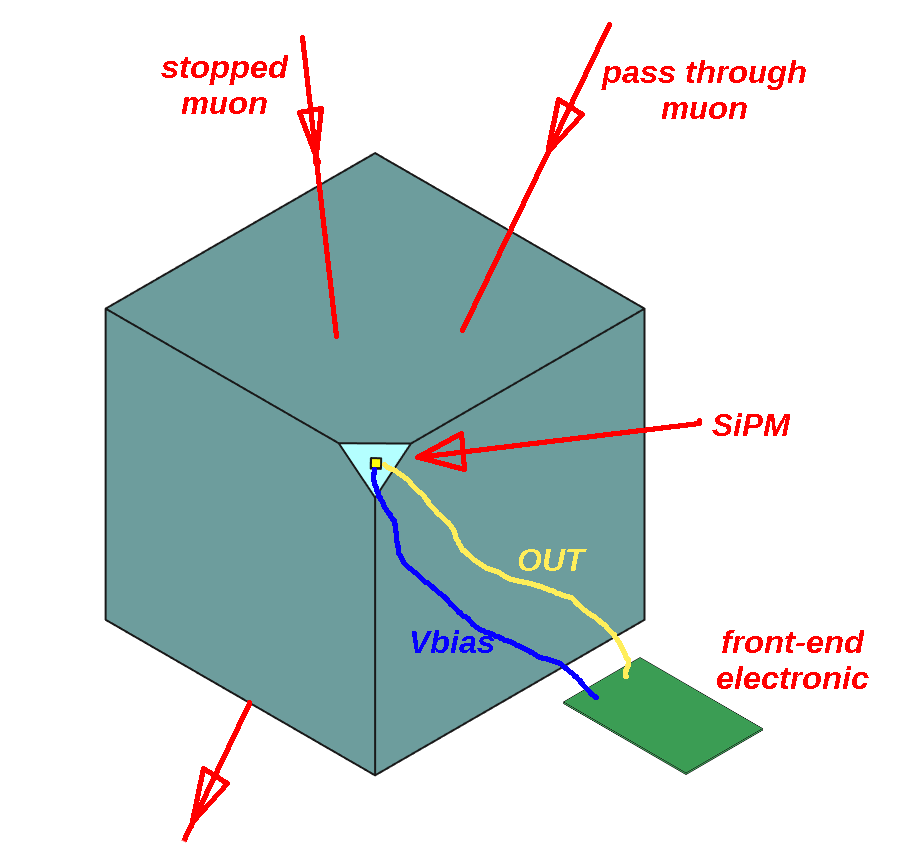} 
  \caption{Scheme of the detector setup. The apparatus is based on a 
           cube of scintillator detector of dimensions $15 \times 15 \times 15 \; cm^3$. The $6 \times 6 \; mm^2$ SiPM detector is placed on
           one corner cut at $45^o$.}
  \label{Fig:detector1}
\end{wrapfigure}

The experimental setup, shown in Fig.~\ref{Fig:detector1}, 
consists of a cube of plastic scintillator of dimensions $15 \times 15 \times 15 \; cm^3$, 
coupled to a solid state Silicon Photo Multiplier (SiPM). 
  
The size of the scintillator is a trade-off the counting time to get reliable 
results and the cost of the scintillator. Size can be increased to benefit 
data acquisition time, increasing cost.
Instead of the plastic scintillator, a large volume of liquid scintillator, 
enclosed in an airtight light-proof bin, can be used. 
The same SiPM, facing the liquid scintillator, would constitute the 
photon detector.

 The scintillator is a general purpose plastic scintillator like PS-923 produced by DETECT~\cite{Detec}.
 
 The Silicon Photo Multiplier (SiPM) is a new sensor \\ MICROFICHE-60035-SMUT-TR ,
 produced by ON SEMI~\cite{ONSEMI}.

\begin{figure}[hbt]
\centering
  \begin{minipage}[]{.32\textwidth}
     \centering
     \includegraphics[width=.98\textwidth]{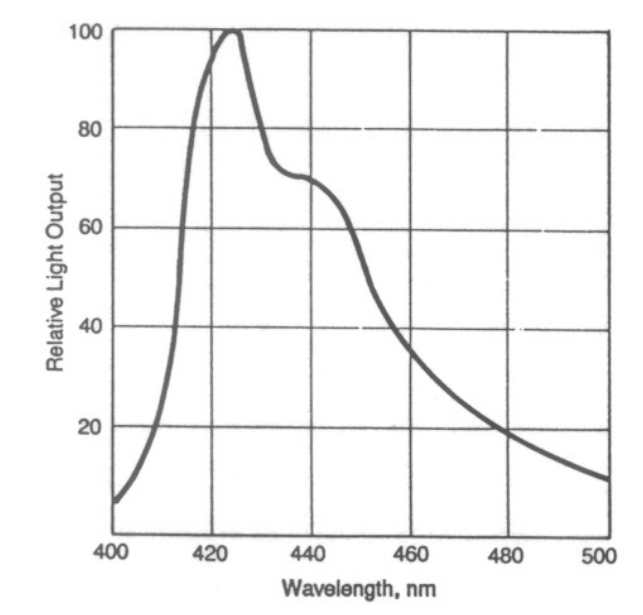}
  \end{minipage} 
  \hfill
  \begin{minipage}[]{.32\textwidth}
     \centering
     \includegraphics[width=.98\textwidth]{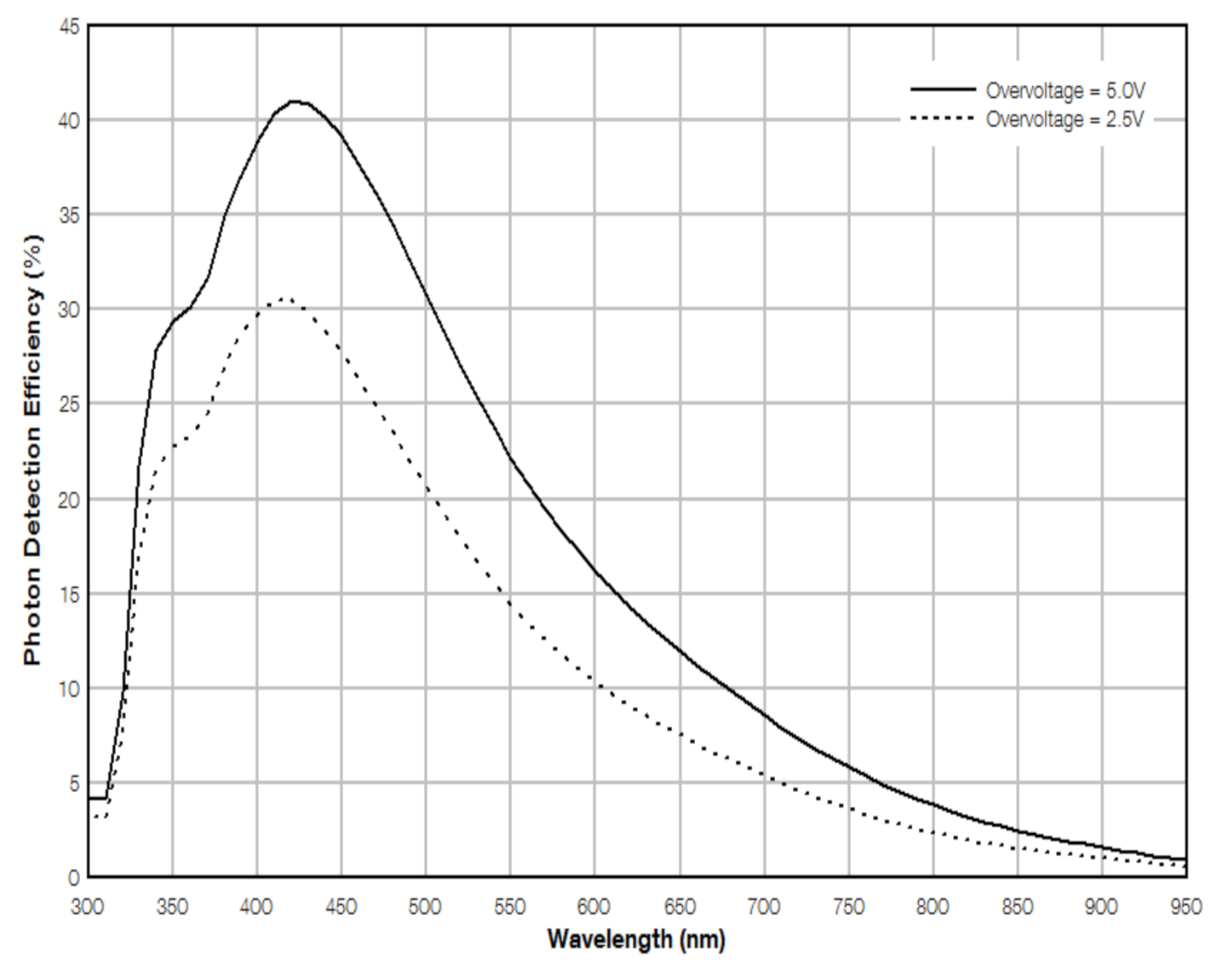}
  \end{minipage} 
  \hfill
  \begin{minipage}[]{.32\textwidth}
     \centering
     \includegraphics[width=.95\textwidth]{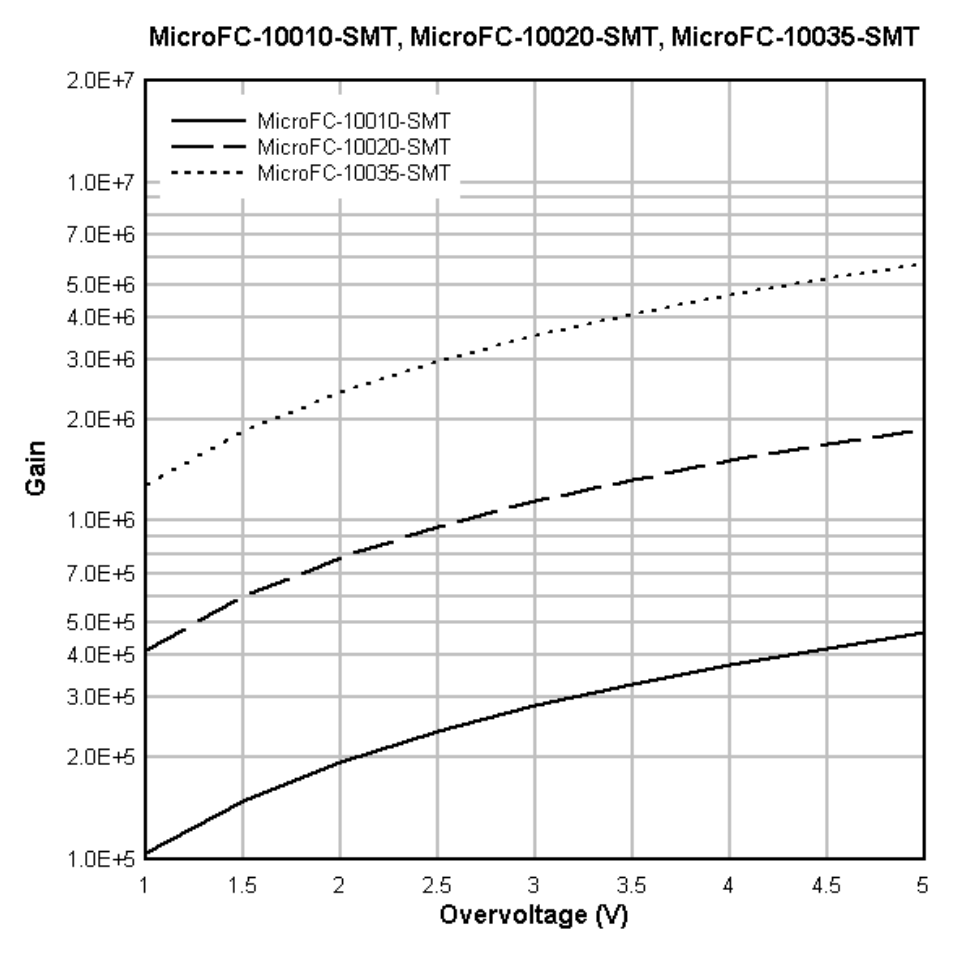}
  \end{minipage} 
\caption{Left: relative light output of BC-400 scintillator; 
         Center: Quantum efficiency of Onsemi 60035 SiPM;
         Right: SiPM gain versus overvoltage $V_{bias} = B_{reakDown} + O_{verVoltage}$} 
\label{Fig:SiPM_QE}
\end{figure}

Its active area is $6 \times 6 mm^2$, high quantum efficiency and gain are shown in Fig.~\ref{Fig:SiPM_QE}.
 From the figures, it is clear that the quantum efficiency of the SiPM matches perfectly with that of the scintillator.
 
 Plastic scintillator and SiPM have been covered by a thin sheet of aluminized mylar, covered, in turn,
 by a black box made from black cardstock and black adhesive tape. 
 The aluminized mylar sheet reflects light inwards to maximize the number of photons hitting the SiPM
 in order to maximize light collection.
 It is placed on a edge surface cut at 45$^o$.
 This position maximizes the uniformity of light collection throughout the volume.
 
 The output of the small frontend SiPM card consist of 4 wires: 1) $V_{bias}$;
 2) SiPM output; 3) temperature sensor; 4) ground.
 
 The front-end electronic, described in detail in Sec.~\ref{Sec:FrontEndElectronic},
 is placed on the top of the scintillator block.
 It contain a $V_{bias}$ DC-DC converter, threshold and Sample\&Hold circuits. 
 The frontend electronic is controlled by a microcontroller based on ESP32. 
 The data are shown on a OLED display and recorded on a $\mu$SD card for off-line 
 evaluation.

Although the detector is small in size, compact and cheap, it has all 
the characteristics of any modern physics experiment, such as those 
at CERN on the LHC accelerator.

In particular:

\begin{itemize}
 \item uses detectors of innovative design, widely used in modern 
       physics experiments;
 \item coupled with state-of-the-art sensors;
 \item modern, high-performance front-end electronics and 
       digitization of incoming data;
 \item has a microcontroller, which is a computer with which 
       essential parts of an experiment are carried out:
 \begin{itemize}
        \item data recording on $\mu$SD, the modern version of
              magnetic tapes, for off-line data analysis;
        \item slow control, i.e., cyclic control of system 
              operating parameters;
        \item display and system monitor during data taking;
        \item ability to set and change detector operating parameters.
 \end{itemize}
\end{itemize}

\subsection{Scintillator detectors.}
\label{Sec:ScintDetector}

Scintillators are charged particle detectors widely used in modern and past 
physics experiments in the following detectors:

\begin{itemize}
 \setlength\itemsep{0.1em}
 \item hadron and electron calorimetry;
 \item time of flight measurements;
 \item trigger counters;
 \item veto counters;
 \item particles identification by number of photons.
\end{itemize}

There are three types of organic scintillators: 1) crystalline; 2) liquid; 
3) plastic.

In Amelie, the plastic scintillator was used, but the system can be easily 
adapted to a liquid scintillator.

\begin{figure}[hbt]
\centering
  \begin{minipage}[]{.34\textwidth}
     \centering
     \includegraphics[width=.95\textwidth]{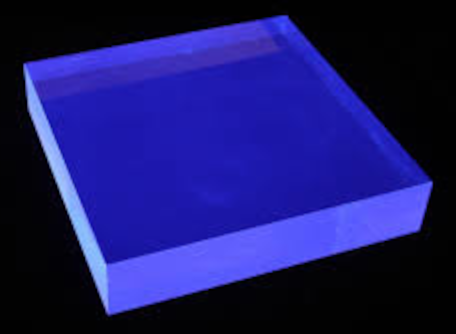}
  \end{minipage} 
  \hfill
  \begin{minipage}[]{.64\textwidth}
     \centering
     \includegraphics[width=.95\textwidth]{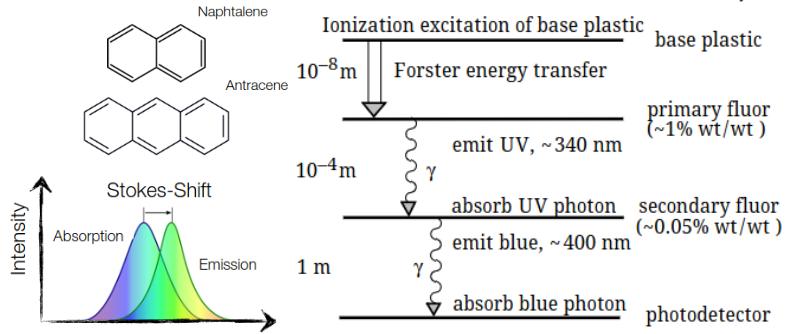}
  \end{minipage} 
\caption{Left: Plastic scintillator; Right: excited level of organic compounds} 
\label{Fig:OrganicScint}
\end{figure}

The working principle is the same for all.
Charged particle through a medium generates a track of excited molecules.
It, loses energy by ionization, emission of delta rays etc.,
see Fig.~\ref{Fig:OrganicScint}.
Some type of molecules releases a small fraction ($\approx$3\%) of energy as 
optical photons.
The scintillation process is particularly marked in substances containing aromatic
rings (polystyrene, polyvinyltoluene etc.)

The excitation is generated by the absorption of a photon while de-excitation
generates the emission of a longer wavelength photon.
The effect of wavelength difference between absorption and emission peaks is called
Stokes’ shift (greater Stokes’ shift is better as it minimize self-absorption).
Fluors are used as “waveshifter”.

The dE/dx converted into visible light detected by photosensors.

The most common bases used in plastic scintillators are the aromatic plastics, 
polymers with aromatic rings as pendant groups along the polymer backbone, 
amongst which polyvinyltoluene (PVT) and polystyrene (PS) are the most prominent.

With a density range of a plastic scintillator from 1.03 to 1.2 gr cm$^3$,
the photon yields are about 1 photon per 100 eV of energy deposit~\cite{AMS1}.
This means that a muon of 50 MeV that is arrested in the block, 
see Chap.~\ref{Sec:simulations}, emits 500000 photons in all directions.

\begin{table}[h!]
  \centering
    \begin{tabular}{ |l|c|} \hline
      light output                  & 56 \% of antatacene  \\
      rise time                     & 0.9 ns               \\
      decay time                    & 3.3 ns               \\
      bulk attenuation length (BAL) &  250-450 cm          \\ 
      technical attenuation length (TAL) & 150-250 cm      \\
      wavelength of max emission    & 418 nm               \\ 
      n. of H atoms per cm\textsuperscript{3} & $5.17 \times 10^{22}$     \\
      n. of C atoms per cm\textsuperscript{3} & $4.69 \times 10^{22}$     \\ 
      n. of electrons per cm\textsuperscript{3} & $3.33 \times 10^{23}$   \\ 
      density                       & 1.023 g/cm\textsuperscript{3}       \\ \hline
   \end{tabular}
\caption{Physical characteristic of DETEC P923A plastic scintillator.}
\label{Tab:DetecP923A}
\end{table}

\subsection{FLUKA simulations.}
\label{Sec:simulations}
The FLUKA code~\cite{Fluka} is a general purpose Monte Carlo code for the 
interaction and transport of hadrons, leptons, and photons from keV 
(with the exception of neutrons, tracked down to thermal energies) to cosmic 
ray energies in any material. 
It has many applications in high energy experimental physics and engineering, 
shielding, detector and telescope design, cosmic ray studies, dosimetry, 
medical physics and radio-biology.

The history of FLUKA goes back to 1962-1967, when J. Ranft was working at CERN 
on hadron cascades.
Its development has never ceased since then and, today, continues as part 
of a collaboration between CERN and INFN.

Fluka has been used extensively in Compass to evaluate the environmental irradiance 
caused by the beam and secondary particles in the various measurements made from 2000 
to the present.

As part of the AMELIE project, Fluka has been used for:
\begin{itemize}
 \item optimization of the detector;
 \item evaluation of counting rate by simulation of primary cosmic rays and muon shower.
\end{itemize}

\subsubsection{Optimization of the detector}
In order to optimize the experimental setup and to find the optimal 
configuration, the experimental setup, shown in Fig.~\ref{Fig:detector1}, 
has been extensively simulated using Fluka.

The following simulations has been done:

\begin{itemize}
 \item optimization of the collection of optical photons by the SiPM placed 
       on one corner of the scintillator cube;
 \item study of the energy loss of low energetic muons on the scintillator cube;
 \item study of the number of optical photons vs muon energy and muon stopped in 
       the scintillator.
\end{itemize}

The material of the scintillator is P923A, with a density of $1,032 gr/cm^3$, with a composition of 
H 0.084757, C 0.915243.

The $\mu^-$ source simulating cosmic muons was placed along the Z axes at Z = -10 cm.

\vspace{4mm}
\noindent \emph{Limit of Fluka simulations}

However, one must keep in mind that Fluka, like any simulation program, 
has inherent limitations.
The dependence of the number of photons on wavelength
in the scintillator, shown in Fig.~\ref{Fig:SiPM_QE} left, is not considered. 
It is simplified with a mono energetic photon of 2.9 nm.

In any case, although the absolute values on the number of photons collected by 
the SiPM cannot be considered reliable, we believe that the relative ratios are sufficient for the purpose of this study.

\subsubsection{Number of optical photons vs energy of the muon.}
For this simulations, the source of $\mu^-$ was placed at the center of the cube, X = 0 ; Y = 0 ;
Z = -2.5cm from the top surface.

\begin{figure}[hbt]
\centering
  \begin{minipage}[]{.32\textwidth}
     \centering
     \includegraphics[width=.98\textwidth]{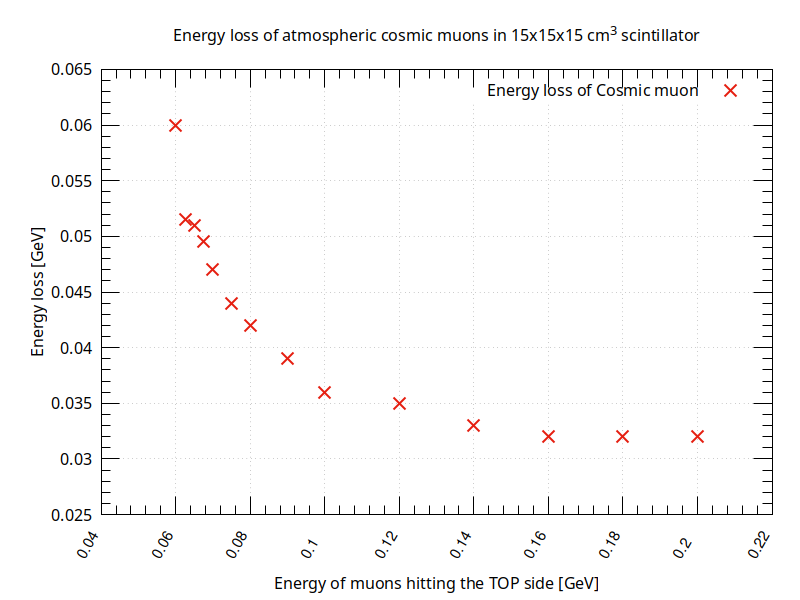}
  \end{minipage} 
  \hfill
  \begin{minipage}[]{.32\textwidth}
     \centering
     \includegraphics[width=.98\textwidth]{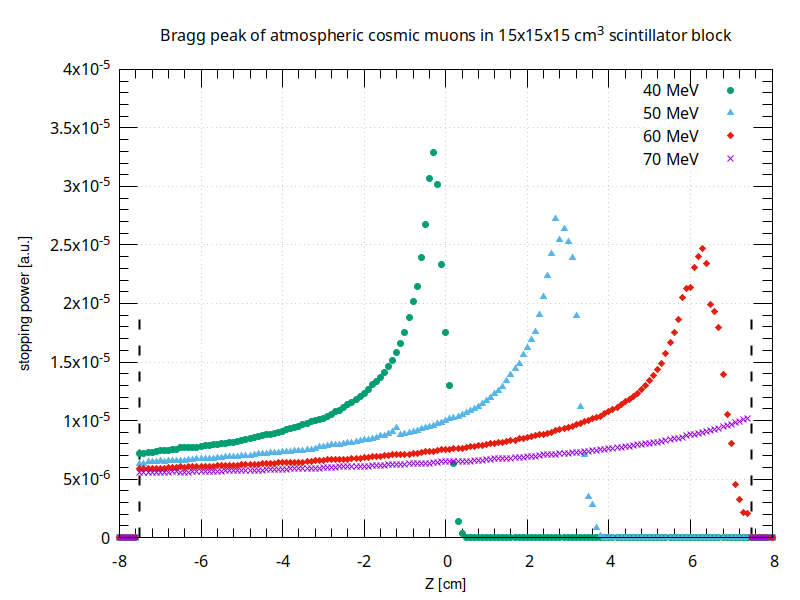}
  \end{minipage} 
  \hfill
  \begin{minipage}[]{.32\textwidth}
     \centering
     \includegraphics[width=.98\textwidth]{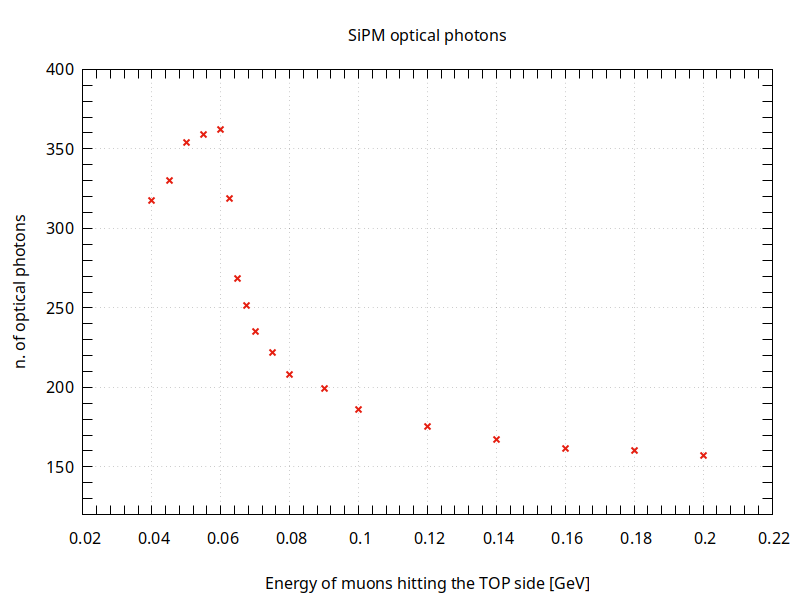}
  \end{minipage} 
\caption{Left: energy loss in the scintillator cube for muons that cross 
               completely the cube. 
               Below 60 MeV the muons are stopped in the cube;
         Center: Bragg peak for muons stopped in the cube, 40, 50, 60 MeV, 
               and crossing the cube 70 MeV ray flux;
         Right: number of photons collected by SiPM.} 
\label{Fig:Fluka1}
\end{figure}

In agreement with the increase of Stopping Power at low energies, shown in Fig.~\ref{Fig:BetheBlock}, the energy loss of muons during the path increase 
at low energies.

Below an energy of 60 MeV, all $\mu^-$ are stopped in the scintillator block and the Bragg peaks are clearly shown in Fig.~\ref{Fig:Fluka1}, center.

Consequently, also the number of optical photons generated by the muon increases at low energies, increase from about 150 photons/primary per muon at the minimum of ionization (MIP) to 300-360, as shown in Fig.~\ref{Fig:Fluka1}, right.

It is evident that, by suitably raising the threshold with respect to the MIP muons, it is possible to distinguish the muons that stop in the scintillator block from those that cross it. 
These are the muons of interest for the measure of their average life.

\subsubsection{Uniformity of optical photons collection.}

For this simulations the energy of the was 50 MeV, low enough to stop the muon 
in the block.

The position of the muon beam was at the center and at the four corner of the block.

The number of photons collected by the SiPM in each position is shown in Tab.~\ref{Tab:PhoPosition}

\begin{table}[tb]
  \begin{center}
    \begin{tabular}{ |c|c|c| } \hline
      X [cm]  &  Y [cm] & n. of SiPM optical photons \\ \hline
      0  &  0  & 352.6 \\ 
      5  &  5  & 315.5 \\ 
      5  & -5  & 302.3 \\
     -5  &  5  & 300.3 \\
     -5  & -5  & 628.7 \\ \hline
   \end{tabular}
  \end{center}
\caption{Number of optical photons collected by SiPM vs position of muon hitting 
         the TOP side, the same side of SiPM}
\label{Tab:PhoPosition}
\end{table}

With the exception of the last position, just in front of the SiPM, the number 
of photons is uniform and above 300 photons.

\subsubsection{Muon decay.}
\label{Sec:MuonDecay}

The decay process of the muon is a three body decay shown in Eq.~\ref{Eq:mu-} and \ref{Eq:mu+}.
The energy corresponding to the mass of the muon at rest, 105.66 $MeV/c^2$, is shared between the 
two neutrinos and the electron/positron.
Of course we cannot see the two neutrinos as they are not charged. We only see the light 
emitted by the shower produced by the charged lepton.
This means that the energetic spectrum of the three particles are continuous 
with a maximum of 53 $MeV/c^2$ for the charged lepton.

\begin{figure}[hbt]
\centering
  \begin{minipage}[]{.48\textwidth}
     \centering
     \includegraphics[width=.98\textwidth]{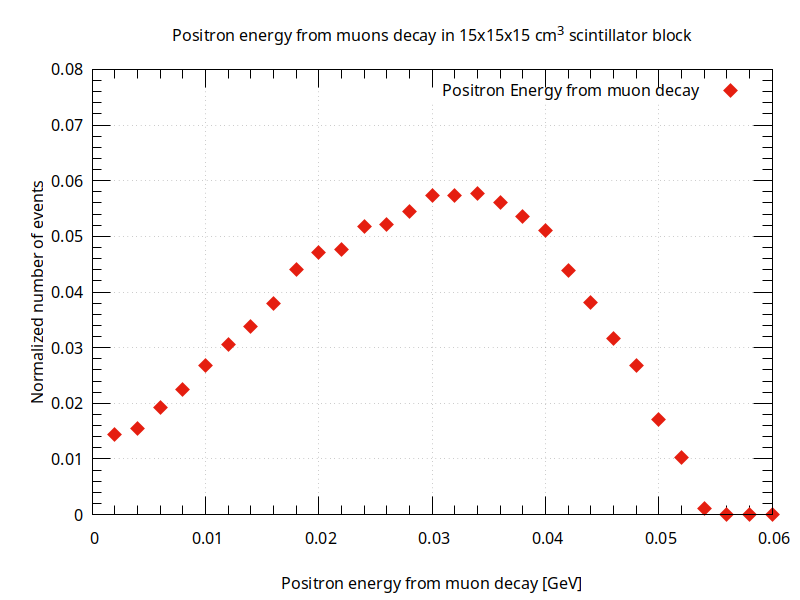}
  \end{minipage} 
  \hfill
  \begin{minipage}[]{.48\textwidth}
     \centering
     \includegraphics[width=.98\textwidth]{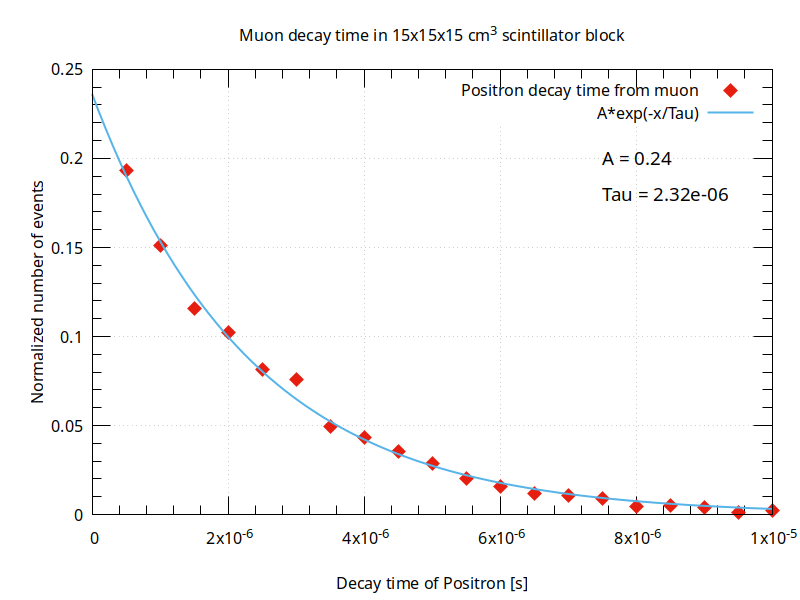}
  \end{minipage} 
\caption{Fluka simulations. $E_{\mu^+}$ = 50 MeV at x = y = 0.
         Left: Energy of positrons from muons decay;  
         Right: Muon decay time.} 
\label{Fig:ElectronEn}
\end{figure}

\begin{figure}[hbt]
\centering
  \begin{minipage}[]{.48\textwidth}
     \centering
     \includegraphics[width=.98\textwidth]{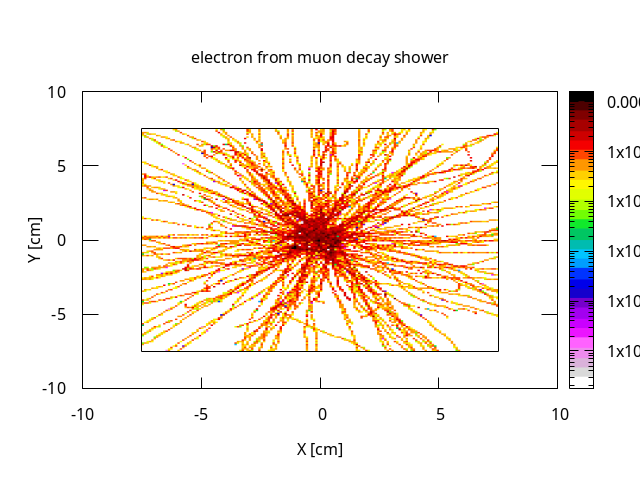}
  \end{minipage} 
  \hfill
  \begin{minipage}[]{.48\textwidth}
     \centering
     \includegraphics[width=.98\textwidth]{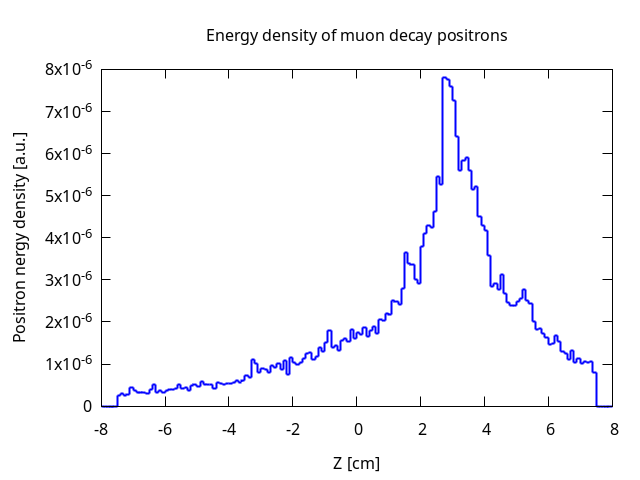}
  \end{minipage} 
\caption{Left: Energy density of muons decay positrons; muons energy = $50 MeV/c^2$; position X = 0, Y = 0;
         Right: cut of the left histogram. The peak is the Bragg peak of Fig.~\ref{Fig:Fluka1}} 
\label{Fig:Fluka3}
\end{figure}

In Fig.~\ref{Fig:ElectronEn} left, the energy of electron/positron is shown. 
The electron or positron will produce a shower until the energy is completely absorbed.
In Fig.~\ref{Fig:Fluka3} the energy of the positron shower is shown.

Again, the energy released in the scintillator block is high and the SiPM threshold can be left
higher than minimum ionization particles.

The Fig.~\ref{Fig:ElectronEn} right shown the muon decay time from fluka simulations.
The fit of decay time is 2.32$\mu$s

\subsubsection{Primary cosmic rays and the muon shower simulation.}
This point is of great importance in the project. If the events rate is high, this must 
be taken into account in the design of the front-end and acquisition electronics. 
If the rate is too low, the measurement may not be feasible in a reasonable time 
by the students.

The primary cosmic rays, the shower of secondary particles on Torino (Italy)
(35$^\circ$N , 7.7$^\circ$E) produced by them, were simulated using Fluka code.

Fluka provides a complete calculation to determine particle fluxes in the atmosphere at
ground level at the geographical location of the chosen location.
It take care of:

\begin{itemize}
\setlength\itemsep{0.1em}
 \item the spectrum and composition of primary cosmic rays at the local interstellar medium;
 \item the determination of changing conditions in the solar wind magnetic field and the 
   resulting interaction with the inward flow of galactic cosmic rays from the local 
   interstellar medium;
 \item the determination of the trajectories of cosmic rays through the Earth's geomagnetic field;
 \item the transport of the surviving incident cosmic rays through the Earth's atmosphere to 
   various depths.
\end{itemize}
   
\begin{figure}[hbt]
\centering
  \begin{minipage}[]{.48\textwidth}
     \centering
     \includegraphics[width=.98\textwidth]{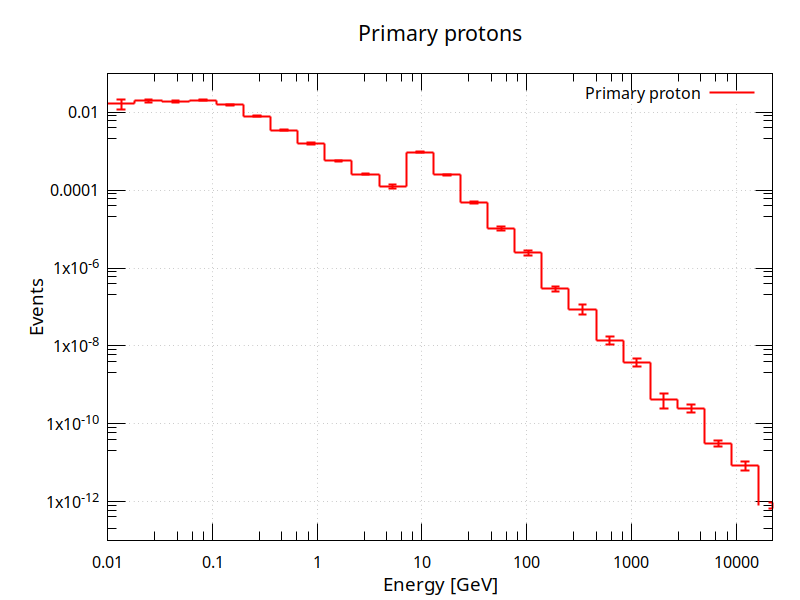}
  \end{minipage} 
  \hfill
  \begin{minipage}[]{.48\textwidth}
     \centering
     \includegraphics[width=.98\textwidth]{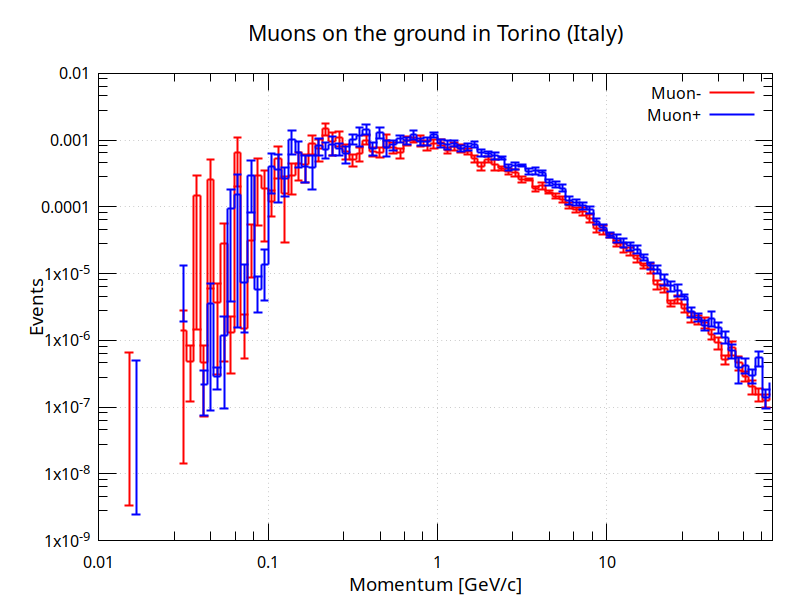}
  \end{minipage} 
\caption{Left: Fluka simulations of primary protons; 
         Right: Muos at ground;
         The simulation is centered at the Torino (Italy) position.
         To be compared with Fig.~\ref{Fig:Energy}.} 
\label{Fig:FlukaCosmicRay}
\end{figure}

Fig.~\ref{Fig:FlukaCosmicRay} shows histograms of primary cosmic and muon shower.

From the simulations, the average $\mu^+$ / $\mu^-$ ratio, for energy below 70 Mev/c is:
\begin{equation}
  R_{\mu^+/\mu^-} = \frac{N_{\mu^+}}{N_{\mu^-}} = 0.240 = 24\%
\end{equation}

and the percentage of $\mu^+ + \mu^-$ with energy below 70 MeV/c over the total is:

\begin{equation}
 N_{\mu} = 2,2\%
\end{equation}

70 MeV/c is the limit beyond which muons pass through the scintillator block, as 
shown by the simulations in Cap.~\ref{Sec:simulations}.

Taking into account the cosmic rate at sea level is about (rule of thumb) 
$r = 0.01 s^{-1} cm^{2} sr^{-1}$, and the AMELIE cube is $15 \times 15 cm^2$,
the total mouns rate is $2.25 s^{-1}$ and the rate of low energetic muons stopped in the cube
is in the order of 2 - 3 per minute.

Altough, in any case, these numbers should be taken with a grain of salt, they are, however, 
reasonably indicative of the feasibility of the measure.
The percentage depends strongly on the materials that can slow down muons that overlie 
or surround the detector.
The simple simulation performed involves only air above the detector up to the stratosphere.

\subsection{Front-end electronic.}
\label{Sec:FrontEndElectronic}

\begin{figure}[hbt]
  \begin{center}
    \includegraphics[width=0.95\linewidth, frame]{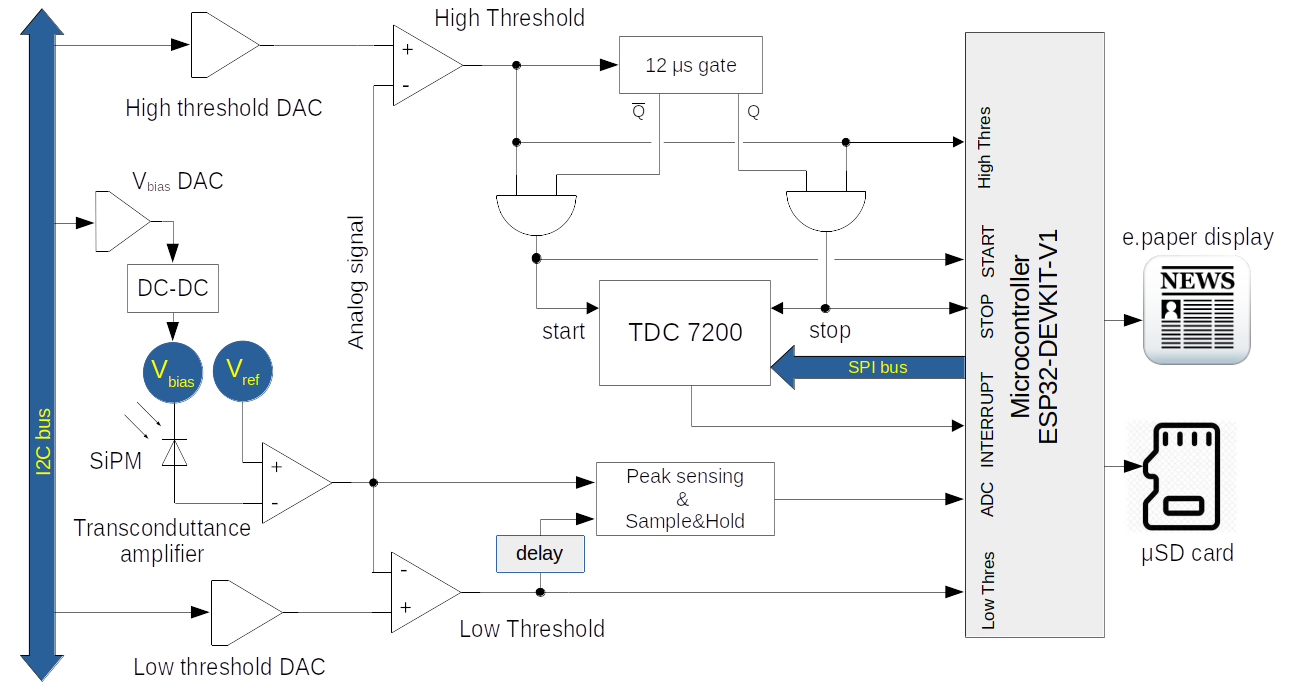}    
  \end{center}
  \caption{Scheme of electronic}
  \label{Fig:Scheme}
\end{figure}

The front-end electronics is divided into three parts~\ref{Fig:Scheme}. 
A small PCB that houses the SiPM attached directly 
to the scintillator and a second PCB that houses the analog circuitry, 
the part that transforms the analog signal provided by the SiPM to digital, 
and a third PCB with the remaining electronics, microcontroller and TDC 
(Time to Digital Converter).

The front-end electronic is placed on top of the scintillator cube, see 
Fig.~\ref{Fig:AmelieAssembly}.

The microcontroller, based on ESP32 chip, will manage the full front-end electronic.
The current setup and data are shown on a e-paper display. 
The working parameters can be modified by a knob and/or push-buttons.
Finally, the data are recorded on a $\mu$SD card for off-line analysis.

The optical photons from stopped muon or decay lepton shower are collected by a SiPM. 
A fast current pulse is generated by each photons burst by a fast trans-conductance
amplifier. 
Since the trans-conductance amplifier is keep to a reference signal
V\textsubscript{ref}, the output will therefore be negative with respect to 
V\textsubscript{ref}.
The transconductance analog signal is converted into a logic signal
by two fast discriminators, V\textsubscript{ThHigh} and V\textsubscript{ThLow}.

The charge of the SiPM signal is proportional to the number of photons collected. 
Consequently, the peak of the signal is also proportional to the number of photons collected.
A pick-sensing circuit, followed by sampling-and-hold, adapts the fast SiPM 
signal to the microcontroller's internal ADC. 

The gain of the SiPM is controlled by V\textsubscript{bias},
generated on-board by DAC and DC-DC converter, under control of microcontroller. 
A DC-DC converter will generate the required V\textsubscript{bias} = 25-32 V. 
The DAC is controlled by the ESP32 using the I2C bus. 
The noise generated by DC-DC oscillator is filtered by a onboard dual T filter.
The reference voltage V\textsubscript{ref} is controlled by a stable ``voltage reference'' onboard.
In order to have a large dynamic range, V\textsubscript{ref} will be 4.096 V.

From the Fluka simulations, see Sec.~\ref{Sec:simulations}, the number of 
photons collected by SiPM generated by stopped muons or leptons from muon 
decay are, at least, two time the photons generated by a high energetic 
muon that cross the scintillator block.

To take advantage of this typical feature of muon decay, two discriminators 
were installed, one with a low threshold and a second with a higher threshold. 

The low threshold discriminator, is suitable to detect minimum ionization particles,
and the second, with high threshold, to detect stopped muon or decay lepton. 
The high threshold signal is detected by the microcontroller by polling 
the input pin. It is used for time measurements using the TDC.

\subsubsection{TDC characteristics and measure of muon decay time.}
\label{Cap:TDCtest}
In order to measure the decay time of the muon, we plan to use the TDC7200 chip
produced by Texas Instruments~\cite{TDC7200}.

The chip is designed to be used in LIDAR applications or ultrasonic sensing 
measurements such as water flow meter, gas flow meter, and heat flow meter.

Its main features are: 1) resolution 55 ps; 2) two ranges: 12 to 500 ns and 250 ns to 8 ms; 3) readout through a SPI bus, TDC time is stored in internal registers.

The chip require a simple but very stable 8 MHz clock and the device has an 
internal self-calibrated time base which compensates for drift over time and temperature. 
Self-calibration enables time-to-digital conversion accuracy in the order of picoseconds. 
This accuracy makes the TDC7200 ideal for our application, indeed, in reality, 
it exceeds the requirements, and the time resolution is downgraded 
to 1 ns.

The TDC has separated START and STOP input, while, in our case, we have 
a single SiPM which generates both the START and STOP signals.

A little logic had to be built to separate the two signals, as shown in Fig.~\ref{Fig:Scheme}.

\begin{figure}[hbt]
  \begin{center}
    \includegraphics[width=0.95\linewidth, frame]{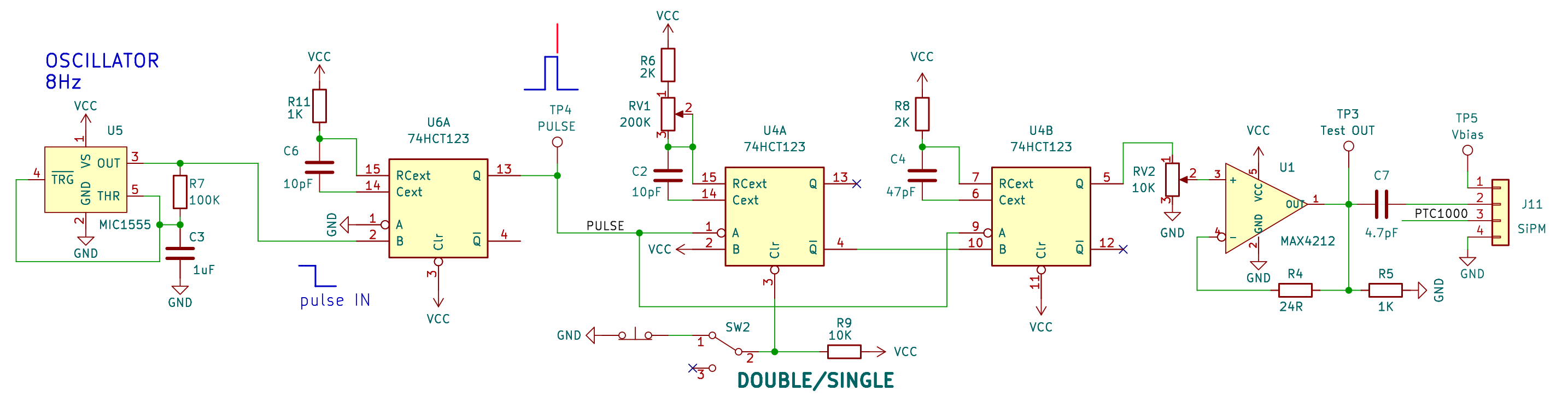}
  \end{center}
  \caption{Test circuit capable of generating a double pulse at the input 
           of the front-end board.}
  \label{Fig:DoublePulser}
\end{figure}

It is evident that the measurement of time between two signals constitutes the distinguishing feature of this detector.
Special care was taken in the testing and calibration of the TDC by designing 
and constructing a signal pulser capable of generating a double pulse in the 
range provided by the gate. The circuit, shown in Fig.~\ref{Fig:DoublePulser}
emulates the SiPM by injecting a charge equivalent to the transconductance 
amplifier.

The TDC is designed to generate an interrupt signal to the microcontroller only 
when it receives a STOP signal. This interrupt signal is not used by 
microcontroller.

Only the high threshold signal is used to generate TDC START and STOP signals.

The START will also trigger a not-retriggerable flip-flop gate with a signal width of 12 $\mu$s. 
This signal inhibits the generation of a second START signal and allows the generation of a STOP signal if it arrives in the gate time.

The TDC and charge ADC are then readout by microcontrolled and the system is set for 
the next event.

\subsection{PSpice simulations.}
\label{Sec:PSpice}
The front-end electronics has been extensively simulates using LTspice
in order to optimize the active and passive components.
LTspice was developed by Linear Technology, now acquired by Analog Devices
~\cite{LTspice}.
It is a graphical interface to the well known SPICE, 
Simulation Program with Integrated Circuit Emphasis, a circuit simulation 
program developed in 1975 by Electronics Research Laboratory California 
University in Berkeley.

\subsubsection{Simulation of SiPM signal.}
The current generated from 1 to 40 photo electrons in the scintillator,
varies between 13 and 512 $\mu$A.

\begin{figure}[hbt!]
\centering
  \includegraphics[width=0.98\linewidth, frame]{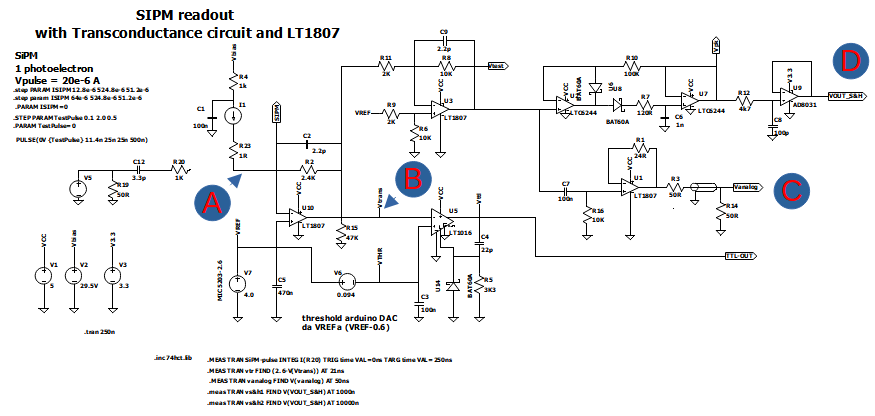} \\
  \begin{minipage}[]{.24\textwidth}
     \centering
     \includegraphics[width=.98\textwidth]{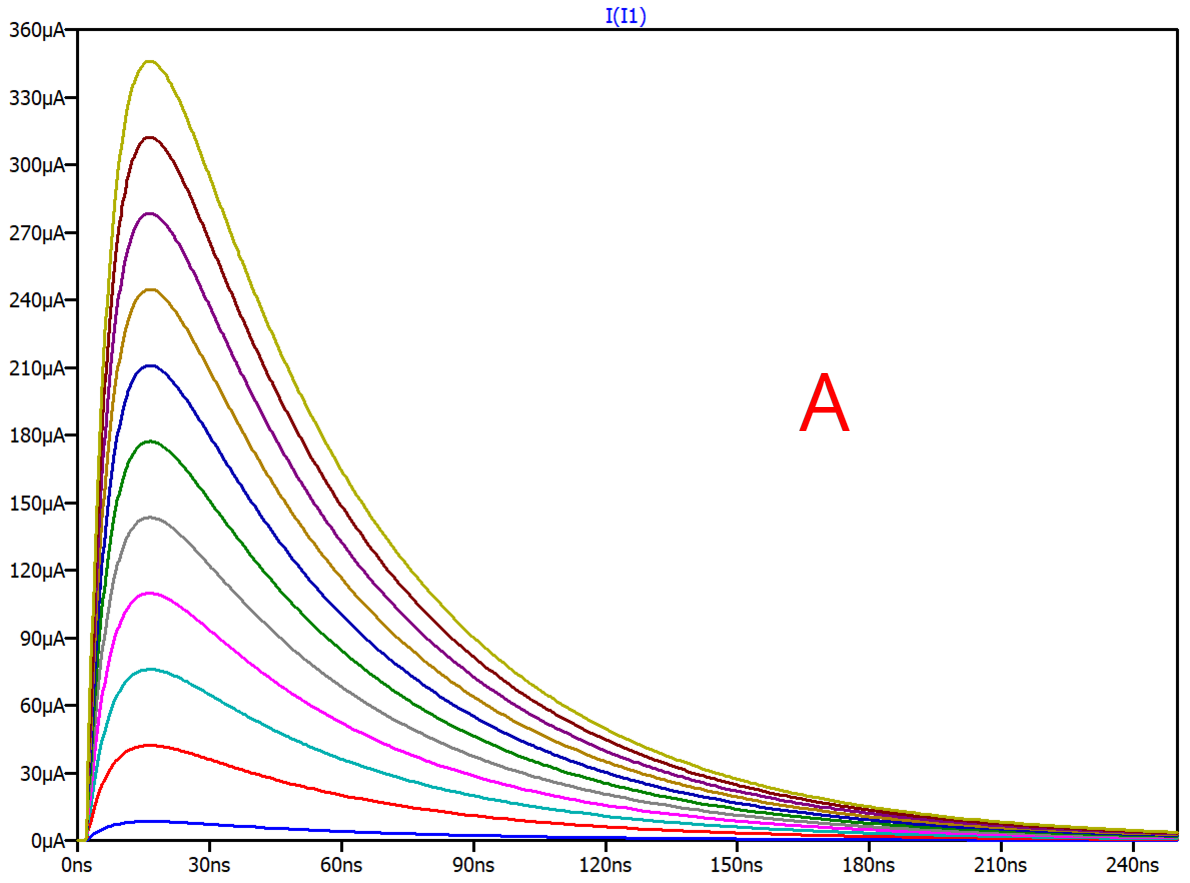}
  \end{minipage} 
  \hfill
  \begin{minipage}[]{.24\textwidth}
     \centering
     \includegraphics[width=.98\textwidth]{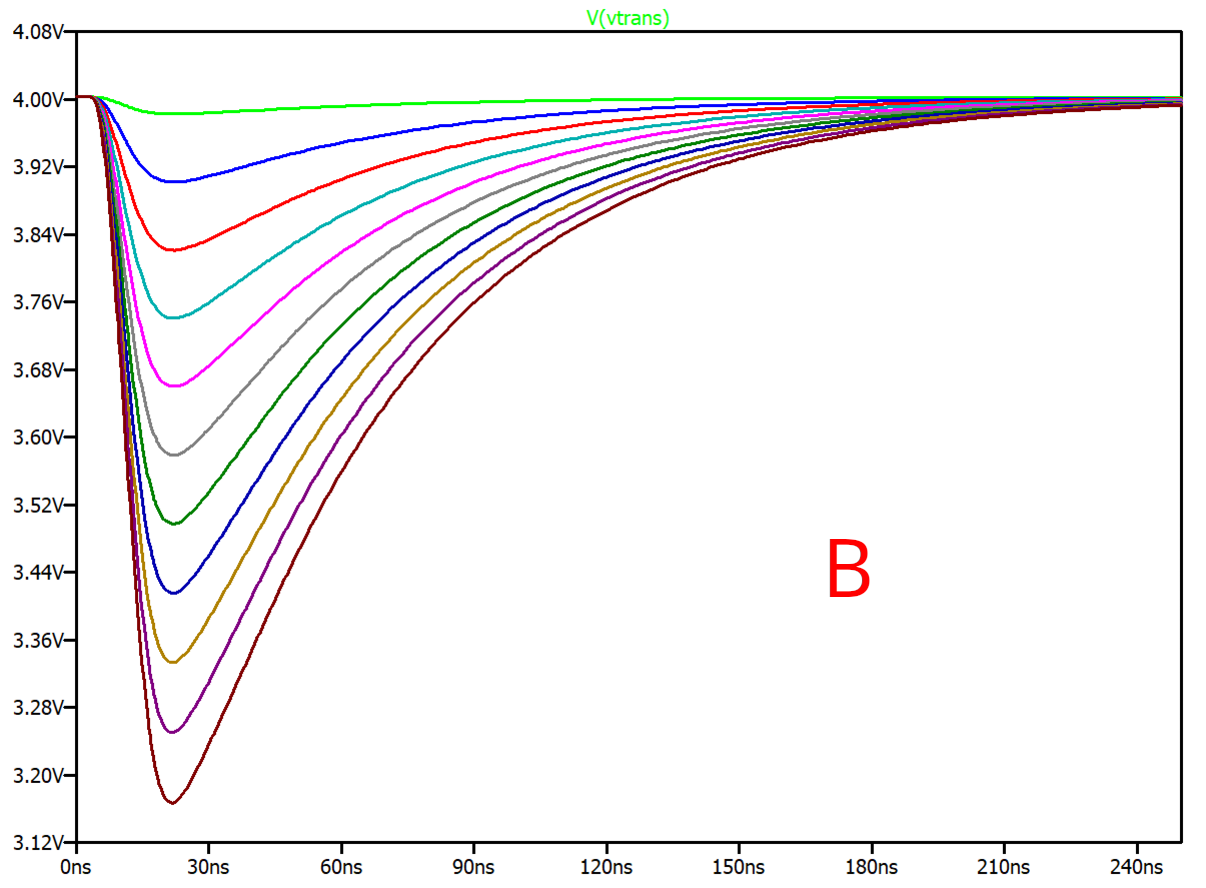}
  \end{minipage} 
  \hfill
  \begin{minipage}[]{.24\textwidth}
     \centering
     \includegraphics[width=.98\textwidth]{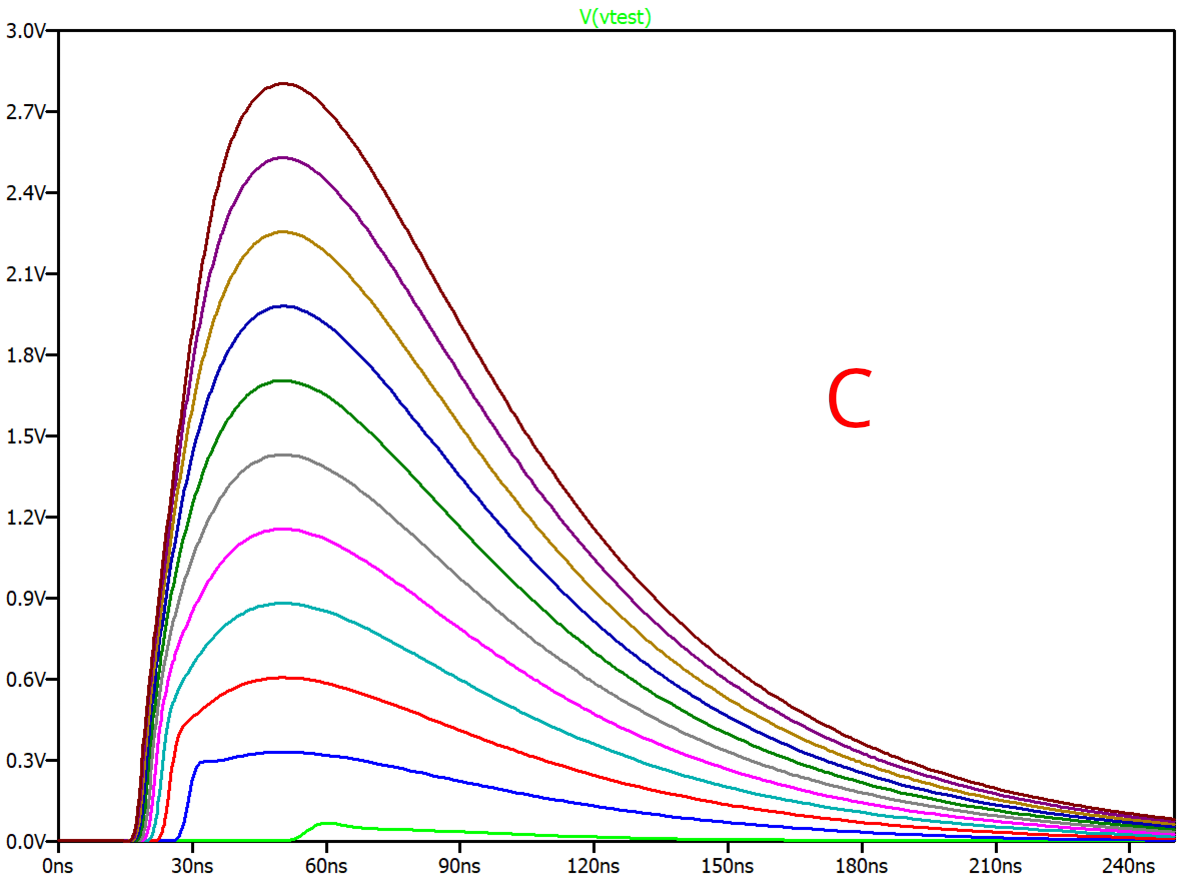}
  \end{minipage} 
    \hfill
  \begin{minipage}[]{.24\textwidth}
     \centering
     \includegraphics[width=.98\textwidth]{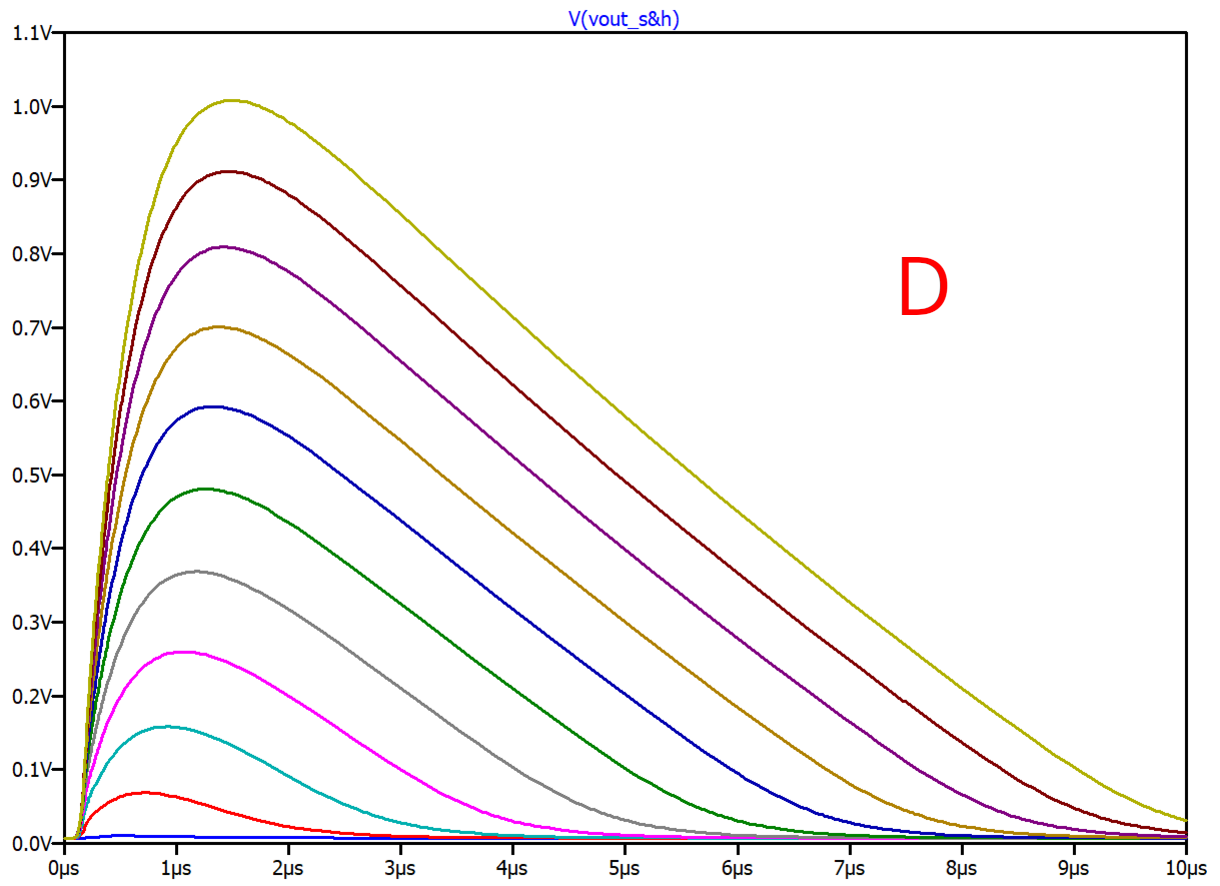}
  \end{minipage} 

\caption{PSpice simulations of front-end electronics. 
         The curves relate to a variable number of photoelectrons from 1 to 40.
         Top: Scheme of PSpice simulation;
         (A): SiPM current generated from 1 to 40 photoelectrons;
         (B): output of transconduttance amplifier with $V_{ref} = 4V$;
         (C): output of fast inverter amplifier;
         (D): output of peak-sensing amplifier, the peak is delayed by 1.2 $\mu$s.} 
\label{Fig:PSpice}
\end{figure}

The analog parameters of the various components are optimized.
The simulated circuit simulated with PSpice is shown on the top of Fig.~\ref{Fig:PSpice}.

The SiPM photoelectron response is simulated by a exponential current pulse, 
shown Fig.~\ref{Fig:PSpice} (A):

\begin{equation}
 I(t) = I_{pulsed} \times ( 1 - e^{-\frac{t - T_{RiseDelay}}{\tau_{Rise}}}) - 
        I_{pulsed} \times ( 1 - e^{-\frac{t - T_{FallDelay}}{\tau_{Fall}}}) \\
\end{equation}
  where:
\begin{center}
  $I_{pulsed} = 12.8 \mu A$ to $524.9 \mu A$ i.e. 1 to 40 photoelectrons; \\
  $T_{RiseDelay} = 2 ns$  \\
  $\tau_{Rise} = 6 ns$    \\
  $T_{FallDelay} = 2 ns$  \\
  $\tau_{Fall} = 50 ns$
\end{center}

The photoelectrons collected by SiPM are amplified by the optical detector. 
The gain of SiPM is shown in Fig.~\ref{Fig:SiPM_QE}.
Then, the charge generated by photoelectrons is:

\begin{equation}
    Q = N_{phot} \times G \times 6.241 \times 10^{18} =  \int_{0}^{\infty} I(t) \, dt \hspace{10mm} [C]
\end{equation}

In the PSpice simulations the gain of the SiPM was set to $3.5 \times 10^6$ and
the integral of the current signal show in Fig.~\ref{Fig:PSpice} (A) is
the charge for $N_{phot} = 1 \; to \; 40$.

The results of the simulations are shown in Fig.~\ref{Fig:PSpice}.
When compared to the measured values of Fig.~\ref{Fig:SiPManalog}, the simulations are quite 
realistic and enough for this letter of intent.

\subsubsection{SiPM Amplitude/Charge measurement.}
In addition to the TDC time measurement, we want to monitor also the amplitude of the:

\begin{itemize}
 \item MIP muons crossing the scintillator block;
 \item the energy released by the muons stopped in the scintillator block;
\end{itemize}

The SiPM signal is a short and unstable signal, see Fig.~\ref{Fig:PSpice}-A . 
The ADC of the microcontroller, conversely, requires a stable signal 
throughout the conversion time. To adapt the signal to the ADC, 
a pick-sensing circuit followed by sample\&hold was designed.

\begin{figure}[hbt]
  \includegraphics[width=0.95\linewidth, frame]{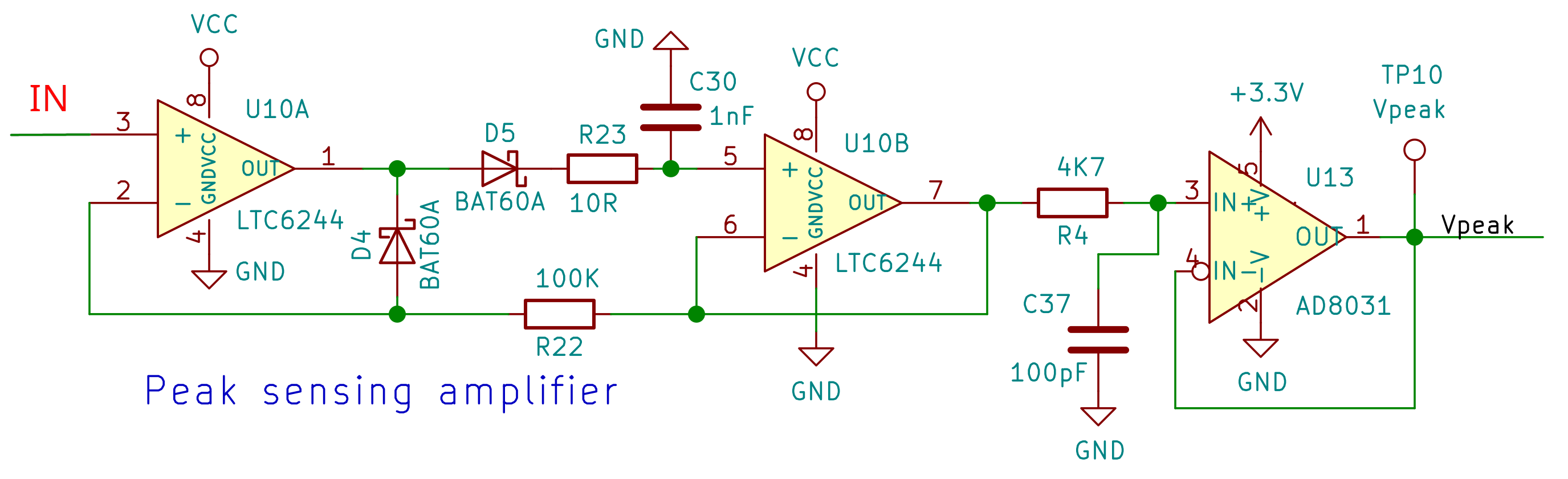} 
  \caption{Scheme of peak sensing amplifier, simulated with PSpice. See Fig.~\ref{Fig:PSpice}}
  \label{Fig:PeakSensing}
\end{figure}

The peak sensing output signal is proportional to the charge collected by SiPM.
This signal is delayed by 1.2 $\mu$s from fast SiPM signal.
The input signal of Sample\&Hold chip is the delayed by 12. $\mu$s and will stabilize 
its output for a time required by microcontrolled ADC to digitize the fast input.
The scheme of the peak sensing amplifier, which use two Schottky diodes and capacitors 
to integrate the charge, is shown in Fig.~\ref{Fig:PeakSensing}.

\begin{figure}[hbt]
\centering
  \begin{minipage}[]{.32\textwidth}
     \centering
     \includegraphics[width=.98\textwidth]{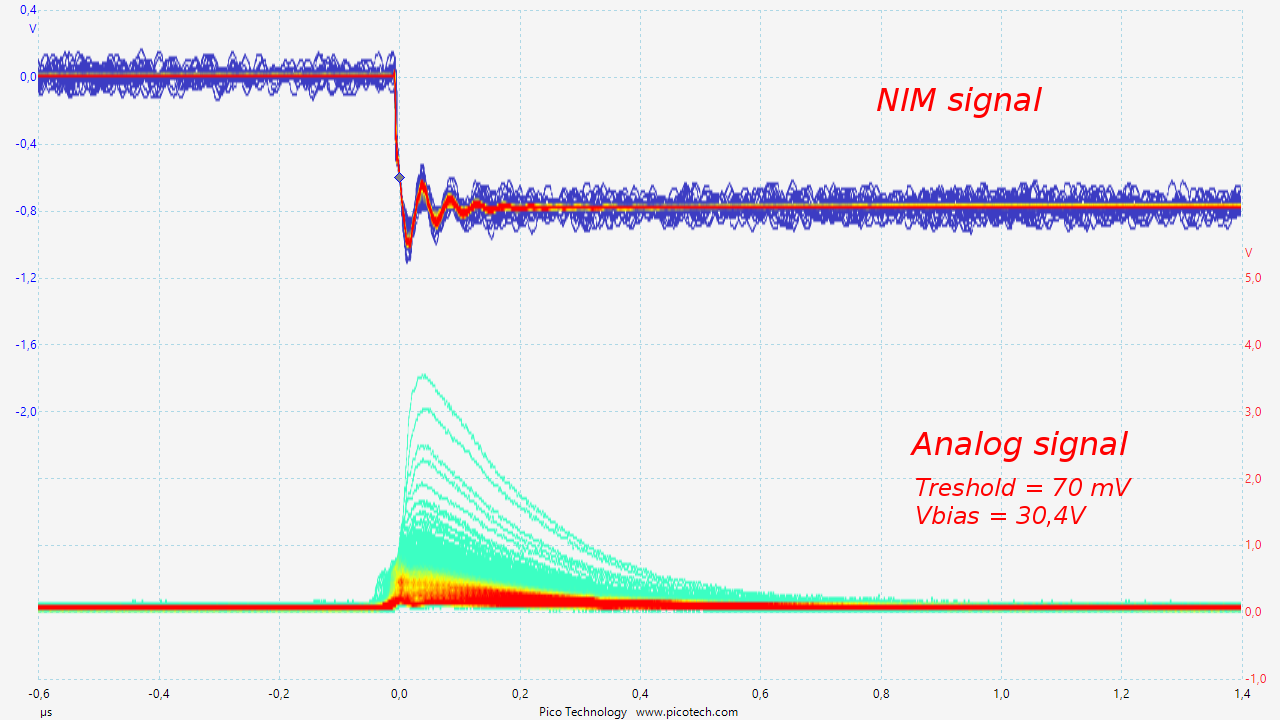}
  \end{minipage} 
  \hfill
  \begin{minipage}[]{.32\textwidth}
     \centering
     \includegraphics[width=.98\textwidth]{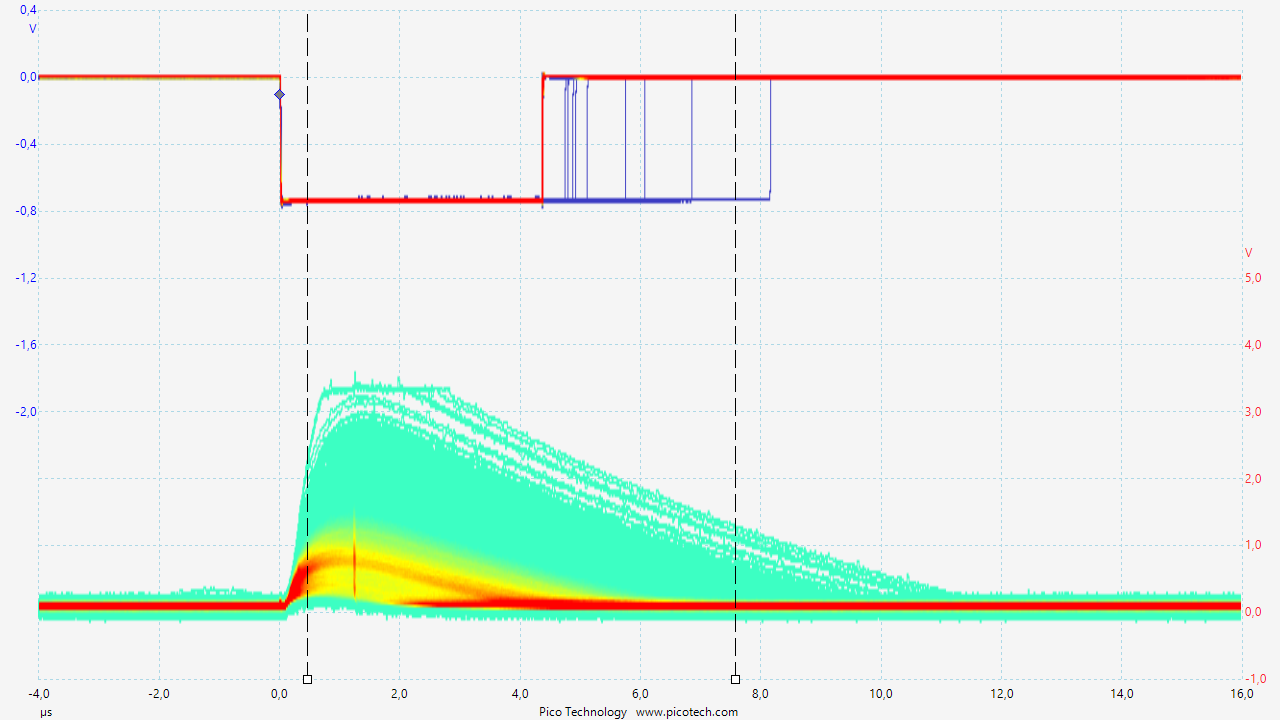}
  \end{minipage} 
  \hfill
  \begin{minipage}[]{.32\textwidth}
     \centering
     \includegraphics[width=.98\textwidth]{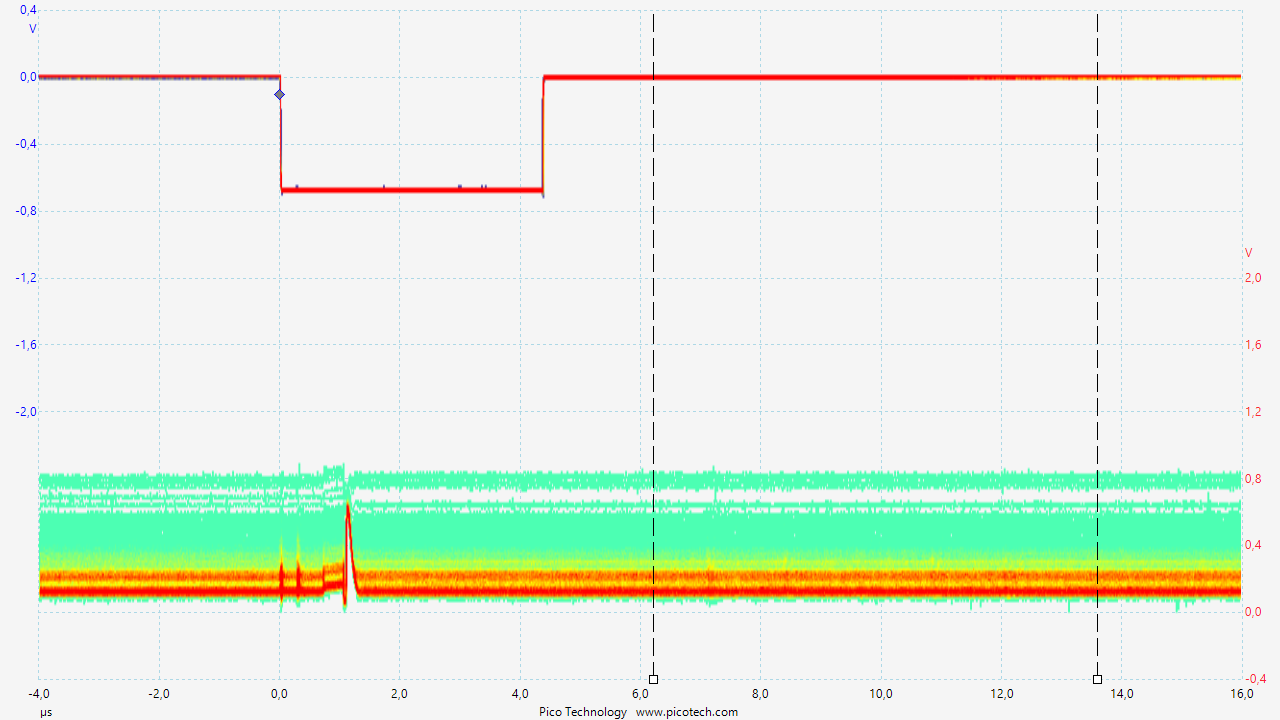}
  \end{minipage} 
\caption{Left: output signal of SiPM transconductance amplifier for a minimum ionizing particle;
         Center: output of the peak sensing circuit;
         Right: output of Sample\&Hold DS1843. The two vertical lines are the start and stop of 
         microcontroller ADC conversion. The 300 ns Sample\%Hold input signal is generated on the top of 
         peak sensing circuit, shown by the small peak.
         The upper signal is always the output of the discriminator (NIM level).} 
\label{Fig:SiPManalog}
\end{figure}

In Fig.~\ref{Fig:SiPManalog} the oscilloscope signal for a minimum ionization muon is shown.
The detector used is a scintillator tile of dimensions 15x15x1 cm$^3$ with the same 
SiPM of this proposal~\cite{ONSEMI}.

On the left, the fast signal of SiPM is shown; In the center, the output of peak sensing amplifier
is shown and, on the right, is the output of the Sample\&Hold circuit is shown.
The sample\&hold is triggered by the High Threshold discriminator signal 
delayed by 1.2 $\mu$s in order to hold the peak of the signal. 
The start of ADC readout is further delayed by 7 $\mu$s by microcontroller 
in order to have a stable signal from the sample\&hold. 
The ADC readout time is between the two vertical lines where the analog
signal is stable as shown on the oscilloscope.

\subsection{Particles Identification.}
It would be possible to identify the type of incident particle and/or $\mu$ charge 
by adding time of flight detector or magnetic fields.
This is routinely done in physics experiments at large laboratories such as CERN in Geneva.
This would have greatly complicated the detector without adding anything to 
the educational purposes for which it was built.
The solid angle and, consequently, the counting rate would also be greatly reduced.
Therefore, it was decided not to identify the incident particles but to 
adopt some techniques in data analysis that allow, nevertheless, 
to achieve the goal, indeed, showing the data analysis approach techniques 
used in real experiments. 

\section{Description of AMELIE detector.}
\label{Sec:GenDescription}

The main aim of the project is the measurement of the muon lifetime in a school environment with the use of simple and cheap instruments.

In other words, the idea is to bring students to the knowledge of modern physics 
by allowing them to carry out real scientific experiments, collect the data, 
make any corrections, analyze the results and compare them with existing 
models or data.

The purpose of this apparatus is to measure the lifetime of the muon, 
one of the elementary particles that reach the ground produced by cosmic rays.

\begin{figure}[hbt]
\centering
  \begin{minipage}[]{.34\textwidth}
     \centering
     \includegraphics[width=.98\textwidth]{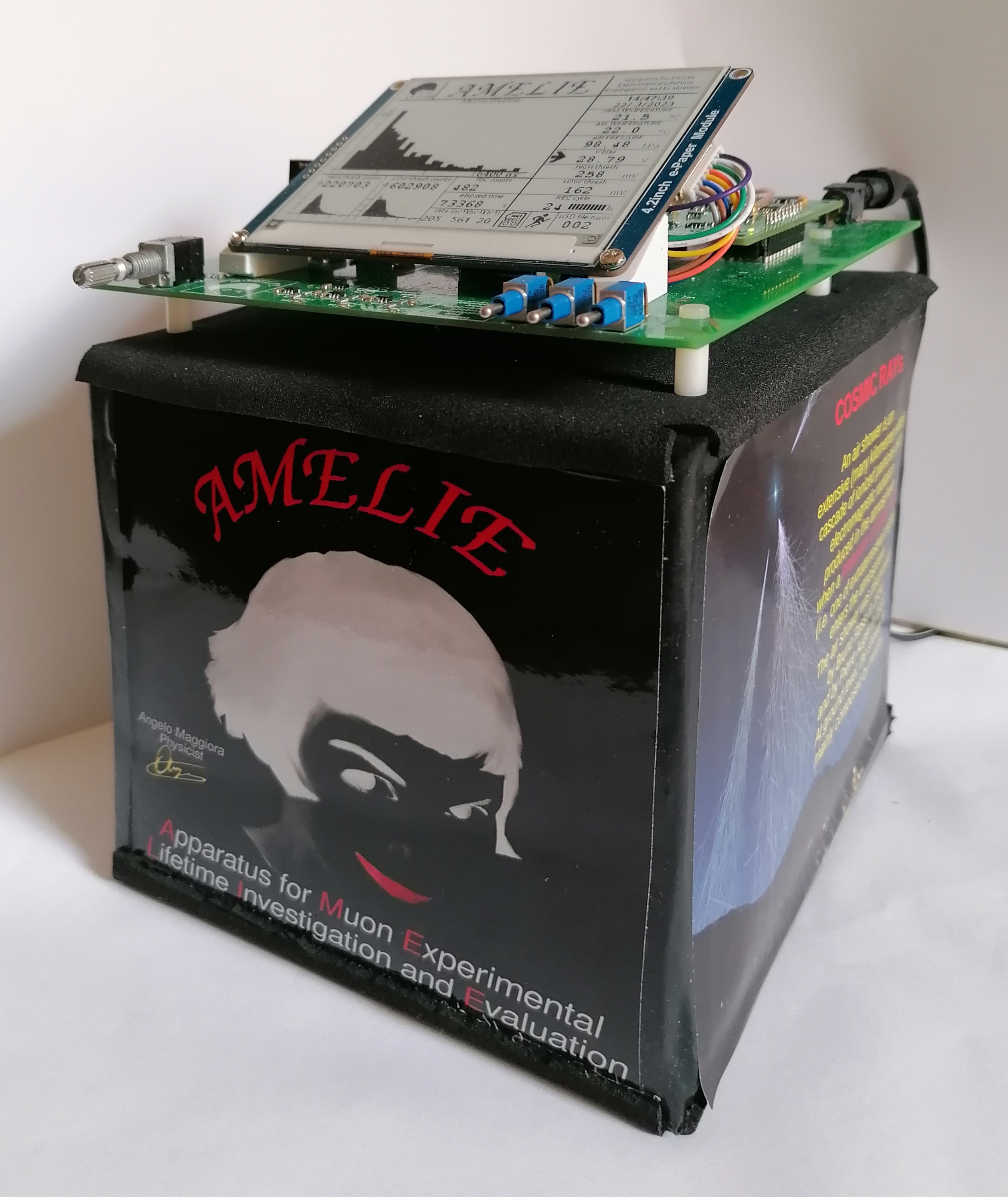}
  \end{minipage} 
  \hfill
  \begin{minipage}[]{.64\textwidth}
     \centering
     \includegraphics[width=.98\textwidth]{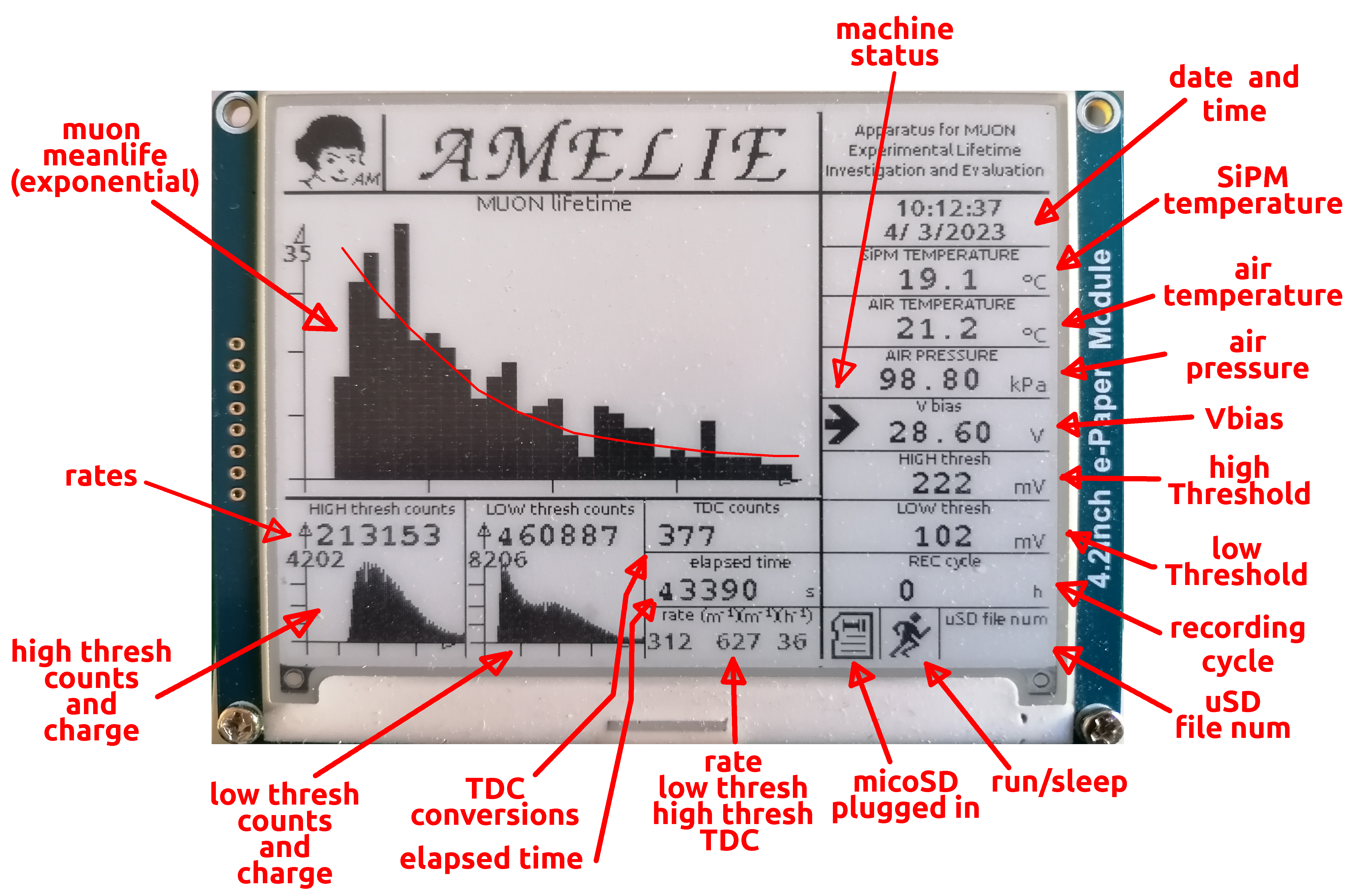}
  \end{minipage} 
\caption{Detector assembly and the display} 
\label{Fig:AmelieAssembly}
\end{figure}

The Amelie apparatus and its display is shown in Fig.~\ref{Fig:AmelieAssembly} left.
In the same figure, the e-paper display is also shown.

The system handles four operating parameters, essential for the set-up of
the detector:

\begin{itemize}
  \setlength\itemsep{0.1em}
  \item SiPM V\textsubscript{bias} [V];
  \item V\textsubscript{HighThresold} [mV]:
  \item V\textsubscript{LowThresold} [mV];
  \item Recording cycle [h].
\end{itemize}

In addition, some other parameters as atmospheric pressure, atmospheric 
temperature, and of the SiPM, are only informative of the date and time, 
environmental conditions. 
Absolute and relative rates as well as acquisition time also belong 
to this category.
This parameters are very important for the monitoring of data acquisition and 
for data quality check.

Finally, three histograms are shown, two containing the charge signal for 
low and high threshold and the last, the most important one, the time of TDC.
The y-axis of histograms is autoranging. When a value reaches the maximum 
value, the range of y is doubled.

The display used is a 400 x 200 pixels e-paper. This type of display is very 
visible even without back lighting, does not consume anything except during 
refresh, and the data is not erased even if the power is removed.
The problem with this display is the long refresh times. 
Specifically, this display takes 0.4s to partially refresh the data. 
In this time the microcontroller is dedicated full time to refresh and any 
data from the SiPM would be lost.

To avoid having a high dead time during the acquisition, refresh is done every 
second during the first 10s after the power-on. 
Then it is done every 10m, unless any button is 
operated, especially the arrow indicating machine status.
In this case, the refresh is done every second for the next 10s.

This configuration is perfectly compatible with the detector's purpose of 
long acquisition times, hours or days, without changing any parameters.

\begin{figure}[hbt]
\centering
  \begin{minipage}[]{.48\textwidth}
     \centering
     \includegraphics[width=.98\textwidth]{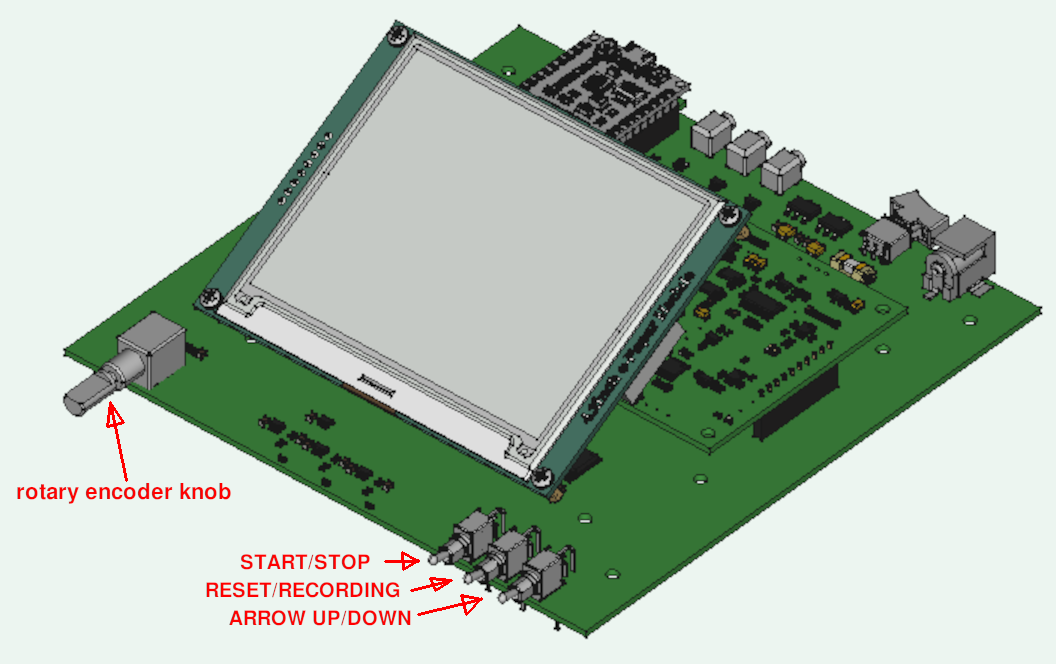}
  \end{minipage} 
  \hfill
  \begin{minipage}[]{.48\textwidth}
     \centering
     \includegraphics[width=.98\textwidth]{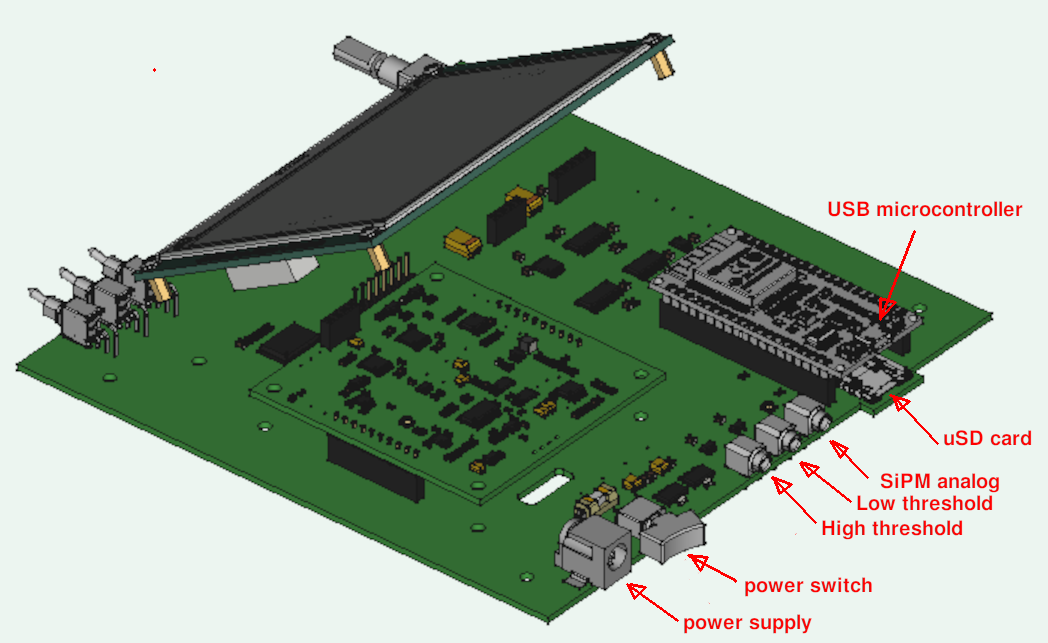}
  \end{minipage} 
\caption{Left: front view with commands; Right: back view with output signals,
          power switch and $\mu$SD card} 
\label{Fig:CommandAndOutput}
\end{figure}

System commands and outputs are shown in Fig.~\ref{Fig:CommandAndOutput}.

\subsection{Front panel.}
The front panel is shown in Fig.~\ref{Fig:CommandAndOutput} at left.

Three switches on the right and a rotary encoder on the left are provided for 
changing parameters.
The encoder also functions as a switch. 
When pressed, the microcontroller stores all operating parameters 
in the internal e-prom memory. 
This memory is read when the detector is turned on, and in operating 
parameters, at the time of power off, are reset. This procedure is used 
to restart the system in case of power failure at the previous conditions.

Each of the three switches has three positions. One central, neutral, 
one upward and one downward.
How they work is shown in Tab.~\ref{Tab:Switches}.

\begin{table}[ht]
  \begin{center}
    \begin{tabular}{| c | c | c |}
       \hline
       switch  & position & function                                \\ \hline \hline
       \multirow{2}{*}{left}           & up       & START           \\ \cline{2-3}
                                       & down     & STOP            \\ \hline
       \multirow{2}{*}{central}        & up       & RESET           \\ \cline{2-3}
                                       & down     & RECORD DATA on $\mu$SD OR CLOSE THE FILE \\ \hline
       \multirow{2}{*}{right}          & up       & MACHINE STATUS ARROW+  \\ \cline{2-3}
                                       & down     & MACHINE STATUS ARROW-  \\ \hline
       \multirow{2}{*}{rotary encoder} & push     & STORE CONFIGURATION on e-PROM \\ \cline{2-3} 
                                       & rotation & SET PARAMETER UNDER ARROW \\ \hline
    \end{tabular}
  \end{center}
  \caption{Operating switches}
  \label{Tab:Switches}
\end{table}

\begin{figure}[hbt]
\centering
  \hfill
  \begin{minipage}[]{.18\textwidth}
     \centering
     \includegraphics[width=.4\textwidth]{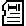}
     \caption*{\small $\mu$SD plugged}
  \end{minipage} 
  \hfill
  \begin{minipage}[]{.18\textwidth}
     \centering
     \includegraphics[width=.4\textwidth]{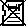}
     \caption*{\small $\mu$SD unplugged}
  \end{minipage} 
  \hfill
  \begin{minipage}[]{.18\textwidth}
     \centering
     \includegraphics[width=.4\textwidth]{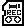}
     \caption*{\small $\mu$SD recording}
  \end{minipage} 
  \hfill
  \begin{minipage}[]{.18\textwidth}
     \centering
     \includegraphics[width=.4\textwidth]{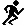}
     \caption*{\small DAQ running}
  \end{minipage} 
  \hfill
  \begin{minipage}[]{.18\textwidth}
     \centering
     \includegraphics[width=.4\textwidth]{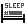}
     \caption*{\small DAQ stopped}
  \end{minipage}
  \hspace{20mm}
  \caption{Icons showing the status of the system configuration.}
  \label{Fig:icons}
\end{figure}

The bottom right part of the display shows the current status of the system 
via icons. The icons are shown in Fig.~\ref{Fig:icons}

In addition, the system has a $\mu$SD memory card where data from each event 
are stored.
In the recording state, two files are provided on $\mu$SD. The first one stores 
the data of each event above the high threshold, the second one, in addition 
to this data, also the time given by the TDC in ns.

The REC button is a toggle switch. The REC state changes each time it is pressed.
Also, when the recording is turned off, the file on $\mu$SD is closed.
This action is very important in case of manual operation of the system for proper 
prevention of recorded data.

Files written to $\mu$SD are automatically numbered at each run.
The maximum number of runs is 120. 
After that, you have to empty the $\mu$SD. 
The current run number is shown in the right pane at the bottom of the display.

\subsection{Back panel.}
The rear panel is shown in Fig.~\ref{Fig:CommandAndOutput} at right.

In the rear panel contains the power supply jack (V = 7 $\div$ 12 Vdc), 
the main switch, the digital outputs of low and high threshold signals, 
the analog signal of SiPM, the $\mu$SD, and the micro-USB connector of the microcontroller. The analog and digital signals must be terminated at $50 \Omega$.

\subsection{Cyclical recording.}
The detector allows two types of recording on $\mu$SD card: 
1) totally manual; 2) automatic cyclic recording.

The selection is made by setting the REC cycle parameter in the window at 
the bottom right.
If set to zero, the user decides whether to start data acquisition, stop it, 
and record the data to $\mu$SD, using the switches on the front panel. The process 
has no set time limit.

Cyclic recording on $\mu$SD is enabled by setting values $>$0 in the window 
and setting START and REC with the switches.
The recording cycle can range from 1 to 100 hours. At the end of each cycle 
the files are closed, all counts, graphs etc. are reset and two new files 
are opened with the increased number.
The "REC cycle" box shows a bar indicating the percentage of time in the 
current cycle, and the bottom box shows the number of files open in that cycle.

The cycle time can be changed by bringing the arrow indicating "machine status" 
to the proper box using the right switch. Turning the "rotary encoder" 
clockwise or counterclockwise, sets the desired cycle time.

\subsection{Adjustment of the operating parameters.}
The operating parameters, V\textsubscript{bias} and high or low threshold, 
can be changed by bringing the arrow indicating "machine status" to the 
desired box.
Then, they are increased or decreased by turning the "rotary encoder" 
clockwise or counterclockwise.

It is advisable to memorize the chosen status by pressing the "rotary encoder" button.

\section{Data acquisition and analysis.}
\label{Sec:analisys}

Before starting data taking, preliminary checks were made on the apparatus to 
have confidence in the quality of the data.

\subsection{Data quality check.}

Using the test pulser circuit in Cap.~\ref{Cap:TDCtest} it is possible to 
simulate a single or double pulse of the SiPM.
The time between two pulses given by the TDC7200 was compared with the 
time between two output pulses using a digital oscilloscope. 
Both gave the same result within 1 ns.

\begin{figure}[hbt]
\centering
  \begin{minipage}[]{.48\textwidth}
     \centering
     \includegraphics[width=.98\textwidth]{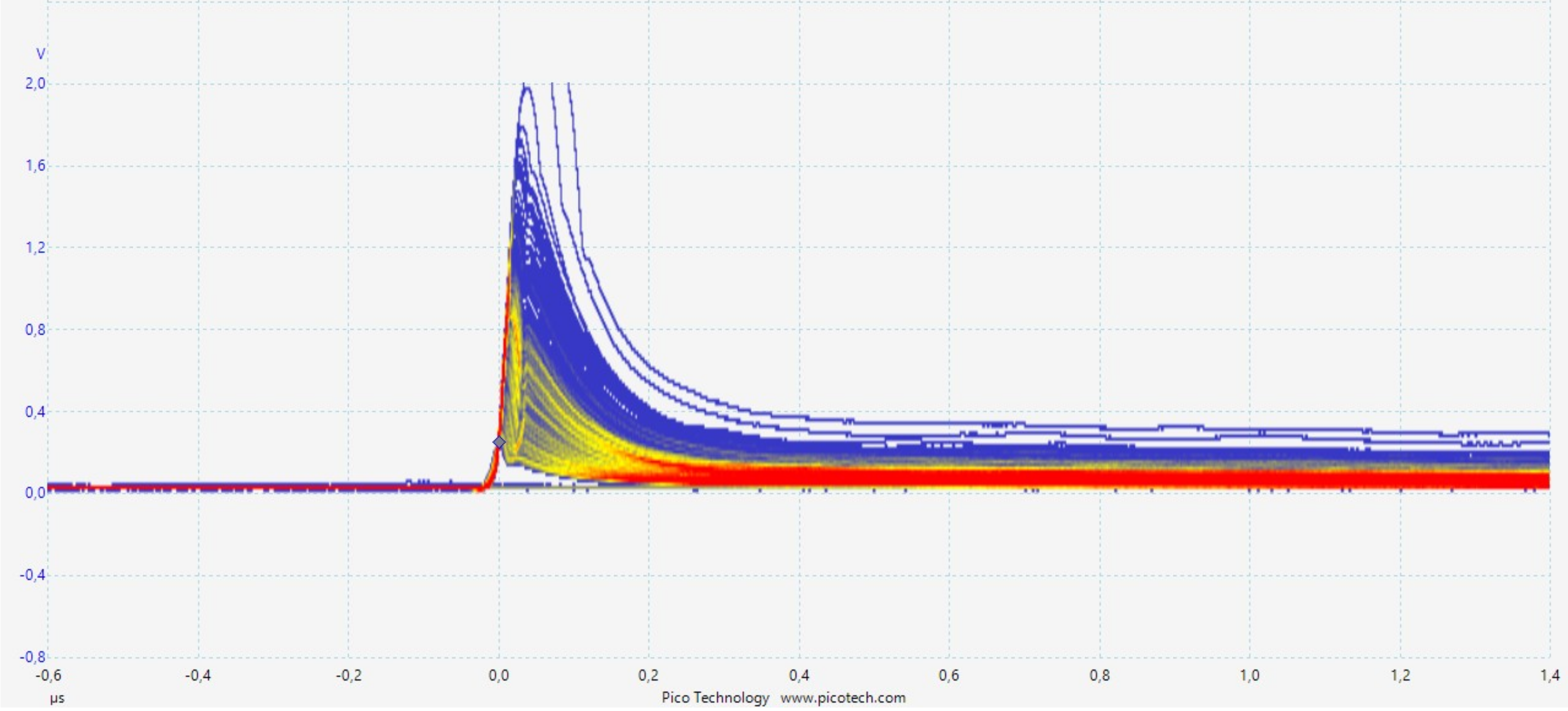}
  \end{minipage} 
  \hfill
  \begin{minipage}[]{.48\textwidth}
     \centering
     \includegraphics[width=.98\textwidth]{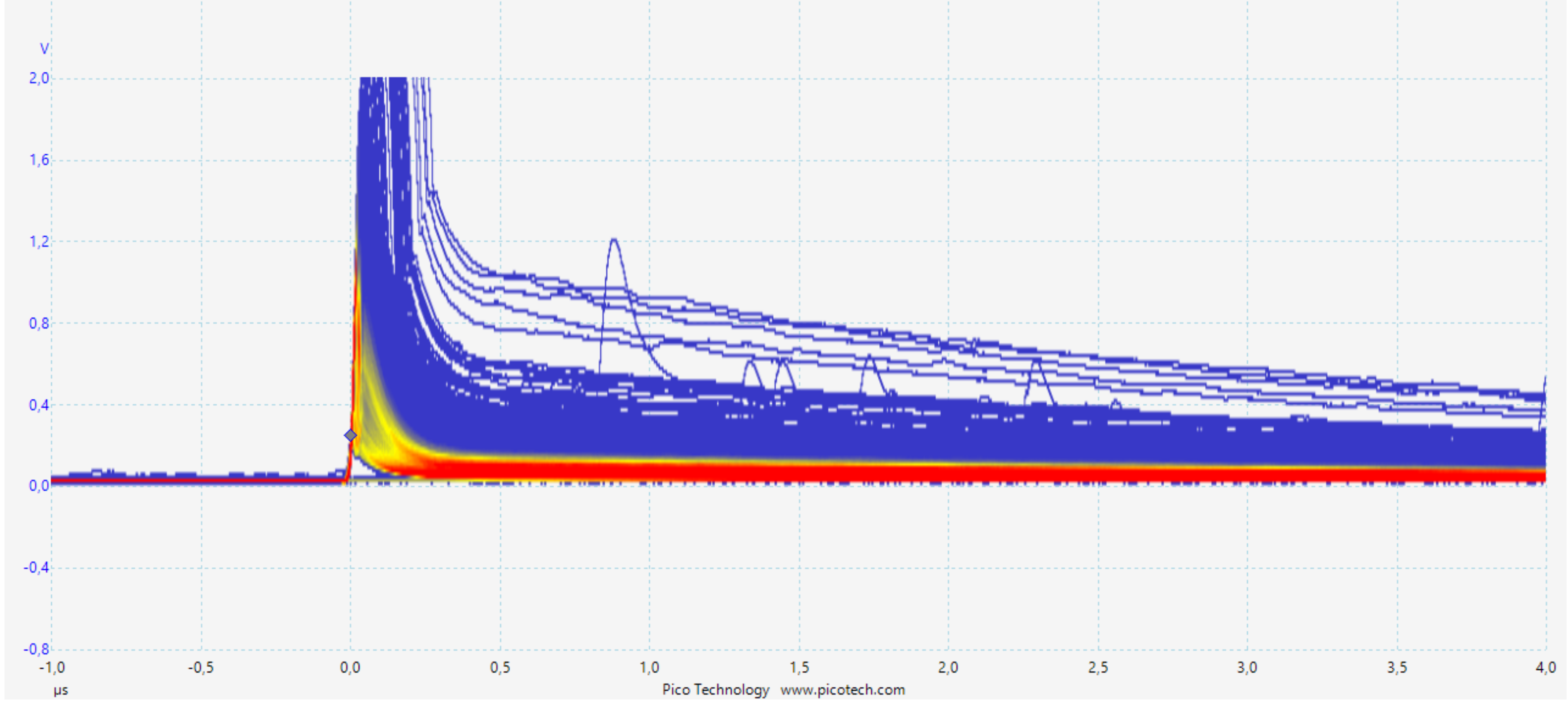}
  \end{minipage} 
\caption{Output signal of the SiPM} 
\label{Fig:SiPManalog2}
\end{figure}

By connecting the SiPM sensor to the front-end circuit, it was possible 
to verify the analog output of the sensor.
The AMELIE analog output signal of the SiPM was connected to an oscilloscope
that can integrate the input signal for a long time.
The output signal is shown in Fig.~\ref{Fig:SiPManalog2}.
Also visible, on the right, some second signals that follow 
the first one within the gate time.

\subsection{Data tacking for SiPM gain $G = 5 \times 10^6$.}

In order to check for systematic errors, three data sets were collected at different SiPM gains, $5 \times 10^6$, $6 \times 10^6$ and $7 \times 10^6$.
The procedure for collecting the data was the same in all cases.
Data were collected in runs of 24 hours each, using the cycling recording facility,
for a totoal of 20 runs for each SiPM gain has been recorded.

For each run, using the data recorded on $\mu$SD, several tests were
carried out for each run to check the stability of the system during the 
24h of data taking and the quality of the data collected. 

\begin{figure}[hbt]
\centering
  \begin{minipage}[]{.48\textwidth}
     \centering
     \includegraphics[width=.95\textwidth]{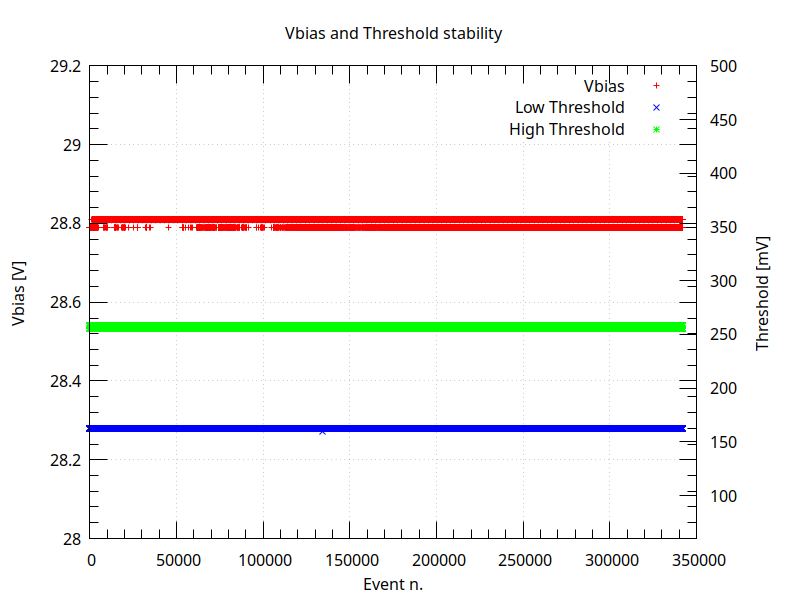}
  \end{minipage} 
  \hfill
  \begin{minipage}[]{.48\textwidth}
     \centering
     \includegraphics[width=.95\textwidth]{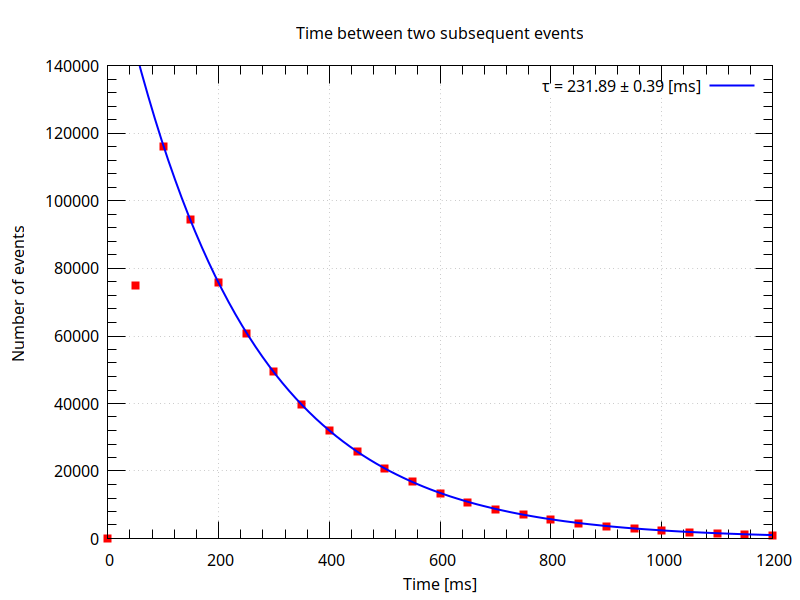}
  \end{minipage} 
\caption{Left: Data quality check; test of V\textsubscript{bias}, 
        V\textsubscript{LowThresh} and V\textsubscript{HighThresh} stability;
        Right: measured time between two cosmic rays.} 
\label{Fig:QualityCheck1&charge}
\end{figure}

In particular, V\textsubscript{bias}, V\textsubscript{HighThresh} and
V\textsubscript{LowThresh} signals has been carefully checked. 
The stability is optimal as shown by the graphs in 
Fig.~\ref{Fig:QualityCheck1&charge} left.

\begin{figure}[bt]
\centering
  \begin{minipage}[]{.48\textwidth}
     \centering
     \includegraphics[width=.95\textwidth]{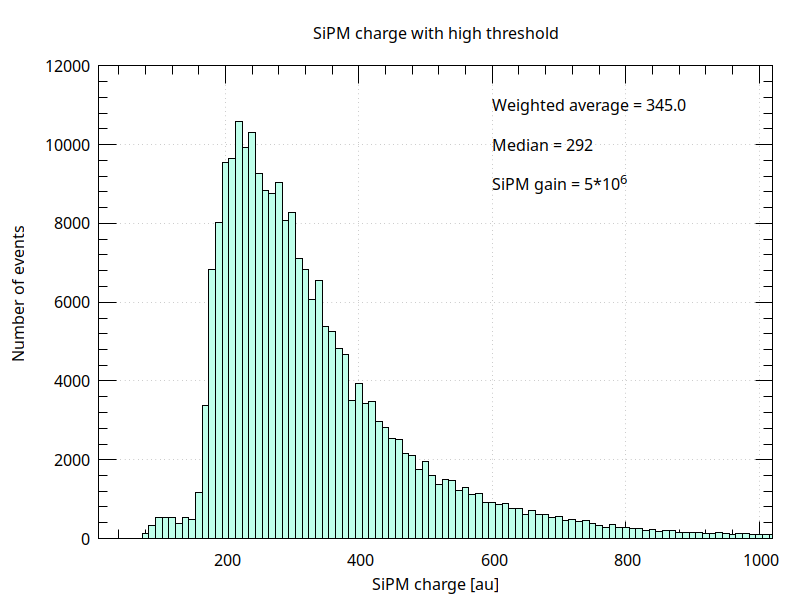}
  \end{minipage} 
  \hfill
  \begin{minipage}[]{.48\textwidth}
     \centering
     \includegraphics[width=.95\textwidth]{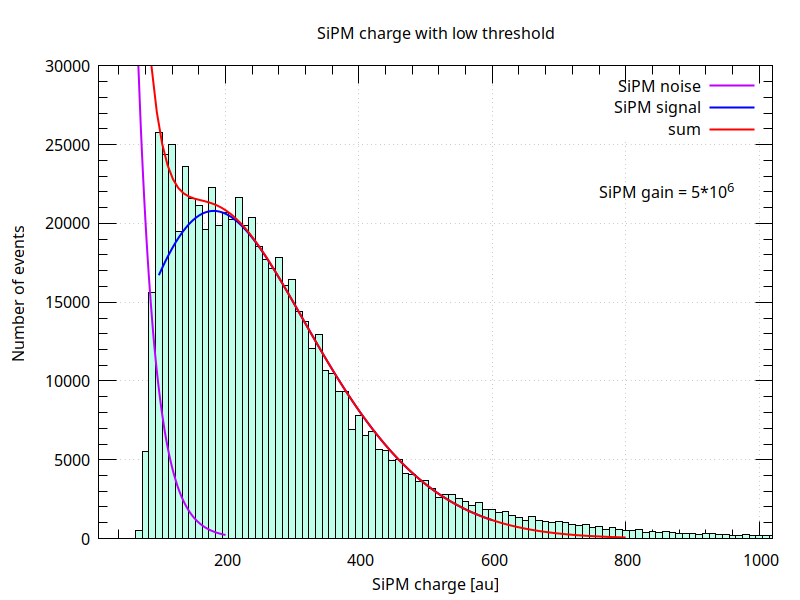}
  \end{minipage} 
\caption{Cosmic rays charge for a SiPM gain of $5 \times 10^6$: 
         Left: High Threshold; Right: Low Threshold} 
\label{Fig:CosmicCharge_g5}
\end{figure}

Using the recorded data, the histogram of the charge released by each 
cosmic ray was made for high and low threshold, as shown in 
Fig.~\ref{Fig:CosmicCharge_g5}.

In the right figure, you can clearly see the two components of the signal: 
1) SiPM electronic noise (in purple); 2) SiPM signal (in blue).

In the left figure, it can be seen that the electronic noise is cancelled 
by raising the threshold.

\begin{figure}[bt]
\centering
  \begin{minipage}[]{.48\textwidth}
     \centering
     \includegraphics[width=.95\textwidth]{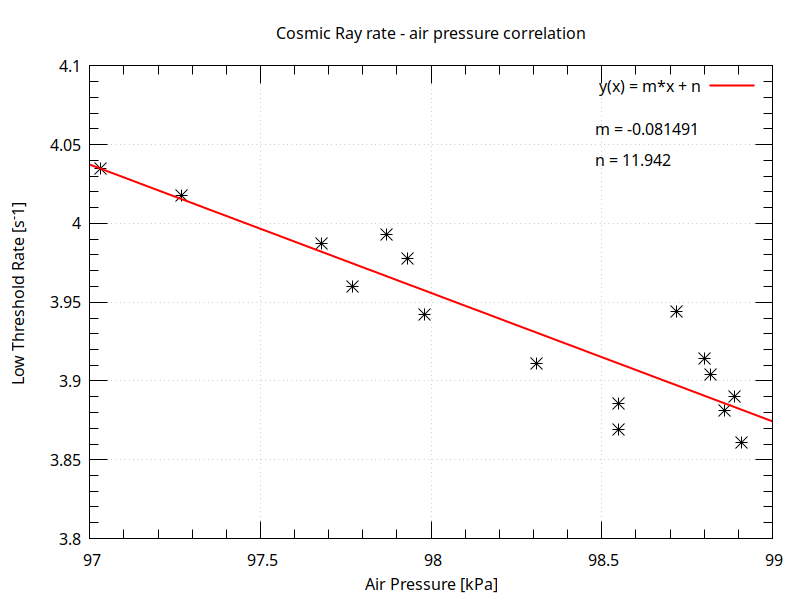}
  \end{minipage} 
  \hfill
  \begin{minipage}[]{.48\textwidth}
     \centering
     \includegraphics[width=.95\textwidth]{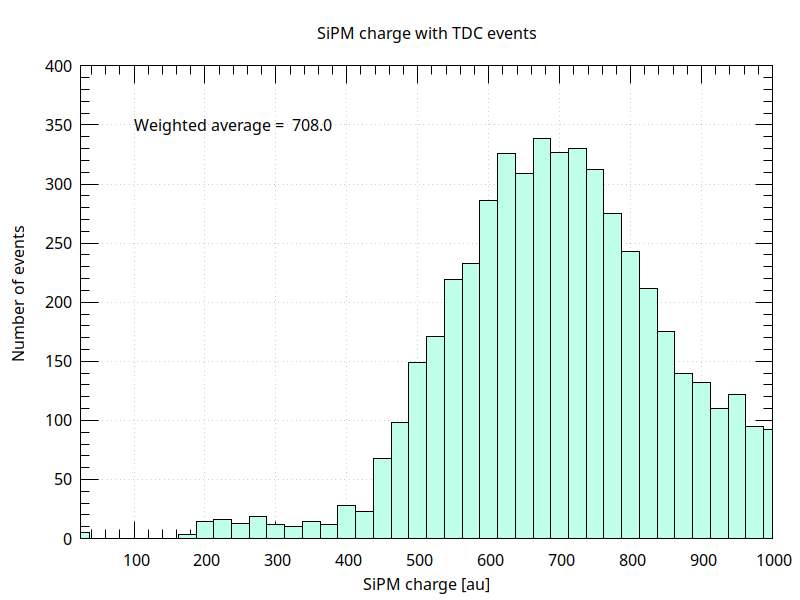}
  \end{minipage} 
\caption{Left: Rate of cosmic rays VS. air pressure;
          Right: SiPM charge for TDC events} 
\label{Fig:TimePressCharge}
\end{figure}

It is interesting to verify experimentally the dependence of the rate 
of cosmic rays on atmospheric pressure, as shown in the histogram in 
Fig.~\ref{Fig:TimePressCharge} left. 
The correlation between cosmic ray frequency and atmospheric pressure is
clearly visible. 
As atmospheric pressure increases, the frequency of cosmic rays decreases.

It is then possible to make the histogram of the times between the arrival 
of one cosmic ray and the following, Fig.~\ref{Fig:TimePressCharge} the left.

This measure has already been discussed theoretically in 
Chap.~\ref{Sec:TimeArrival}.

From this histogram one can deduce the frequency of cosmic rays as an alternative to the simple division $R = N_{cosmic} / T_{time}$.

The frequency from the graph appears to be: $R = 1 / \tau $.

In Fig.~\ref{Fig:TimePressCharge} right, the SiPM charge is shown for 
only those events that provided a signal from the TDC, that is, 
with a double signal in the gate time.
Comparing with Fig.~\ref{Fig:CosmicCharge_g5} it is evident that 
for these events the SiPM signal is much larger than that of the 
single muons.
In Chap.~\ref{Sec:MuonDecay} this phenomenon was discussed.

At the conclusion of all consistency checks, you can proceed with the 
analysis of the TDC data.

Tab.~\ref{Tab:DataSummary} shows the average measured values of a some paramenters.
It is interesting to compare the rate of cosmic muons interacting with the 
detector, about 4 $s^{-1}$, with those giving a TDC interrupt, START, and STOP 
signal in gate time, which is about 1 every 2 minutes (30 per hour).

\begin{figure} [hbt]
  \centering
  \includegraphics[width=.9\textwidth]{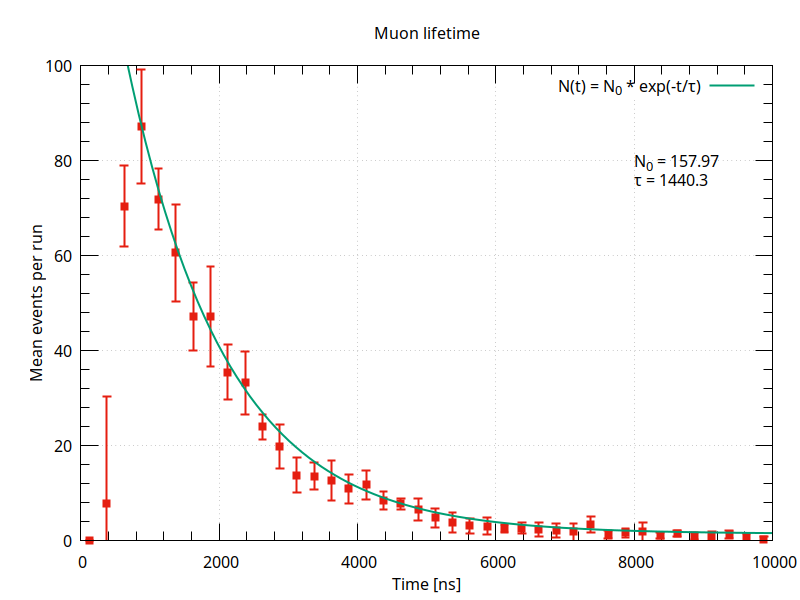}
  \caption{TDC time}
  \label{Fig:TDC_g5}
\end{figure}

For each run, a histogram was made of the timing provided by the TDC between 
0 and 10 $\mu$s.
The mean value and standard deviation were calculated for each bin. 
The result is plotted in Fig.~\ref{Fig:TDC_g5}.

The histogram clearly shows the negative exponential trend,
typical of particle decays. 
The data were fit using a function shown in Eq.~\ref{Eq:MuonLifeFormula},
as shown in Chap.~\ref{Sec:physmuon}:

\begin{equation}
  N(t) = N_0 \exp^{-t/\tau}
\end{equation}

from 0.6 to 10 $\mu$s.
Below 500 ns, the TDC data are unreliable both because of the overlapping 
widths of the START and STOP signals and because the TDC7200 is used in 
"Mode 2," between 250 ns and 8 ms, while "Mode 1" is usable between 12 
and 500 ns.

The decay time given by the fit is 1.406 $\mu$s, far from the official 
muon decay value of 2.197 $\mu$s, as indicated in Eq.~\ref{Eq:MuonLifeTime}.

This means that some physical phenomena stated in Chap.~\ref{Sec:MuDecAbsorp}
and Chap.~\ref{Sec:TimeArrival} were not taken into account.

\subsection{Background \& background subtraction.}
As discussed in previous chapters on muon physics, 
there may be physical contributions that alter the lifetime measurement even significantly.
In particular:

\begin{itemize}
\setlength\itemsep{0.1em}
 \item ``fake events'';
 \item $\mu^-$ absorption in nuclear matter.
\end{itemize}

\subsubsection{Fake events.}
The time distribution of cosmic muons is flat. This means that the 
probability of having two cosmic muons in gate time is not zero and 
varies with the frequency of them.

``Fake events'' are defined as events that simulate decay but due to the 
subsequent arrival of two cosmic muons in the gate time = 12 $\mu$s.

Unfortunately, as shown in Chap.~\ref{Sec:TimeArrival}, the trend of these 
events is exponential as the decay of muon end can be easily confused 
with the lifetime events.

This correction is, however, small. 
If we look at the events above $3 \sigma = 6.6 \mu$s the muon lifetime,
where the probability to have a real muon decay is only $0.3\%$, and,
therefore we expect to have only ``fake-events'', 
from Fig.~\ref{Fig:TDC_g5} left, we see that this number of events is 
negligible.

However, this correction was taken into account by using a fit with two 
time coefficients:

\begin{equation}
  f(t) =  D \times e^{-t/T} + n \times e^{-t/\tau}
\end{equation}

Where the coefficients D and T are taken from Fig.~\ref{Fig:TimePressCharge} left.

\begin{figure}[hbt]
\centering
  \begin{minipage}[]{.48\textwidth}
     \centering
     \includegraphics[width=.95\textwidth]{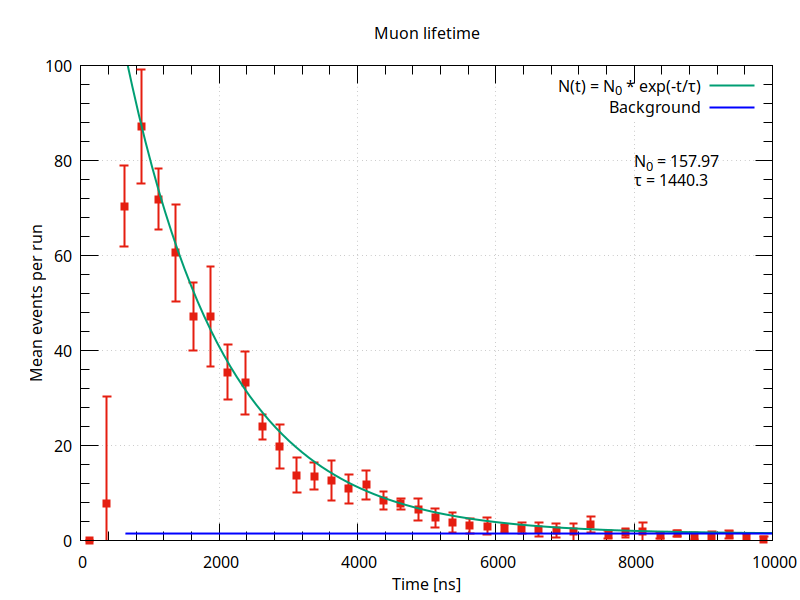}
  \end{minipage} 
  \hfill
  \begin{minipage}[]{.48\textwidth}
     \centering
     \includegraphics[width=.95\textwidth]{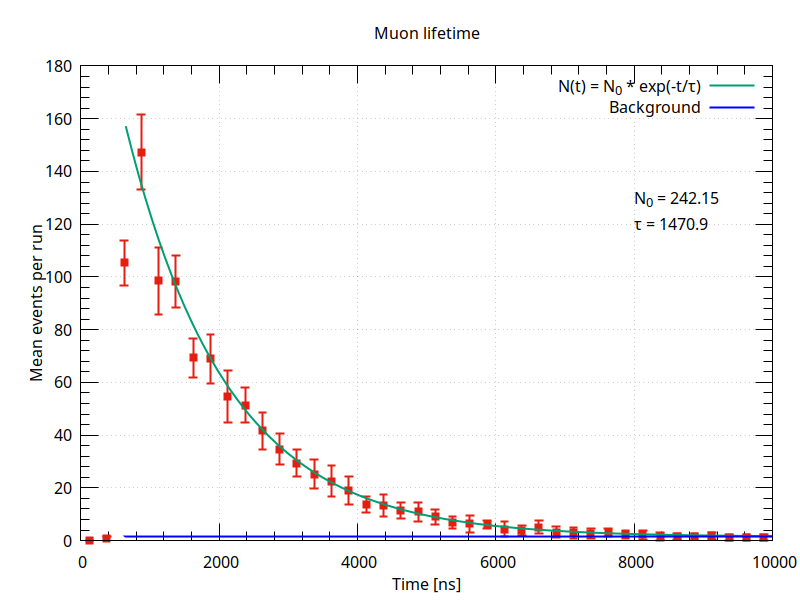}
  \end{minipage} 
\caption{Left: TDC fit with fake events for SiPM gain $G = 5 \times 10^6$;
         Right: the same bat for $G = 6 \times 10^6$.} 
\label{Fig:TDCfit}
\end{figure}

The result is shown in Fig.~\ref{Fig:TDCfit} at left.
The blue line is the fake events correction.

``Fake-events'' could be eliminated by surrounding the cube with scintillation
detectors in anti-coincidence.
Cosmic rays, in fact, originate from outside instead of inside the cube.
This would lead to detector complication, increased cost, and complicated
operation, which cannot be justified by the small measured contribution.

It was chosen, therefore, to use the technique of background subtraction
too save on hardware cost and complication, and to allow the students 
to ``see''- and understand the random background.

But, it is evident from Fig.~\ref{Fig:TDCfit}, that such this small noise 
rate is statistically insignificant for background considerations.
Then this contribution will be ignored in subsequent analyses.

\subsubsection{Negative Muons absorption in nuclear matter.}

The configuration of the AMELIE detector does not include the possibility 
of separating $\mu^-$ from $\mu^+$, a possibility offered only by complex 
and expensive experiments on accelerator machines or balloons or satellites.

As discussed in Chap.~\ref{Sec:MuDecAbsorp}, in nuclear matter, the behavior
of $\mu^-$ differs from that of $\mu^+$.
In particular, it should be kept in mind that $\mu^-$ tend to replace 
electrons in atomic orbitals and be absorbed by the nucleus.

Not only that, it must be taken into account that, at the ground level, 
the number of $\mu^-$ differs from that of $\mu^+$:

\begin{center}
   $R ( \mu^+ / \mu^-) = 1.30 \pm 0.05 $ at 0.86 GeV/c \\[2mm]
   $R ( \mu^+ / \mu^-) = 1.27 \pm 0.01 $ at 0.52 GeV/c
\end{center}

See Chap.~\ref{Sec:muoninmatter} and Fig.~\ref{Fig:mu+mu-}.
and also~\cite{Vulpescu}.

The competition between absorption and decay processes, reduce the meanlife
of the obseved decay of the negative muon as compared with the free decay.

Expressed in terms of process rates $\Lambda = 1/\tau$, 
the total disappearance probability is written as the
sum of the capture and of the decay probabilities:

\begin{center}
  $\Lambda_{total} = \Lambda_{capture} + \Lambda_{decay}$ \\[2mm]
  $\Lambda_{decay} = Q \times 1 / \tau_{\mu^+}$           \\[2mm]
  $\Lambda_{total}^{-1} = \tau_{\mu^{-}} < \; \tau_{\mu^{+}} = 2.197703 \; \mu s$
\end{center}

Using the different lifetimes of muonic atoms compared with the lifetime of
the positive muon, it is possible to measure the ratio of positive to negative
muons in a flux, without using a magnetic spectrometer.

In general, if a device detects decay electrons coming from muons stopped in 
different materials, the total decay curve will be a superposition of several 
decay laws:

\begin{equation}
 \frac{dN}{dt} = N_+ c_0 \frac{1}{\tau_0} e^{- \frac{t}{\tau_0}} +
 N_- \sum_{j=1}^{m} c_j \frac{1}{\tau_j} e^{- \frac{t}{\tau_j}}
\end{equation}

where:

\begin{table} [!ht]
 \begin{tabular}{m{1.5cm} m{12cm}}
    $N_+$ , $N_-$ & - number of $\mu^+$, $\mu^-$ which enter the detector       \\
    m             & - number of materials in the detector                       \\
    $c_0$         & - detection efficiency for $\mu^+$ stopped in all materials \\
    $c_j$         & - detection efficiency for $\mu^-$ stopped in material      \\
    $\tau_0$      & - mean lifetime of $\mu^+$ (2.19703 $\mu$s)                 \\
    $\tau_j$      & - mean lifetime of $\mu^-$ in material j
 \end{tabular} 
\end{table}

This method is also called ``delayed coincidences'' from historical reasons.

Taking into account the composition of the plastic scintillator, (H-11, C-10),
(Chap.~\ref{Sec:ScintDetector}), the above formula become:

\begin{equation}
 \frac{dN}{dt} = \frac{N_0}{R + 1} \left[ R c_0 \frac{1}{\tau_0} e^{- \frac{t}{\tau_0}}  +  c_h \frac{1}{\tau_h} e^{- \frac{t}{\tau_h}}
 +  c_c \frac{1}{\tau_c} e^{- \frac{t}{\tau_c}} \right]
 \label{Eq:MuInMaterial}
\end{equation}

where:

\begin{table} [!ht]
 \begin{tabular}{m{1.5cm} m{12cm}}
    $N_0$         & = $N_+ + N_-$                          \\
    $R(\mu^+ / \mu^-)$ & = $N_+ / N_- \approx 1.27$        \\
    $c_h$         & = detection efficiency for $\mu^+$ stopped in hydrogen \\
    $\tau_h$      & = 2.19490 $\mu$s (mean lifetime of $\mu^+$ in hydrogen) \\             
    $c_c$         & = detection efficiency for $\mu^+$ stopped in carbon \\
    $\tau_c$      & = 2.028 $\mu$s (mean lifetime of $\mu^+$ in carbon)               
 \end{tabular} 
\end{table}

In this material, at the absorption of $\mu^-$ shown in Eq.~\ref{Eq:mu-capture}
on hydrogen, we must add the absorption on carbon: 

\begin{equation}
  \mu^- + \: ^{12}C \rightarrow \nu_\mu + \: ^{13}B^{*}
  \label{Eq:mu-capture-C}
\end{equation}

where $^{13}B^{*}$ has at unknown excitation energy.
The ``mean'' excitation energy in the nuclear muon capture is around 
15 to 20 MeV.
This is well above the nucleon emission threshold in all complex nuclei. 
Thereby, the neutron emission is the preferred channel, compared to other channels 
like: emission of charged particles, gamma rays or delayed fission.

Using the formula [\ref{Eq:MuInMaterial}] and keeping the value of the average 
$\mu^+$ life fixed at the value [\ref{Eq:MuonLifeTime}], using the data acquired 
in the $5 \times 10^6$ gain runs, the average lives of $\mu^-$ in the scintillator 
can be determined, particularly on hydrogen and carbon, as shown in 
Fig.~\ref{Fig:TDC_g5}.

\begin{figure} [hbt]
  \centering
  \includegraphics[width=.9\textwidth]{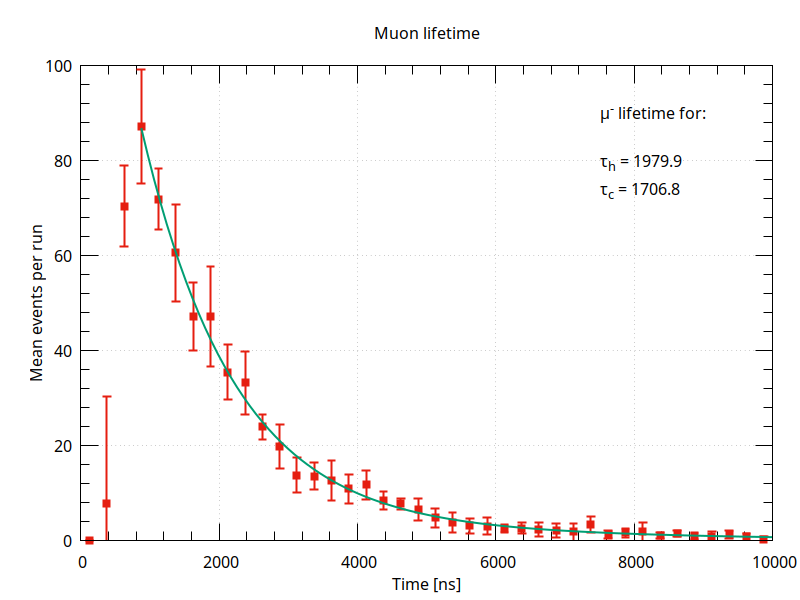}
  \caption{$\mu^-$ and $\mu^+$ mean lifetime in plastic scintillator, 
           consisting primarily of hydrogen and carbon.
           20 runs of 24h each, with SiPM gain $5 \times 10^6$.}
  \label{Fig:TDC_g5_scint}
\end{figure}

Statistics is a very important parameter for improving any measurement.
In this analysis, 24h runs were made but, given the stability of the system, 
taking into account that muons stopping in the cube are about one every 
two minutes, much longer time runs can also be implemented.

\subsection{Data analysis for SiPM gain $G = 6 \times 10^6$.}

The same procedure used for the analysis of data acquired with a SiPM gain of 
$5 \times 10^6$, shown in the previous chapter, 
was used by increasing the gain to $6 \times 10^6$.

\begin{figure}[hbt]
\centering
  \begin{minipage}[]{.32\textwidth}
     \centering
     \includegraphics[width=.98\textwidth]{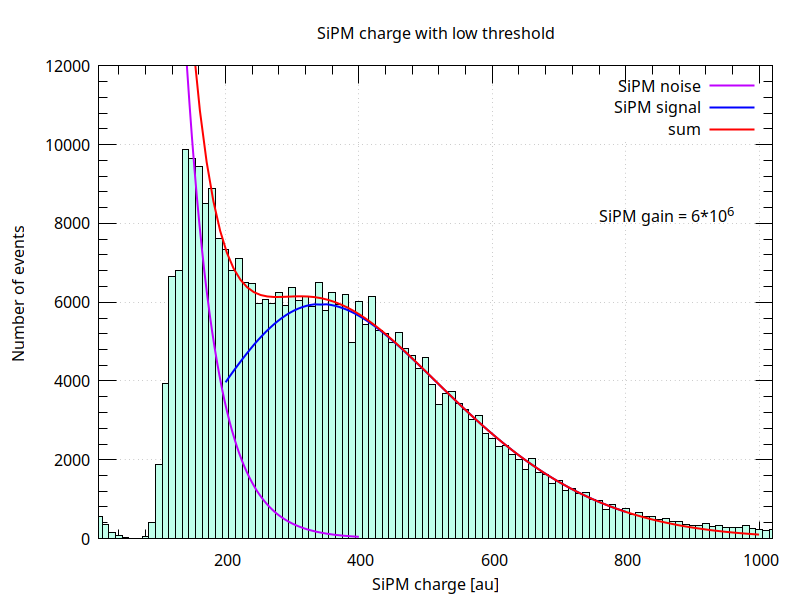}
  \end{minipage} 
  \hfill
  \begin{minipage}[]{.32\textwidth}
     \centering
     \includegraphics[width=.98\textwidth]{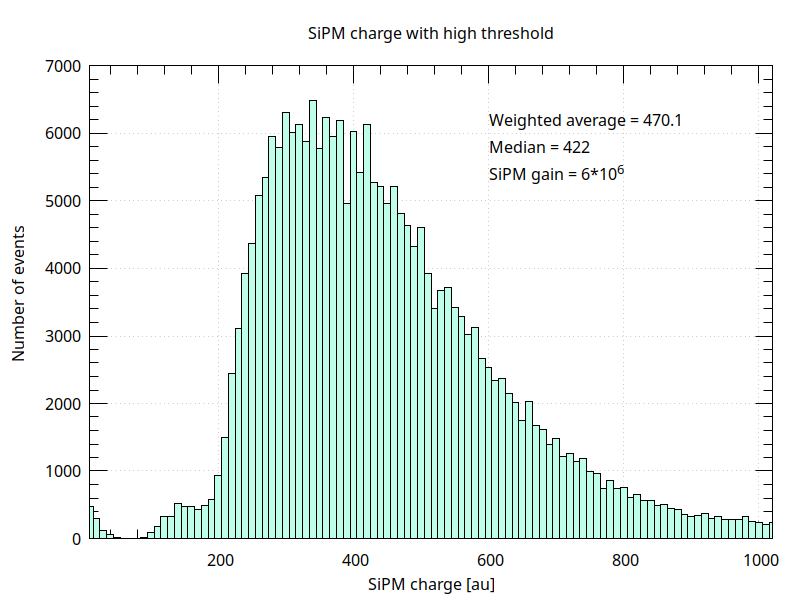}
  \end{minipage} 
  \hfill
  \begin{minipage}[]{.32\textwidth}
     \centering
     \includegraphics[width=.98\textwidth]{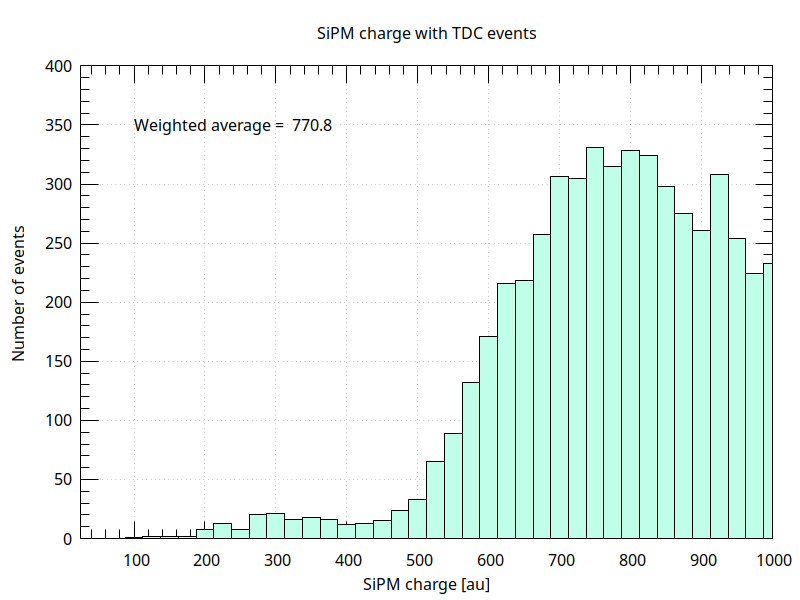}
  \end{minipage} 
\caption{Left: SiPM charge signal for low threshold; 
         Center; SiPM charge signal for high threshold;
         Right: SiPM charge signal for TDC events. SiPM gain $G = 6 \times 10^6$} 
\label{Fig:SiPMcharge_g6}
\end{figure}

Fig.~\ref{Fig:SiPMcharge_g6} shows the SiPM charge signal at low threshold, 
high threshold and for events that gave a TDC signal.

In Fig.~\ref{Fig:TDCfit} , the muon decay times for different SiPM 
gains are shown.

It is evident that the exponential curvature given by the $\tau$ coefficient, 
does not depend on the SiPM gain.
This means that the measurement is not affected by systematic errors dependent 
on the operating parameters of the system but only on the times provided by 
the TDC, already calibrated and tested as described in Chap.~\ref{Cap:TDCtest}.

\begin{figure} [hbt]
  \centering
  \includegraphics[width=.9\textwidth]{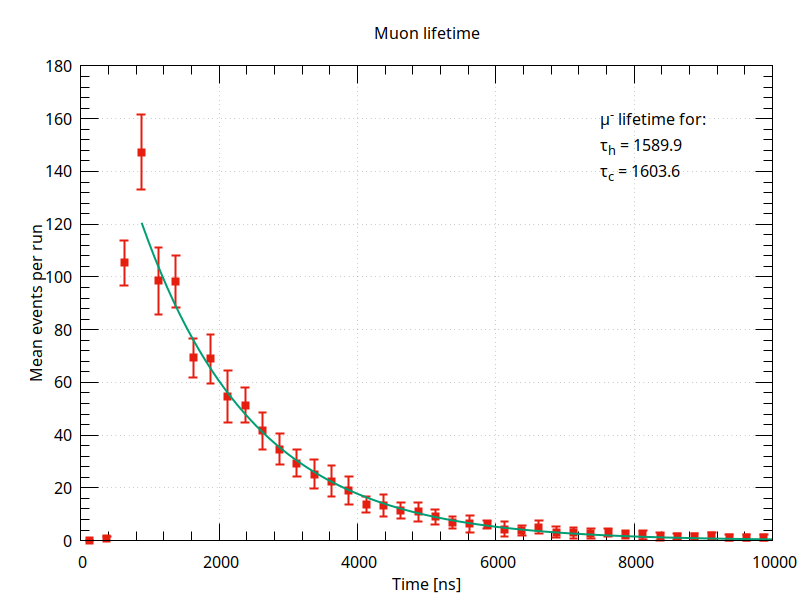}
  \caption{$\mu^-$ and $\mu^+$ mean lifetime in plastic scintillator, 
           consisting primarily of hydrogen and carbon.
           20 runs of 24h each, with SiPM gain $6 \times 10^6$.}
  \label{Fig:TDC_g6_scint}
\end{figure}

Fig.~\ref{Fig:TDC_g6_scint} shows the average muon lifetime measured with a SiPM 
gain of $6 \times 10^6$. 
The same figure shows the fit made with the formula [\ref{Eq:MuInMaterial}].

The muon lifetimes in hydrogen and carbon are similar to those shown 
in Fig.~\ref{Cap:TDCtest} with a gain of the SiPM of $5 \times 10^6$.

\subsection{Data analysis for SiPM gain $G = 7 \times 10^6$.}
In order to verify the stability of the system and the consistency of the collected data, 
a third round of 20 runs was made with a SiPM gain of $G = 7 \times 10^6$.

\begin{figure}[hbt]
\centering
  \begin{minipage}[]{.32\textwidth}
     \centering
     \includegraphics[width=.98\textwidth]{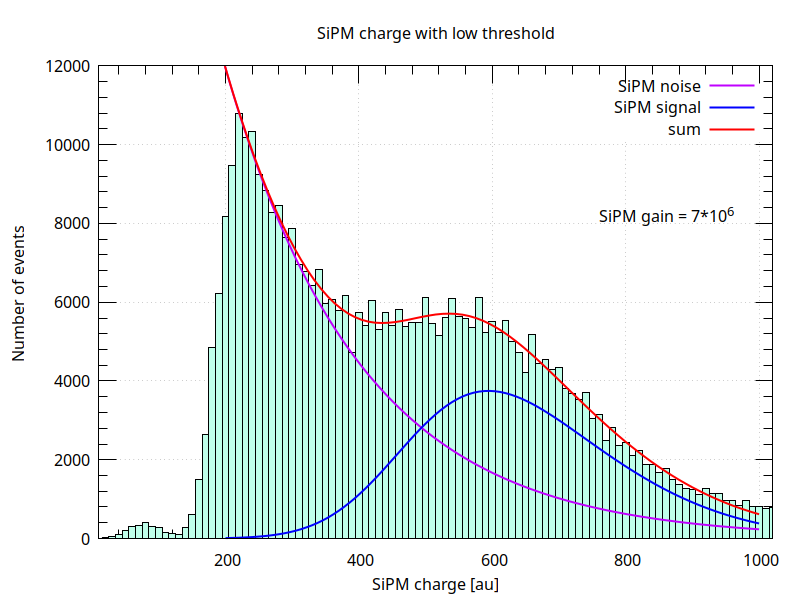}
  \end{minipage} 
  \hfill
  \begin{minipage}[]{.32\textwidth}
     \centering
     \includegraphics[width=.98\textwidth]{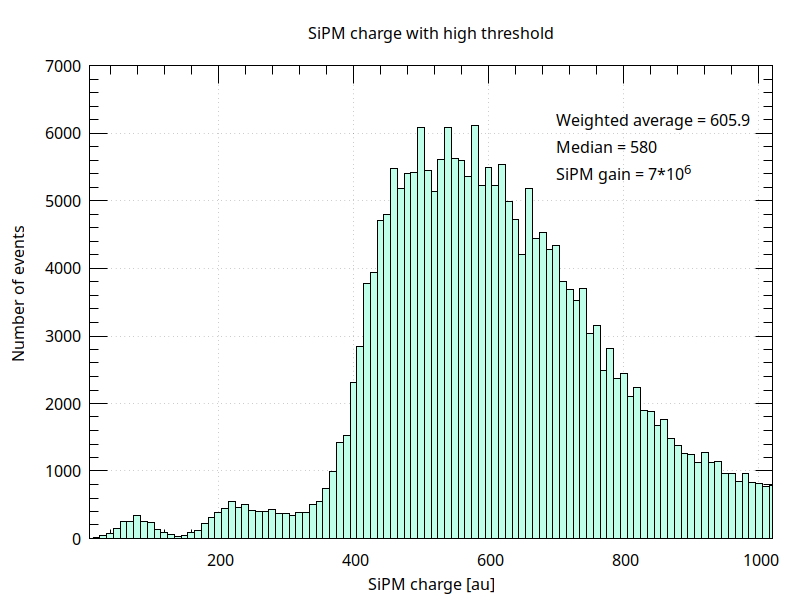}
  \end{minipage} 
  \hfill
  \begin{minipage}[]{.32\textwidth}
     \centering
     \includegraphics[width=.98\textwidth]{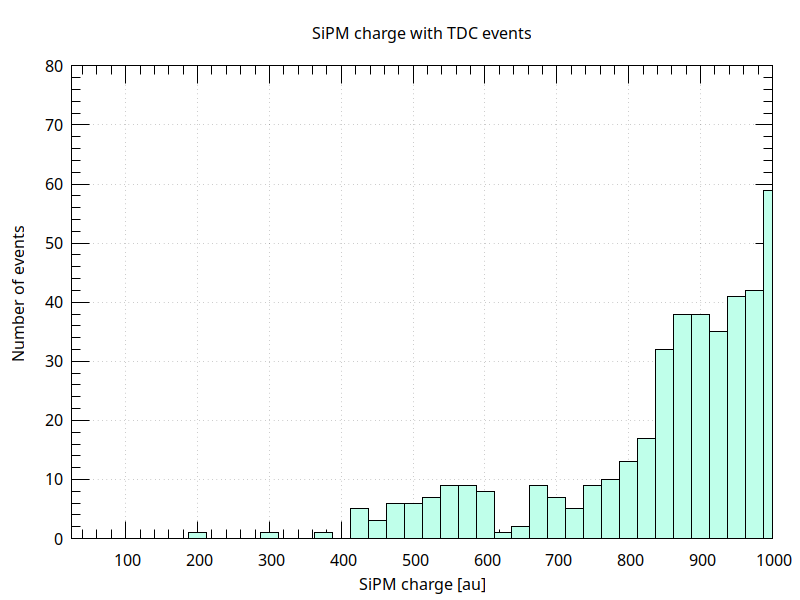}
  \end{minipage} 
\caption{Left: SiPM charge signal for low threshold; 
         Center; SiPM charge signal for high threshold;
         Right: SiPM charge signal for TDC events. SiPM gain $G = 7 \times 10^6$.} 
\label{Fig:SiPMcharge_g7}
\end{figure}

Fig.~\ref{Fig:SiPMcharge_g7} shows the SiPM charge signal at low threshold, 
high threshold and for events that gave a TDC signal.

\begin{figure}[hbt]
\centering
  \begin{minipage}[]{.48\textwidth}
     \centering
     \includegraphics[width=.95\textwidth]{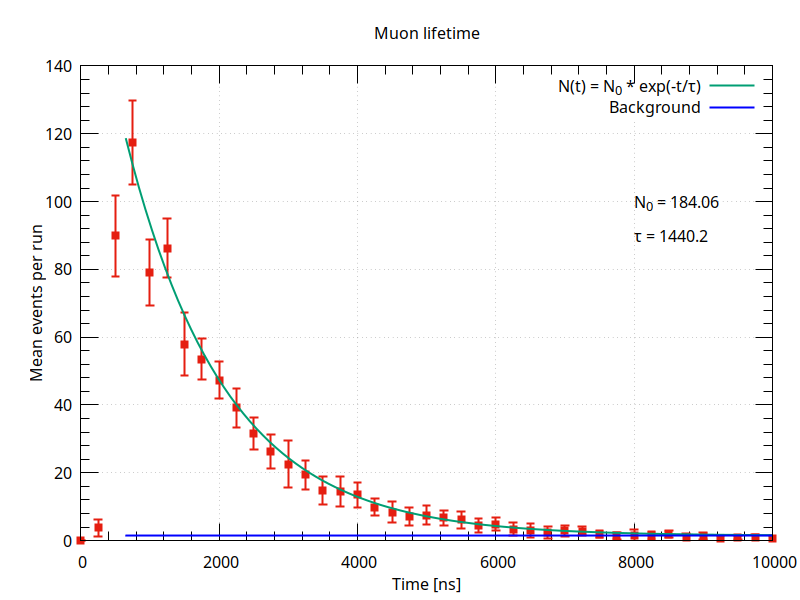}
  \end{minipage} 
  \hfill
  \begin{minipage}[]{.48\textwidth}
     \centering
     \includegraphics[width=.95\textwidth]{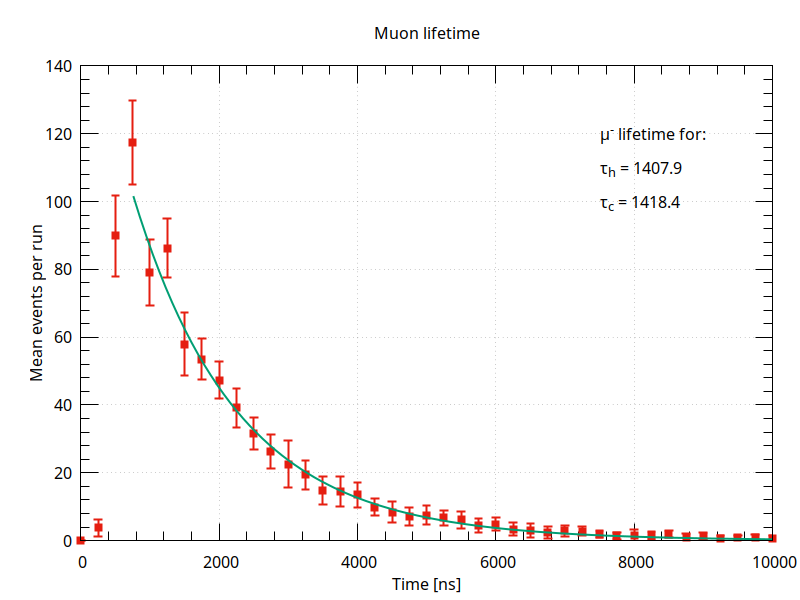}
  \end{minipage} 
\caption{Left: TDC fit with fake events for SiPM gain;
         Right: $\mu^-$ and $\mu^+$ mean lifetime in plastic scintillator.
         SiPM gain $G = 7 \times 10^6$.} 
\label{Fig:TDCfit_g7}
\end{figure}

Conversely, Fig.~\ref{Fig:TDCfit_g7} shows the time measured by the TDC , raw data on the 
left and with the time measurement in Carbon and hydrogen on the right.

\subsection{Summary of the three rounds of data taking.}
As mentioned above, three cycles of data taking were done at three different values of the SiPM
gain $5 \:, 6 \:, 7 \times 10^6$.
For each cycle, 20 runs of 24 hours each were made.

Tab.~\ref{Tab:DataSummary} summarizes all relevant parameters of each cycle.

\begin{table}[hbt]
  \centering
  \small
    \begin{tabular}{ |l|c|c|c|} \hline
        & \multicolumn{3}{c|}{SiPM gain}                          \\ \cline{2-4}
        & $5 \times 10^6$  & $6 \times 10^6$ & $7 \times 10^6$    \\ \hline
      Average n. of low thresh. ev. per run  & 343350 $\pm$ 4716 & 392222 $\pm$ 5134  & 433560 $\pm$ 3580 \\
      Average n. of high thresh. ev. per run & 249759 $\pm$ 4439 & 276409 $\pm$ 2360  & 253269 $\pm$ 3121\\
      Average n. of TDC events per run          & 632 $\pm$ 31      & 990 $\pm$ 42    & 804 $\pm$ 22 \\
      Low Thresh. rate  [$s^{-1}$]              & 3.97 $\pm$ 0.05   & 4.49 $\pm$ 0.06 & 4.97 $\pm$ 0.03 \\
      High Thresh. rate [$s^{-1}$]              & 2.79 $\pm$ 0.05   & 3.16 $\pm$ 0.03 & 2.90 $\pm$ 0.03 \\
      $\mu$ decay rate  [$min^{-1}$]            & 0.43 $\pm$ 0.02   & 0.68 $\pm$ 0.03 & 0.55 $\pm$ 0.02 \\
      $\Delta$t between two siccessive events [$s^{-1}$] & 0.252 $\pm$ 0.003 & 0.223 $\pm$ 0.003 & 0.201 $\pm$ 0.001 \\
      row TDC data decay constant $\tau$ [ns]   & 1440.3 $\pm$ 30.3 & 1470.9 $\pm$ 48.2 & 1440.2 $\pm$ 48.0 \\ 
      $\mu^-$ lifetime in hydrogen [ns]       & 1979.9 & 1589.9 & 1407.9 \\ 
      $\mu^-$ lifetime in carbon [ns]         & 1706.8 & 1603.6 & 1418.4 \\ 
      SiPM charge median for TDC ev. [au] & 708.0  &  770.8 & \\ \hline
   \end{tabular}
\caption{Summary of the three data tacking rounds at different SiPM gains.}
\label{Tab:DataSummary}
\end{table}

From the Tab.~\ref{Tab:DataSummary}, it can be stated that, while the decay constant 
of the raw data from the three rounds lies in the statistical error, the convergence fit of
negative muon lifetime in hydrogen and carbon is not stable and, then, a much higher statistics
is required.

The frequency of TDC events is affected by the environment in which the detector is placed. 
For example, the frequency increases greatly if the detector is very close to thick concrete 
walls that tend to slow down cosmic muons, instead of in the center of the room.

\section{Conclusions.}
\label{Sec:conclusions}

An innovative detector has been designed.
Its main purpose is to measure the average lifetime 
of muons and to study the absorption of negative muons in nuclear matter.
The detector is designed for high school students but can also be used in 
university physics laboratories.

The idea of muons lifetime measurement with stoppend cosmic muons is not new. 
Many setups of this type have been successfully implemented in the physics laboratories of many universities around the world.
They require, however, expensive detectors, photomultipliers using high voltages 
of 1,000 volts or more, modular NIM electronics, expensive and complex to handle.

AMELIE is innovative compared to the others setup because it use SiPM 
optical detectors instead of the usual photomultipliers. 
This avoids having high voltages generators, cables, connectors, and offers an additional 
safety margin for students.
In addition, use a Time to Digital Converter (TDC) that are to state of the art 
of this type of converter.

AMELIE is a detector that is compact, easy to use and records data for off-line 
data analysis in universal .csv format.
Is a complete system, requires no additional equipment other than common PCs with 
free Open Source programs like ``Libre Office'' or ``gnuplot''.

Taking into account the AMELIE simplicity, low cost and educational purpose of 
this apparatus, the results obtained do not differ much from the values measured
with complex and expensive experimental apparatus shown in 
Tab.~\ref{Tab:MatLifeTime}.
It should be kept in mind that the purpose of the apparatus is not to make 
publishable measurements in peer-reviewed scientific journals, measurements 
widely available in the literature, but educational and didactic.

The detector is very stable, reliable and the measurement does not appear to have
systematic errors due to the particular setting of functional parameters
or relevant background that could invalidate the measures.

In addition, the detector uses cosmic rays which are present all over the world 
24h/7d, free of charge and consistent over time.

It allows professors to introduce modern physics, the world of elementary particles with simple, didactic but very formative measurements, data analysis and data
correction. 
Other accessible fields are statistics, data analysis, corrections and counting over time.

\bibliographystyle{unsrt}
\bibliography{AmelieBib.bib}

\end{document}